\documentclass[aps,twocolumn,amsmath,amssymb,preprintnumbers]{revtex4}
\usepackage{amsmath} \usepackage{amsfonts} \usepackage{amssymb}
\usepackage{titlesec}
\usepackage{bbm}
\usepackage{epsfig}
\usepackage{graphics}
\usepackage{graphicx}
\newcommand{\be}{\begin{equation}}
\newcommand{\ee}{\end{equation}}
\newcommand{\ba}{\begin{eqnarray}}
\newcommand{\ea}{\end{eqnarray}}
\newcommand{\nn}{\nonumber}
\newcommand{\kl}{\langle}
\newcommand{\kr}{\rangle}
\newcommand{\h}{{\cal H}}
\newcommand{\g}{\tilde G}
\newcommand{\kk}{(k,\eta,\eta')}
\newcommand{\e}{e^{-\bar ek(\eta-\eta')}}
\newcommand{\w}{w^\pm_k}
\newcommand{\ww}{w^+_k}
\newcommand{\www}{w^-_k}
\newcommand{\et}{(\eta)}
\newcommand{\y}{(y)}

\newcommand{\kt}{(k,\eta)}
\newcommand{\ev}{(\eta,\vec k)}
\newcommand{\eev}{(\eta,\vec k')}
\newcommand{\va}{_{\varphi\varphi}}
\newcommand{\C}{{\cal C}}
\newcommand{\kkl}{(k)}

\titleformat{\subsection}[block]{\normalfont\bfseries}{\thesubsection.}{1ex}{}
\titlespacing{\subsection}{0pt}{10pt}{1pt}[0pt]
\titleformat*{\section}{\large\bfseries}
\renewcommand{\thesubsection}{\arabic{subsection}}

\begin{document}

\title[ ]{Cosmic fluctuations from quantum effective action}

\author{C. Wetterich}
\affiliation{Institut  f\"ur Theoretische Physik\\
Universit\"at Heidelberg\\
Philosophenweg 16, D-69120 Heidelberg}

\begin{abstract}
Does the observable spectrum of cosmic fluctuations depend on detailed initial conditions? This addresses the question if the general inflationary paradigm is sufficient to predict within a given model the spectrum and amplitude of cosmic fluctuations, or if additional particular assumptions about the initial conditions are needed. The answer depends on the number of e-foldings $N_{in}$ between the beginning of inflation and horizon crossing of the observable fluctuations. We discuss an interacting inflaton field in an arbitrary homogeneous and isotropic geometry, employing the quantum effective action $\Gamma$. An exact time evolution equation for the correlation function involves the second functional derivative $\Gamma^{(2)}$. The operator formalism and quantum vacua for interacting fields are not needed. Use of the effective action also allows one to address the change of frames by field transformations (field relativity). Within the approximation of a derivative expansion for the effective action we find the most general solution for the correlation function, including mixed quantum states. For not too large $N_{in}$ the memory of the initial conditions is preserved. In this case the cosmic microwave background cannot disentangle between the initial spectrum and its  processing at horizon crossing. The inflaton potential cannot be reconstructed without assumptions about the initial state of the universe. We argue that for very large $N_{in}$ a universal scaling form of the correlation function is reached for the range of observable modes. This can be due to symmetrization and equilibration effects, not yet contained in our approximation, which drive the short distance tail of the correlation function towards the Lorentz invariant propagator in flat space.   
\end{abstract}

\maketitle
\section{Introduction}

Are amplitude and spectrum of the primordial cosmic fluctuations uniquely determined by the inflaton potential or is there some influence of the initial state of the universe? Is there a good reason to assume a quantum vacuum state during inflation and is this unique? Is it necessary to postulate or understand particular properties of the initial conditions in order to explain the simple successful picture of the almost scale invariant spectrum? Or could it be that the dynamics is such that all memory of the correlation function about the initial state is lost, such that the two point function during inflation is essentially uniquely fixed for a given solution of the field equations? Answers to these questions are crucial  in order to assess if the inflationary paradigm by itself is sufficient for a unique prediction of the properties of primordial fluctuations within a given model, or if additional ingredients are needed for their understanding.

Conceptually, it is rather obvious that the fluctuation spectrum depends in some way on the initial state. The question is how strongly. We need to understand quantitatively how much of the initial information survives at the time when the relevant primordial fluctuations go out of the horizon. This general type of problem is addressed within non-equilibrium quantum field theory. We undertake here first steps within the conceptual framework of the quantum effective action. This allows us to discuss interacting fields and to go beyond investigations \cite{VF,AEH,BM1,BM2,BMR,AHR,DGPR} of free fields in a pure quantum state. Interactions are crucial for possible symmetrization and equilibration processes.

The quantum effective action $\Gamma$ for gravity, coupled to scalars and other fields, is the basis for the cosmological field equations. Indeed, the field equations derived by variation of $\Gamma$ are exact, in contrast to those derived from the ``classical action'', which need to be corrected by fluctuation effects. The issue of quantum gravity can be viewed as the problem of computing the quantum effective action. String theory or the functional renormalization group \cite{CW1}, \cite{CWMR}, \cite{MR} among others, have made important progress in this direction. 

The effective action provides essential information beyond the exact field equations for the mean values of fields. It describes all aspects of fluctuations around the mean field as well. The second functional derivative $\Gamma^{(2)}$ of the effective action constitutes the exact inverse propagator. If it is invertible within a suitable space of field functions, the inversion of $\Gamma^{(2)}$ provides for a unique connected two-point correlation function (propagator, Green`s function). This relation is exact. In a cosmological setting the equal time Green's function comprises the full information about the power spectrum of the cosmic fluctuations. Furthermore, the third functional derivative $\Gamma^{(3)}$ of the effective action is the one-particle irreducible three point function, from which the bispectrum can be extracted. Indeed, the connected three point function (bispectrum) obtains from $\Gamma^{(3)}$ by multiplication with three propagators and amounts therefore to a tree calculation. All loop effects are already incorporated in the computation of $\Gamma$. Higher order correlations can be obtained from higher functional derivatives of $\Gamma$.

Once $\Gamma$ is computed or assumed, the propagator is determined uniquely only if $\Gamma^{(2)}$ is invertible. For massless fields invertibility typically requires a regularization. This may be provided by a small mass term or, more generally, by an effective infrared cutoff as, for example, within the effective average action \cite{CW1}. Special care is needed in case of local continuous symmetries. Typically, local symmetries need some gauge fixing procedure. For non-compact geometries or spaces with boundary invertibility needs a specification of boundary conditions. Once the issue of invertibility of $\Gamma^{(2)}$ is taken care of, its relation to the propagator $G$ is an exact identity,
\be\label{ABA}
\Gamma^{(2)}G=1.
\ee
In this work we will assume a given form of the effective action and compute the propagator by solving eq. \eqref{ABA}. This equation constitutes a differential equation for the time evolution of the Green's function which is solved as an initial value problem. All our computations are uniquely based on the solution of eq. \eqref{ABA}, no other input is employed. 

We emphasize that no further corrections of any kind are needed once $\Gamma$ is given. All what is usually considered as higher order corrections in a perturbative framework, or quantum corrections in a semi-classical approach, is already contained in the effective action. The ability of finding the Green`s function with a certain precision is directly 
connected to the ability of computing $\Gamma$ to the required precision. Powerful functional methods are available for the computation of the effective action. Often general aspects, such as such symmetries and a derivative expansion, are sufficient to determine its main features.

For a given effective action cosmology reduces to a setting of classical field theory. This concerns both the field equations for mean values and the fluctuations. We will not encounter any operators, commutation relations or quantum vacua in our discussion. On the level of the effective action there is no distinction between classical and quantum fluctuations. Only one object describes all aspects of Gaussian fluctuations - it is the Green's function as obtained from the second functional derivative of $\Gamma$. This raises interesting questions on the ``quantum origin'' of the cosmic fluctuations. In quantum field theory $\Gamma$ is regarded as a quantum object - hence the name ``quantum effective action''. It can also be viewed, however, as a generating functional obtained by a Legendre transform of the logarithm of the partition function in a rather general setting of statistical physics. This poses the question if and which particular quantum properties are needed for the understanding of the origin of cosmic fluctuations.

Field relativity \cite{Wetterich:2013aca} states that the expectation values of physical observables do not depend on the choice of fields used for their description. Observations are the same for all ``frames'' obtained from each other by arbitrary non-linear field transformations. This strong principle is only practically valid on the level of the effective action, where observables are simply expressed as functionals of fields or correlation functions. (On the level of functional integrals a field transformation involves a Jacobian that is most often too complicated to be handled in practice.) In order to establish the equivalence of different frames for fluctuation properties it is therefore crucial that those can be extracted from the effective action. This is an additional motivation for the present note. On the level of the effective action the expression of a given fluctuation observable in terms of fields transforms according to the general rules of field transformations. For example, the correlation 
function is 
a field bilinear and transforms as such. Typically, the correlation functions themselves are not frame-invariant, but the physical observables describing the properties of fluctuations are. Within the framework of the effective action the mapping between different frames becomes much easier than the question which quantum state in one frame corresponds to the vacuum (or other quantum state) in another frame. 

The standard treatment of primordial cosmic fluctuations \cite{STAT,MUK,RSV,STA,GP,BST,AW} employs the linearized field equations for small deviations from a background cosmology. They permit to compute the time evolution of the power spectrum, which is associated to the squared fluctuations in momentum space. As for any linear homogeneous problem an initial value is needed in order to fix the amplitude. This is usually provided by invoking the quantum fluctuations in a given ``vacuum'' for a given cosmological geometry, as for example the ``Bunch-Davies'' vacuum \cite{BD} for de Sitter space. There is an ongoing discussion on the selection of the ``correct vacuum'' for de Sitter space, \cite{EM,BA,AP,FSS}. More generally, the question has to be answered why the fluctuations are described by the vacuum and not by some arbitrary excited state, mixed quantum state or classical state. From a pure conceptual viewpoint an arbitrary consistent ``initial'' state is allowed. Properties of the fluctuation spectrum may depend strongly on the assumed initial state. 

Within the operator formalism the amplitude of the fluctuations is (partially) fixed by commutation relations for creation and annihilation operators. In our treatment the equivalent properties follow from the inversion of $\Gamma^{(2)}$. The defining equation \eqref{ABA} is not homogeneous due to the non-vanishing right hand side, such that the normalization of $G$ is not arbitrary.

Within the concept of the functional integral underlying the effective action, the ``initial values'' are set as boundary values of fields and correlation functions. They may be set at the ``beginning of the universe'', which actually corresponds to infinite physical time in the past for typical inflationary cosmologies \cite{CWEU}. We are less ambitious here and set boundary conditions or ``initial values'' at the beginning of the inflationary epoch. For certain models this may coincide with the ``beginning of the universe'' in the infinite past \cite{Wetterich:2013aca}, \cite{CWVG}. The issue of the influence of initial values can now be addressed in a practical way as the question to what extent information about the boundary values of correlations is still available for some given cosmological time. 

As a general property, such information has a tendency to be ``forgotten'' as time increases. One often finds an approach to a symmetric state. The lower functional derivatives of $\Gamma$ are then largely fixed by a few couplings consistent with symmetries and effective locality (derivative expansion). Assuming that for a given epoch this simple form of $\Gamma$ is already (almost) realized, the information on boundary conditions is no longer encoded in $\Gamma^{(2)}$ on a practical level of accuracy. As a consequence, $\Gamma^{(2)}$ is not invertible and the correlation function is no longer fixed uniquely by $\Gamma^{(2)}$. In this approximation the effective action determines only the time evolution equations for the correlation functions. The correlation functions themselves are uniquely specified only once the initial conditions for the time evolution are fixed. The information about the initial state appears now in the form of initial values for differential equations. This is similar to the evolution of the mean fields. A given $\Gamma$ 
determines the field equations. Initial values are needed to select a cosmological solution. In this paper we assume an effective action that no longer shows a dependence on initial conditions. In particular, it does not explicitly depend on time. We therefore treat the solution of eq. \eqref{ABA} as an initial value problem for a differential evolution equation. 

A central part of our investigation are therefore time evolution equations for correlation functions. They permit us to start with arbitrary initial conditions for the fluctuations and to follow them until horizon crossing and beyond. In principle, the effective action and the initial conditions determine the correlation function for all times. Suitable composite fields for the energy momentum tensor of matter and radiation may be needed for an efficient description of late cosmology. We are concerned here mainly with the early stage of inflationary cosmology where radiation and matter play no role. The effective action is then typically a functional of the metric and a scalar inflaton field. In the present paper we simplify even further by considering the correlation function of a scalar field in a given geometric background. For a given effective action for the scalar field, a given geometric background, and given initial values for the Green's function the propagator is then fixed for all times. 

While the general structure of the evolution equation is exact, our knowledge of the effective action is at best a good approximation. Approximative forms can be computed using appropriate non-equilibration quantum field theoretical methods in the in-in-formalism \cite{BerR,Wei}. Our task is more modest here - we will compute the consequences of a given assumed form of the effective action. For the effective action of the scalar field we mainly consider for practical computations the standard form of a covariant kinetic term plus a potential. These are the lowest order terms in a derivative expansion. In this setting we derive an exact time evolution equation for the Green's function. For a subset of initial conditions this equation is equivalent to the standard approach of computing the linear time evolution of a small fluctuation around the background, but shown to be exact in our context. 

We argue that such an exact evolution equation holds in much more general circumstances far beyond the lowest order derivative expansion. Indeed, we expect the effective action to contain higher order terms, for example reflecting the scale anomaly (see ref. \cite{Sha}). Such terms are typically subleading for our considerations since they involve higher order curvature invariants and higher scalar derivatives. Once additional local or non-local terms for the quantum effective action can be computed reliably, their effect can be included in our treatment in a well defined way.

Within the present approximation the physical outcome of this paper is in many aspects equivalent to previous computations. There is an important practical point beyond previous results, however. We discuss the most general solution for the correlation function. It includes mixed quantum states in addition to the previously considered pure quantum states. As a result, initial states differing from the Bunch-Davies vacuum need not to show oscillatory behavior. Particle or entropy production are not necessary. Conceptually, the inclusion of interactions for the initial conditions and the general evolution of correlation functions allow us to discuss for a first time how possible equilibration and symmetrization processes could occur. Treatments of free quantum fields omit such effects from the beginning. 

The solution of the evolution equation for the simple case of a free massless scalar field in de Sitter space is shown in Fig. 1, where we plot the equal time correlation function $k^3G(k,\eta)/H_0^2$ as a function of $u=k\eta=-k/(aH_0)$ for different initial conditions. Here $G$ is the correlation function, $\eta$ conformal time, $k$ the wave number, $a$ the scale factor and $H_0$ the Hubble parameter. In our units the solutions do not depend on $k$. The solutions for different $k$ are independent. 

\begin{figure}[h!tb]
	\centering
	\includegraphics[scale=0.8]{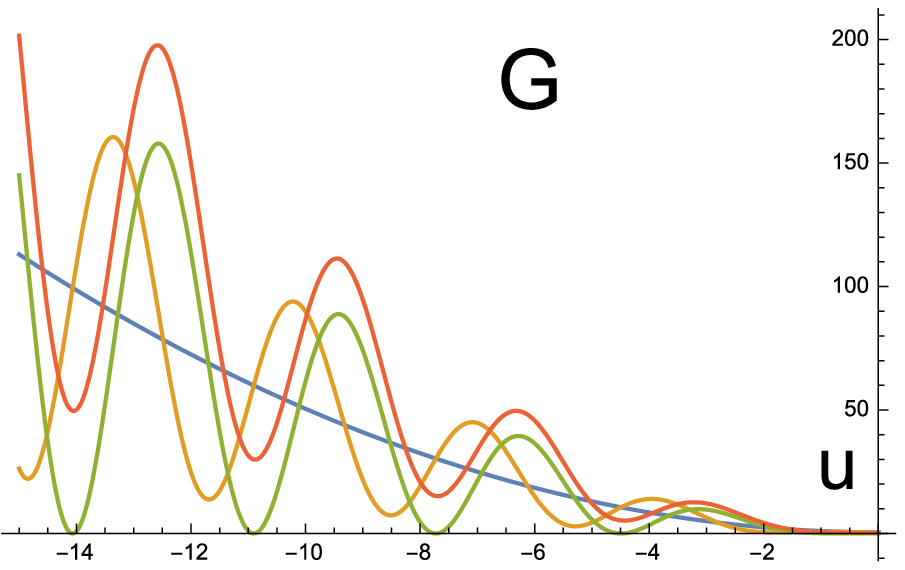}
	
	\vspace{0.8cm}
	\includegraphics[scale=0.8]{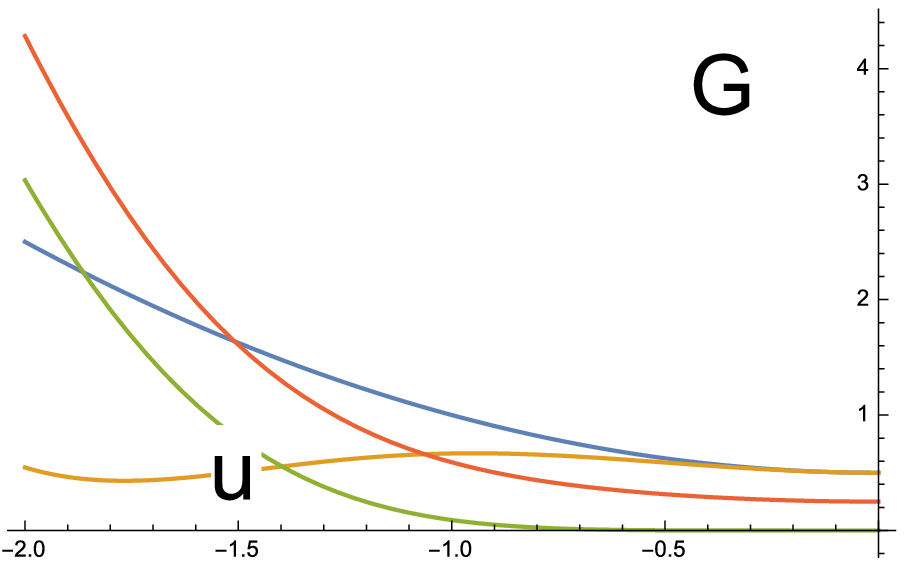}
	 \caption{Time evolution of correlation function for different initial conditions (see text). We show the equal time Green's function $G(k,\eta)$ in units of $H_0$ and multiplied with $-k^3$, as a function of $u=-k/(aH_0)$. In these units the curves do not depend on the comoving wave number $k$. Except for the smooth ``universal de Sitter propagator'' the characteristic behavior shows Hubble-damped oscillations. Memory of initial conditions is kept at horizon crossing for $u=-1$. This is best seen in the lower part of the figure which focuses on a smaller interval in $u$.}	
	 \label{Gsmall_Glarge}
\end{figure}

The smooth curve in Fig. \ref{Gsmall_Glarge} is the scaling correlation corresponding to the Bunch-Davies vacuum \cite{BD}. The oscillating curve that vanishes 
for $u\to 0$ and touches zero at regular intervals 
corresponds to the Green's function which is well defined in position space and consistent with the symmetries of de Sitter space, cf. ref. \cite{ABW1}. The two other curves employ more arbitrary initial 
conditions. The generic behavior shows oscillations whose amplitude decreases due to Hubble damping. At horizon crossing $(u=-1)$ the relation between fluctuation amplitude and Hubble parameter depends on the initial conditions. 

It is obvious from Fig. 1 that information about initial conditions is not lost at the time of horizon crossing. It is this moment that is relevant for observations, since in the presence of metric fluctuations there exists a scalar fluctuation quantity that remains preserved after horizon crossing and transports the information of the primordial spectrum to the density fluctuations in the late universe. Our finding of a preserved memory of initial conditions differs from a tendency towards a loss of memory discussed in refs \cite{AEH}, \cite{AHR}. This is due to the facts that we discuss the fluctuation spectrum rather than the energy momentum tensor, that the relevant time is horizon crossing and not asymptotic time, and that we discuss initial conditions that are less restrictive, not imposing a pure quantum state. The absence of a loss of memory is implicit in the treatment of free quantum fields in ref. \cite{BM1,BM2,BMR}. We show here that memory of initial conditions is preserved for a rather wide setting of models with interactions, provided that explicit boundary terms in $\Gamma$ and backreaction effects for the evolution of the mean fields can be neglected. 

As mentioned already, the overall amplitude of $G$ is set by the initial conditions independently for every $k$-mode. Even if we restrict our attention to non-oscillatory solutions, important memory of initial conditions is kept by the time evolution of correlation functions. This concerns both the amplitude and the shape (spectral index) of the power spectrum. Indeed, the memory of initial conditions has two aspects. We may not impose de Sitter symmetry on the initial conditions because we want to study a possible symmetrization starting from more arbitrary initial conditions. In this case the evolution shown in Fig. 1 holds for each momentum mode separately. For each $k$ the overall amplitude of $G$ can be multiplied by an arbitrary $k$-dependent factor. One finds that no symmetrization occurs in our approximation. In particular, the spectral index of the fluctuations $n_s$ receives a contribution from the shape of the initial spectrum. It is no longer uniquely determined by the shape of the inflaton potential.

One may also study what happens if de Sitter symmetry or other symmetries of the mean field solution are imposed on the initial conditions for the correlation functions. In this case the amplitudes for different $k$-modes are related by symmetry. For Minkowski space the requirement of symmetry is sufficient to make $\Gamma^{(2)}$ invertible and therefore to obtain a unique Green's function. This is not the case for de Sitter space, mainly due to the lack of time translation invariance. The symmetries of de Sitter space are not sufficient to fix the propagator of a scalar field uniquely, even if we regulate by a small mass term \cite{EM,BA}. In the presence of de Sitter symmetry the general solution to the propagator equation involves three free parameters which reflect the dependence on the initial conditions. These parameters can influence the amplitude of the primordial fluctuations substantially. If nothing fixes these parameters, the observable cosmic fluctuations retain information about a particular initial state 
even if de Sitter symmetry of the correlation function is imposed. 

Different criteria for a de Sitter invariant propagator lead to different correlation functions at the time of horizon crossing. If one requires that for de Sitter space a well defined propagator exists in position space and is consistent with the symmetries, the spectrum of cosmic fluctuations is modified as compared to the standard scenario. While the shape of the spectrum (spectral index) reflects mainly the symmetry of de Sitter space and remains unaltered, the amplitude is reduced by a factor around ten. 

The absence of a loss of memory visible in Fig. \ref{Gsmall_Glarge} has been derived with the lowest order derivative expansion for the effective action. It actually holds for a much wider class of effective actions. The basic reason is that eq. \eqref{ABA} restricts the normalization of $G$ only partially. As long as $\Gamma$ does not depend explicitly on $G$, and the evolution of the mean fields is independent of $G$ (neglected backreactions), eq. \eqref{ABA} remains linear in $G$. In this approximation the evolution equation for the equal-time correlation function will turn out to be a homogeneous linear equation, such that the amplitude remains arbitrary. This feature is exact for non-interacting free fields. In the presence of interactions, however, terms non-linear in $G$ will generically appear in the evolution equation. Beyond our approximation (which neglects such terms) they may eventually induce ``equilibration'' and lead to a loss of memory of the detailed initial conditions. We suggest that this loss of memory actually happens if inflation lasts sufficiently long before the horizon crossing of the observable fluctuations, such that equilibration has enough time to occur for the observable modes. This will then select a unique form of the correlation function for the relevant range of scales. 

More precisely, we conjecture that for a sufficiently long time between the beginning of inflation and horizon crossing of the observable fluctuations the time evolution selects a unique ``asymptotic propagator'' or ``scaling correlation''.  
Rather arbitrary initial conditions will be attracted towards the scaling correlation. The precise form of the scaling solution may not always be simple, however. 

A condition for the selection of the scaling correlation is formulated by analytic continuation. We postulate that the analytically continued Green's function should not increase exponentially in the infinite past. Our condition is motivated by the following observations. For Minkowski space it selects uniquely the Lorentz invariant vacuum propagator. For a long enough duration of inflation before horizon crossing the observable modes are far inside the horizon during the early stages of inflation, $k/a\gg H$. Flat space becomes a very good approximation for these modes. Our conjecture holds if there is enough time such that the correlation function for these modes can equilibrate to the Lorentz-invariant propagator. Subsequently, $k/a$ decreases and the correlation function will deviate from the Lorentz invariant form. We will show, however, that our condition concerning the analytic continuation remains preserved by the time evolution. It is therefore sufficient that the scaling correlation is reached at some early stage of inflation. The propagator is then given by the scaling correlation for all later times. For a de Sitter geometry and a derivative expansion of the effective action our criterion leads to the correlation function of the Bunch-Davies vacuum. It can be formulated, however, even if no exact symmetry is present. 

While the description of the equilibration process needs non-linearities in $G$ beyond our approximation, the scaling form of the propagator can be computed for simple approximations of the effective action. One only needs to impose the condition of analytic continuation on the space of functions $G$.  This makes $\Gamma^{(2)}$ invertible (for proper regularization of massless fields and local symmetries) such that eq. \eqref{ABA} has a unique solution. Indeed, if our condition for the scaling correlation is valid, the latter is determined uniquely for a given form of the effective action. For a derivative expansion of the effective action for the scalar field with up to two derivatives, a Bunch-Davies type correlation is selected. Modifications occur for a more complicated form of the effective action. In particular, this may concern the long wavelength tail of the correlation function. For the short wavelength behavior, which is relevant for the observable fluctuations, we expect such modifications to be small. 

Equilibration towards a scaling correlation is not possible for free fields. The inclusion of interactions is therefore crucial for any possible equilibration process. (investigations of the effects of interactions in a non-flat geometry can be found in refs. \cite{Wei,TW,BV,GGG,AKh,ProS,GS,Bur,Poly,KH,One1,One2,Boy}.) In our simple approximation interactions appear in the form of the inflaton potential. Our approach allows to study their effect not only for the evolution of mode functions and correlations but also for the initial conditions (``vacuum''). Interactions are per se not yet sufficient to guarantee equilibration, however. As mentioned already, equilibration needs effective non-linearity in $G$ for the evolution equation. 

The question if an asymptotic attractor exists, and if it obeys our conjecture concerning the properties of analytic continuation, can only be settled by an investigation of the time evolution of correlation functions for a large space of initial conditions. In the absence of explicit boundary terms in $\Gamma$, neglecting backreaction and assuming only terms with up to two time derivatives we find no approach to such a scaling correlation that 
is reached before or at horizon crossing. With these approximations, however, we would not find an approach to the Lorentz symmetric propagator in Minkowski space or to thermal equilibrium either. Even though not found yet explicitly, we expect some type of equilibration process that brings the correlation functions at least for ultra short distances to the Lorentz invariant propagator. For a very long duration of inflation before the observable fluctuations cross the horizon it is the information about the initial ultra short wavelengths that becomes observable. A very long epoch of inflation can bring them to a Lorentz invariant form even if the equilibration time is extremely long. This will be sufficient to establish our conjecture at least for an infinite duration of inflation. 

For a given inflationary model with a finite number of $e$-foldings $N_{\rm in}$ between the beginning of inflation and horizon crossing of the observable modes the practical question not solved in the present paper is the determination of the number of $e$-foldings $N^{(eq)}_{in}$ that is needed in order to achieve ``equilibration'' to the universal scaling correlation. In view of the small scalar self-interactions and the weakness of gravitational interactions we suspect that $N_{in}^{(eq)}$ is rather large. For a given inflationary model $N^{(eq)}_{in}$ is the crucial quantity in order to judge if the memory of initial conditions is effectively lost $(N_{in}>N^{(eq)}_{in}$) or if initial information is still observable $(N_{in}<N^{(eq)}_{in})$ in the CMB-spectrum. 

In this paper we illustrate the considerations above by a computation of the scaling correlation for a scalar field in a cosmological background close to de Sitter space. Several of the concrete results are known - our emphasis lies here on the derivation from the effective action and on statements to which extent results are exact or only approximative. With boundary conditions specified by the scaling correlation, $\Gamma^{(2)}$ is invertible and the Green's function follows from matrix inversion. We find that the equal time correlation function for a massless scalar obeys in momentum space, with scale factor $a(t)$ and Hubble parameter $H(t)$,
\be\label{AI1}
G(k,a)=\frac{1}{2ka^2}\left(\big|1+\frac{i f(y)}{y}\big|^2\right)^{\frac{1}{1+\nu}},
\ee
where
\be\label{AI2}
y=\frac{k}{aH}, \quad \nu=\frac{\dot{H}}{H^2}\quad.
\ee
The function $f(y)$ is close to one for small $\nu$ as appropriate for inflation. Its precise form is computed by a numerical solution of a nonlinear second order differential equation. For de Sitter space, $\nu=0$, one has $f=1$. While for realistic inflation the scalar fluctuation spectrum is modified by the inflaton potential and mixing with gravitational degrees of freedom, eqs. \eqref{AI1}, \eqref{AI2} can actually be used for the tensor fluctuations (up to overall normalization). 

This paper is organized as follows: In sect II we recall briefly the relevant properties of the quantum effective action. We formulate it in a framework that permits a straightforward analytic continuation by changing the value of a ``background vierbein''. We turn in sect. III to the correlation function for a massless scalar field in a homogeneous and isotropic cosmology. The defining ``propagator equation'' relates the Green's function $G$ to the second functional derivative of the effective action $\Gamma^{(2)}$. It is a partial differential equation for $G$ depending on conformal time $\eta$ and comoving space-distance $r$. As an example, we solve it in Minkowski space. In appendix A we relate the general short distance behavior of the propagator to the one in flat Minkowski space and discuss the form of the next to leading terms in position space. 

The propagator in three-dimensional momentum space is addressed in sect. IV. The propagator equation can be solved for each mode independently , with an inhomogeneous term restricting partially the normalization for each mode separately. Within a derivative expansion of the effective action we discuss the general solution of the propagator equation in de Sitter space. It shows no loss of memory. We explore the properties of a possible unique scaling correlation. The symmetries of de Sitter space are not strong enough to select a unique propagator. For selecting the scaling correlation we formulate our conjecture on the properties of the analytically continued propagator. It is assumed to be realized asymptotically as a result of the time evolution of correlation functions. It states that the analytically continued Green's function is restricted not to diverge exponentially towards the past. On the space of functions with such a boundary behavior, the second functional derivative $\Gamma^{(2)}$ becomes invertible. The de Sitter propagator is then unique for the modes with non-zero momentum and corresponds to the Bunch-Davies vacuum. Our conjecture is the analogue of the usual $i\epsilon$-presentation in flat space to arbitrary geometries. In appendix A we discuss that the Fourier transform of this propagator is not well defined due to the lack of regularization for the zero-momentum mode. The proposed unique scaling correlation differs from a possible de Sitter invariant propagator that is well defined in position space. 

Sect.~\ref{Time evolution of correlation function} addresses the central formulation of a time-evolution equation for the correlation function that allows for an investigation of the role of initial conditions. We display an exact evolution equation for a free massless scalar field in de Sitter space and discuss the properties of its general solution. We find that the memory of initial conditions is not lost in this case, cf. Fig. 1. We argue that non-linearities due to explicit boundary terms in $\Gamma$ or backreaction are needed in order to realize our conjectured scaling correlation. The exact evolution equation is extended to arbitrary homogeneous and isotropic cosmologies and interacting scalar fields. The approximations concern then only the precise form of the effective action. We discuss the general structure of the evolution equation for interacting theories in the absence of boundary and backreaction effects and show that their general solution keeps memory of the initial conditions. In appendix B we present the general structure of the solution of the evolution equation. In sect. VI we formulate effective quantum fields for interacting theories and use them to discuss the structure of initial conditions. 

In sect.~\ref{Scalar correlation function in inflationary cosmology} we compute the universal scaling correlation for realistic inflationary geometries, assuming that our conjecture is true. We make contact to the standard results of slow roll inflation. Our results are partly exact, as for constant $\dot H/H^2$ and $H^{-2}\partial^2 V/\partial\varphi^2$, with $V(\varphi)$ the inflaton potential.  The infrared divergence of the de Sitter propagator is regularized by a mass term. In appendix D we turn to the correlation function for a massive scalar in de Sitter space. Appendix E addresses the issue of symmetries, in particular the consequences of the symmetry of de Sitter space and coordinate scale symmetry. The universal propagator is assumed to be consistent with the symmetries of the action and mean field solution. De Sitter symmetry is a strong constraint, but not sufficient to fix the propagator uniquely. 

In sect.~\ref{Impact of initial conditions on primordial fluctuations} we turn to the impact of initial conditions on the spectrum of cosmic fluctuations as observed in the cosmic microwave background. We formulate the evolution equation for correlation functions for an arbitrary inflaton potential and discuss the general solution. The initial spectrum is only processed at horizon crossing by the scale violating effects related to the inflaton potential. If inflation does not last too long between its beginning and the horizon crossing of the observable fluctuations, the neglection of explicit boundary terms in $\Gamma$ as well as backreaction becomes justified. In this case observation can look back to the beginning of inflation. We present an example for an initial spectrum that modifies amplitude and spectral index for the observed fluctuations. Conclusions are drawn in sect.~\ref{Conclusions}. 

\section{Effective action and analytic \newline{continuation}}
\label{Effective action and analytic continuation}

The key element for the description of fluctuations is the propagator $G(x,y)=G(\eta,\vec x;\eta',\vec y)$, where we take for $\eta, \eta'$ conformal time. For a homogeneous background it can only depend on $\vec r=\vec x-\vec y,\quad G(\eta,\vec x;\eta',\vec y)=G(\vec r,\eta,\eta')$, with Fourier transform
\be\label{A}
G(\vec r,\eta,\eta')=\int_k G(\vec k,\eta,\eta')e^{i\vec k\vec r}.
\ee
Isotropy implies $G(\vec k,\eta,\eta')=G(k,\eta,\eta'),k=|\vec k|$. The power spectrum of cosmic fluctuations is related to the equal time propagator, $\eta=\eta'$, 
\be\label{B}
G(k,\eta)=G(k,\eta,\eta).
\ee

The central simple observation of this note recalls that the second functional derivative $\Gamma^{(2)}$ of the quantum effective action yields the exact inverse propagator. If $\Gamma^{(2)}$ is invertible the propagator and therefore the power spectrum can be obtained by matrix inversion. If not, the condition 
\be\label{4A}
\Gamma^{(2)}G=1
\ee
yields a time evolution equation for the correlation function whose solution depends on initial conditions. Invertibility depends on the space of functions that are admitted for $G$. 

\subsection{Effective action}

The quantum effective action arises from a functional integral formulation of quantum field theory, that we recall briefly here for an arbitrary geometrical background and arbitrary signature. The starting point is the partition function, which is defined by a functional integral in the presence of sources
\ba\label{1}
Z[J]&=&\int {\cal D}\tilde \phi \exp \Big(-S+\int_x J\tilde\phi\Big),\nn\\
S&=&\int_x eL[\tilde\phi,e^m_\mu].
\ea
Here the generalized vector $\tilde\phi(x)=\tilde\phi(\eta,\vec x)$ stands collectively for all fluctuating fields, $J(x)$ are the associated sources and we use euclidean sign conventions for the action. The geometry is described by a ``background vierbein'' $e^m_\mu$, with $e=\det(e^m_\mu)$. This vierbein can take complex values. Euclidean signature is realized for real $e^m_\mu$, while a Minkowski signature follows if $e^m_0$ assumes purely imaginary values (with real $e^m_k$). In particular, for flat euclidean space one has $e^m_\mu=\delta^m_\mu,e=1$, whereas flat Minkowski space obtains for $e^m_k=\delta^m_k$, $e^m_0=i\delta^m_0$, $e=i$. 

Analytic continuation can be achieved in a simple way by changing the euclidean value of $e^m_0$ to $e^m_0e^{i\varphi}$, with fixed coordinates $x^\mu=(\eta,\vec x)$ and fixed fields $\tilde\phi$ \cite{CWAC}. (For earlier somewhat different approaches to analytic continuation see refs. \cite{CR1,CR2}.) Arbitrary background metrics
\be\label{2}
\bar g_{\mu\nu}=e^m_\mu e^n_\nu\delta_{mn}
\ee
can be described in this setting. Our fixed coordinates $(\eta,\vec x)$ span ${\mathbbm R}^4$ (or a suitable subspace), with time coordinate $\eta$ corresponding to conformal time. The geometry of a homogeneous and isotropic universe is given by
\be\label{3}
e^m_k=a(\eta)\delta^m_k~,~e^m_0=ia(\eta)\delta^m_0,
\ee
with $a(\eta)$ the scale factor and 
\be\label{4}
ad\eta=dt~,~{\cal H}=\frac{\partial \ln a}{\partial \eta}=Ha,
\ee
where $t$ and $H$ are time and Hubble parameter of the Robertson-Walker metric. Metric fluctuations $\tilde h_{\mu\nu}$ can be included into $\tilde\phi$, defining a fluctuating metric
\be\label{5}
g_{\mu\nu}=e^m_\mu e^n_\nu\delta_{mn}+\tilde h_{\mu\nu}.
\ee

We define the functional
\be\label{6}
W[J]=\ln Z[J],
\ee
such that 
\be\label{7}
\frac{\partial W}{\partial J(x)}=\kl \phi(x)\kr=\phi(x).
\ee
As usual, the quantum effective action $\Gamma[\phi]$ obtains by a Legendre transform
\be\label{8}
\Gamma[\phi]=-W[J]+\int_x J\phi~,~\frac{\partial\Gamma}{\partial\phi(x)}=J(x),
\ee
with $J[\phi]$ obtained by inverting eq. \eqref{7}. The functional $W[J]$ is the generating functional for the connected Green's functions for $\tilde\phi$. The second functional derivative $\Gamma^{(2)}=\partial^2\Gamma/\partial\phi(x)\partial\phi(y)$ and the second functional derivative $W^{(2)}=\partial^2 W/\partial J(x)\partial J(y)$ can be viewed as matrices. The matrix identity
\be\label{9}
\Gamma^{(2)} W^{(2)}=1
\ee
follows directly from the Legendre transformation \eqref{8}. 

\subsection{Correlation function}

The correlation function (connected two point function, Green's function, propagator) is defined as
\be\label{10}
G(x,y)=\kl \tilde\phi (x)\tilde\phi(y)\kr -\kl\tilde \phi(x)\kr\kl \tilde\phi(y)\kr
=\frac{\partial^2 W}{\partial J(x)\partial J(y)}.
\ee
(In the operator formalism this corresponds to a time ordered operator product.) By definition the Green's function is a symmetric 
\be\label{14A}
G(y,x)=G(x,y).
\ee
If $\Gamma^{(2)}$ is invertible the Green's function can be found by matrix inversion of the ``inverse propagator'' $\Gamma^{(2)}$,
\be\label{11}
G(x,y)=\left(\frac{\partial^2\Gamma}{\partial\phi(x)\partial\phi(y)}\right)^{-1}.
\ee
In this case the propagator can be extracted directly from the quantum effective action, without any further assumptions. 

The effective action is well defined for fields and sources for which the relation $\phi[J]$ is invertible. The relation \eqref{9} requires $\Gamma^{(2)}$ to be a regular matrix without zero eigenvalues. This is typically only the case if explicit boundary terms are included in $\Gamma$ or if boundary conditions are imposed on $G$.  Formally, this is achieved by adding to $S$ in eq. \eqref{1} boundary terms that are non-vanishing only at the boundaries. We will assume such boundary terms in the infinite past $\eta\to-\infty$ or for large negative $\eta$. They are equivalent to a specification of the initial values of fields and correlation functions in the infinite past \cite{CSH,CHU}. The boundary conditions affect the form of $\Gamma$ close to the boundary. With restrictive enough boundary conditions $\varphi[J]$ is unique and $\Gamma^{(2)}$ is invertible. 

In this paper we do not deal explicitly with the boundary terms in $S$ or $\Gamma$. We rather consider a finite time $\eta$ for which we assume that $\Gamma$ has already a rather simple form. We will question the validity of this approximation in sect. V. Eq. \eqref{4A} is valid for all $\eta$. Without the explicit boundary terms $\Gamma^{(2)}$ is no longer invertible, however. We have therefore to consider the general solution of eq. \eqref{4A} which involves initial values as free parameters. The effect of a given set of boundary terms translates to the selection of a given set of initial values. At this stage we may still impose generic boundary conditions as the absence of a too strong divergence of correlation functions towards the infinite past. Furthermore, one may require that the symmetry of the background solution is shared by the correlation functions. 

For euclidean signature $\Gamma^{(2)}$ often becomes invertible if one excludes functions that diverge too strongly at the boundaries. This selects a unique ``euclidean correlation function''. For Minkowski signature the justification of boundary conditions is less obvious. Our conjecture for a unique scaling correlation concerns a specification of the boundary conditions for the analytically continued Green's function in the vicinity of Minkowski signature. The scaling correlation is then the analytic continuation of the unique euclidean correlation function. Ultimately, our conjecture has to be justified by a study of the time evolution of correlation functions with arbitrary initial conditions at the boundary.  It holds, at best, asymptotically if sufficient time for ``equilibration'' has passed since the time when initial conditions are set. The choice of boundary conditions is an analogue to the choice of a ``vacuum'' in the operator formalism of quantum mechanics. We will consider here more general initial states, including excited pure quantum states, mixed states and ``classical states''. 

Even the specification of generic boundary conditions does not render $\Gamma^{(2)}$ invertible for all situations. Massless fields often induce zero eigenvalues of $\Gamma^{(2)}$. Obtaining a unique correlation function for massless fields may necessitate the addition of a regulator $R$ to $\Gamma^{(2)}$, such that $\Gamma^{(2)}+R$ is invertible even if $\Gamma^{(2)}$ is not. One may then consider the limit $R\rightarrow0$. The presence of a suitable infrared regulator leads to the concept of the effective average action \cite{CW1}.

\section{Correlation function for free massless scalar field in homogeneous and isotropic cosmology}
\label{Correlation function for free massless scalar}

Before turning later to more realistic inflationary models with an inflaton potential and interactions we investigate in this section the correlation function of a free massless scalar field in a homogeneous and isotropic geometry. We start with flat Minkowski space where Poincar\'{e} symmetry can be employed to define the unique scaling correlation. This scaling correlation corresponds to the propagator in the vacuum. It transforms under scale transformations according to its dimension and contains no parameter with dimension of mass - hence its name. For more complicated geometries scale symmetry remains no longer realized. We conjecture that a possible scaling correlation towards which correlation functions are attracted for arbitrary initial conditions has a high momentum limit given by the universal free propagator. 

Time translation symmetry or Lorentz symmetry will no longer act as a selection principle to find the scaling propagator for more complicated geometries. We can, instead, use the properties of analytic continuation. Indeed, for the analytic continuation to a flat space neighboring Minkowski space one finds that $\Gamma^{(2)}$ is invertible provided one imposes the generic boundary condition that the correlation function should not diverge exponentially towards the infinite past. This criterion for the scaling correlation needs no particular symmetry and can be generalized to a wide class of geometries. For this reason we construct in this section the universal propagator for Minkowski space by analytic continuation. 

In appendix A we discuss the general short distance behavior of the correlation function in position space. For arbitrary geometries it resembles closely the one for Minkowski space. This does not imply, however, that the high momentum behavior of the propagator is the one for Minkowski space.  
\subsection{Propagator equation}
The identity \eqref{4A} is the central equation for the computation of the propagator of a real scalar field $\varphi(x)$ in a homogeneous cosmological background \eqref{3}. For the particular case of a free massless field one has $L=\partial^\mu\varphi\partial_\mu\varphi$ and the effective action reads
\be\label{12}
\Gamma =\int_x\bar e a^2(\bar e^{-2}\partial_\eta\varphi\partial_\eta\varphi+\partial_k\varphi\partial_k\varphi),
\ee
with $\bar e=i$ for Minkowski signature and $\bar e=1$ for the euclidean version, $e=\bar e a^4,~\bar g^{00}=\bar e^{-2}a^{-2},~\bar g^{kl}=a^{-2}\delta_{kl}$. The second functional derivative $\Gamma^{(2)}$ can be written as a differential operator
\ba\label{12a}
&&\Gamma^{(2)}(x,y)=\Gamma^{(2)}(\eta,\vec x;\eta',\vec y)\\
&&\quad =-\bar e\delta(\eta-\eta')\delta^3(\vec x-\vec y)\{\bar e^{-2}\partial_{\eta'} a^2\partial_{\eta'}+a^2\Delta_y\}.\nn
\ea
The ``propagator equation'' \eqref{4A} takes therefore the form
\be\label{17A}
(\partial^2_\eta+2\h\partial_\eta+\bar{e}^{_2}\Delta_x)G(x,y)=
-\frac{\bar e}{a^2}\delta (\eta-\eta')\delta^3(\vec x-\vec y).
\ee
Our task is a solution of this equation for the Green's function $G$. 

The space of functions $G$ is restricted by exchange-symmetry \eqref{14A} and ``reality''. For euclidean signature a real action implies that $G$ is real. For Minkowski signature we write
\ba\label{17B}
G(\eta,\vec x;\eta',\vec y)&=&G_s(\eta,\vec x;\eta',\vec y)\\
&+&G_a(\eta,\vec x;\eta',\vec y)\big[\theta(\eta-\eta')-\theta(\eta'-\eta)\big],\nn
\ea
with step function $\theta(x)=1$ for $x>0,\theta(x)=0$ for $x<0$ and $\theta(0)=1/2$. Symmetry implies that $G_s$ and $G_a$ are symmetric and antisymmetric functions, respectively,
\ba\label{17C}
G_s(\eta',\vec y;\eta,\vec x)&=&G_s(\eta,\vec x;\eta',\vec y),\nn\\
G_a(\eta',\vec y;\eta,\vec x)&=&-G_a(\eta,\vec x;\eta',\vec y).
\ea
For Minkowski signature the action and the argument of the exponential function in eq. \eqref{1} are purely imaginary. The reality condition for Minkowski signature states that $G_s$ is real and $G_a$ purely imaginary,
\be\label{17CC}
G^*_s=G_s~,~G^*_a=-G_a.
\ee

For Minkowski signature solutions of the propagator equation \eqref{17A} can be found if both $G_s$ and $G_a$ obey the homogeneous equation
\be\label{17D}
DG_s=0~,~DG_a=0,
\ee
with 
\be\label{17E}
D=\partial^2_\eta+2\h\partial_\eta-\Delta_x.
\ee
With 
\be\label{17F}
D\Big(G_a\big[\theta(\eta-\eta')-\theta(\eta'-\eta)\big]\Big)=2\partial_\eta G_{a|\eta=\eta'}
\delta(\eta-\eta')
\ee
the solution of the inhomogeneous equation \eqref{17A} requires
\be\label{17G}
\partial_\eta G_{a|\eta=\eta'}=-\frac{i}{2a^2}\delta^3(\vec x-\vec y).
\ee
For Minkowski signature the ``defining equation'' for the Green's function $G$ reads explicitly. 
\ba\label{13}
&&ia^2DG(\eta,\vec x;\eta',\vec y)=\delta(\eta-\eta')\delta^3(\vec x-\vec y),\nn\\
&&D=\partial^2_\eta+2{\cal H}\partial_\eta-\Delta_x,
\ea
such that for purely imaginary $G_a$ both sides of eq. \eqref{13} are real, or both sides of eq. \eqref{17G} purely imaginary. Eqs. \eqref{17B}, \eqref{17C}, \eqref{17CC}, \eqref{17D}, \eqref{17F}, \eqref{17G} generalize to a much wider setting than a free massless scalar field. 

The propagator is uniquely defined only if the operator $D$ is invertible. If not, there exist non-trivial solutions of the differential equation $DH=0$. Any combination $G+H$ solves then eq. \eqref{13} as well as $G$. For Minkowski signature the propagator equation \eqref{13} admits a large family of solutions. This is obvious from eqs. \eqref{17D}, \eqref{17G}. The inhomogeneous term only affects $G_a$, while the differential equation for $G_s$ is a linear equation without further restrictions. Some statement on the initial conditions for $G_s$ is needed in order to specify the propagator. 

We are interested in solutions that are connected to the euclidean setting by analytic continuation. We therefore choose $e^m_0=e^{i\left(\frac\pi2-\epsilon\right)}a=i(1-i\epsilon)a,\epsilon>0$, and take $\epsilon\to 0$ at the end. Eq. \eqref{13} is thus modified
\ba\label{13A}
&&ia^2(1-i\epsilon)DG=\delta(\eta-\eta')\delta^3(\vec x-\vec y),\nn\\
&&D=(1+2i\epsilon)(\partial^2_\eta+2{\cal H}\partial_\eta)-\Delta.
\ea
Our conjecture for the scaling correlation states that for $\epsilon>0$ the Green's function should not increase exponentially for $\eta\to-\infty$. With this conjecture the only remaining solution of $DH=0$ is $H=const$. This degeneracy with respect to a constant shift in $G$ is characteristic for a massless field. To make $\Gamma^{(2)}$ fully invertible a regulator is needed. For example, the addition of a small mass term makes $\Gamma^{(2)}$ regular, as we will discuss later. Then the propagator $G$ and therefore the power spectrum for fluctuations follow from the unique solution of the differential equation \eqref{13A}, without any further assumption about a particular ``vacuum state''. Of course, as mentioned before, our conjecture needs ultimately to be justified by an investigation of general initial conditions for the solutions of eq. \eqref{13}. 

\subsection{Propagator in flat space}
Our task is the computation of the Green's function for the operator $D$. Without loss of generality we take $\eta\geq \eta'$. The symmetry of $G$ under the exchange $\eta\leftrightarrow \eta'$, $\vec x\leftrightarrow \vec y$ can then be used to infer the result for $\eta\leq \eta'$. We first consider flat Minkowski space with constant scale factor $a$ and ${\cal H}=0$. We write the propagator in terms of its Fourier transform $\tilde G$. 
\ba\label{14}
&&G(\eta,\vec x;\eta',\vec y)=\\
&&\int_k\int_{k'}\int_\omega\int_{\omega'}
e^{-i(\omega\eta-\vec k\vec x-\omega'\eta'+\vec k '\vec y)}\tilde G(\omega,\vec k;\omega',\vec k '),\nn
\ea
with 
\be\label{15}
\int_k=\int d^3k/(2\pi)^3~,~\int_\omega=\int d\omega/2\pi.
\ee
For the universal scaling propagator we assume translation symmetry both in space and time, such that $\tilde{G}(\omega, \vec{k}; \omega', \vec{k'})$ is diagonal both in frequency and momentum. For $\epsilon > 0$ $\Gamma^{(2)}$ is invertible and inversion of the operator $D$ is then done easily in Fourier space, 
\be\label{16}
\tilde G(\omega,\vec k;\omega',\vec k ')=
\frac{2\pi i(1+i\epsilon)}{a^2(\omega^2-k^2+2i\epsilon\omega^2)}
\delta(\omega-\omega')\delta(k-k'),
\ee
with $k=|\vec k|$ and $\delta(k-k')=(2\pi)^3\delta^3(\vec k-\vec k ')$. 

The unique scaling correlation in position space obtains from eq. \eqref{14}
\be\label{17}
G(\eta,\vec x;\eta',\vec y)=\int_k\int_\omega
e^{-i\omega(\eta-\eta')+i\vec k(\vec x-\vec y)}
\frac{i(1+i\epsilon)}{a^2(\omega^2-k^2+2i\epsilon\omega^2)}.
\ee
Performing the $\omega$-integration yields
\be\label{18}
G(\eta,\vec x;\eta',\vec y)=\int_k
\frac{1}{2a^2k}
e^{-ik(\eta-\eta')}
e^{i\vec k(\vec x-\vec y)}
e^{-\epsilon k(\eta-\eta')}.
\ee
The three dimensional $\vec k$-integral can be reduced to an integral over $k=|\vec k|$, with $r=|\vec x-\vec y|$,
\ba\label{18A}
G(\eta,\vec x;\eta',\vec y')&=&
\frac{1}{4\pi^2a^2 r}
\int\limits^\infty\limits_0
dk\sin (kr)
e^{-ik(\eta-\eta')}
e^{-\epsilon k(\eta-\eta')}\nn\\
&=&\frac{1}{4\pi^2a^2\big[r^2-(1-2i\epsilon)(\eta-\eta')^2\big]}.
\ea
For $\eta-\eta'\to\infty$ the correlation function vanishes exponentially for $\epsilon>0$. We may therefore also select it by the criterion that it should not diverge exponentially in this limit (which would be the case for $\epsilon<0$).

The result for general $e^m_0=e^{i\varphi}a\delta^m_0,0\leq\varphi<\frac{\pi}{2}$, can be found with similar methods,
\be\label{18B}
G(\eta,\vec x;\eta',\vec y)=
\frac{1}{4\pi^2\Delta x^\mu\Delta x_\mu},
\ee
with $\Delta x^\mu=(\eta-\eta',\vec r),\Delta x_\mu=\bar g_{\mu\nu}\Delta x^\nu$. For small $\epsilon>0$ this coincides with eq. \eqref{18A}. We recognize the analytic continuation from flat euclidean space as well as the associated generalized Lorentz-symmetry. For Minkowski space the requirements of Lorentz symmetry or the analytic continuation are equivalent. They both fix the propagator uniquely (up to a constant). 

In the limit $\epsilon\to 0$ the propagator \eqref{18A} becomes
\ba\label{27A}
\lim_{\epsilon\to 0} G&=&\frac{1}{4\pi^2 a^2}\lim_{\epsilon\to 0}
\frac{r^2-(\eta-\eta')^2}{\big (r^2-(\eta-\eta')^2\big)^2+4\epsilon^2(\eta-\eta')^4}\nn\\
&&-\frac{i}{4\pi a^2}\delta\big (r^2-(\eta-\eta')^2\big).
\ea
The first term corresponds to $G_s$ in eq. \eqref{17B}. For $r^2\neq (\eta-\eta')^2$ the limit $\epsilon\to 0$ can be taken without further complication. In particular, the equal time Green's function takes the simple form
\be\label{18C}
G_s(\vec r,\eta)=G_s(\eta,\vec x;\eta,\vec y)=\frac{1}{4\pi^2 a^2 r^2}.
\ee
It is time-independent and reads in momentum space
\be\label{19}
G(k)=\frac{1}{2a^2k}.
\ee
For $r\neq 0$ the imaginary part of the propagator \eqref{27A} can be written in the form \eqref{17B} with 
\be\label{27B}
G_a(\eta,\vec x;\eta',\vec y)=-\frac{i}{8\pi a^2 r}
\big\{\delta(r-\eta+\eta')-\delta(r+\eta-\eta')\big\}.
\ee
It is straightforward to verify eqs. \eqref{17D} and \eqref{17E}. 

The universal propagator in Minkowski space can be extended to a large class of homogeneous and isotropic cosmologies. Assume that we move continuously away from flat space by switching on a non-trivial $\eta$-dependence of the scale factor $a(\eta)$. The propagator will move continuously away from the free propagator. As long as the operator $D$ in eq. \eqref{13A} remains invertible for $\epsilon>0$ the Green's function remains unique, determined by eq. \eqref{13A}. A unique Green's function would imply the loss of memory of the initial state of the universe. Our conjectured scaling correlation is precisely the one that is continuously connected to the Lorentz invariant correlation function in flat space. 

The issue of the influence of initial conditions concerns the impact of boundary terms for $\epsilon=0$. In our discussion of the propagator in Minkowski space we have not been very explicit on the choice of boundary conditions. The propagator \eqref{27A} is the only one compatible with Poincar\'e-symmetry and therefore singled out. Such a symmetry argument is no longer available if $a(\eta)$ depends on $\eta$. In particular, time translation symmetry is lost. For $\epsilon>0$ we will require that the scaling correlation in momentum space does not diverge exponentially as the past boundary (typically for $\eta\to-\infty$) is approached. This boundary condition is independent of symmetries and sufficient to single out a unique propagator for $\epsilon>0$. 

For $\epsilon=0$, however, there will appear other solutions that do not diverge exponentially at the boundary. They can transport information about initial conditions to any finite time $\eta$. It is a dynamical question if the unique analytically continued Green's function is approached rapidly enough such that memory of initial conditions is effectively lost. We will see that for a free scalar field in a given background geometry this is actually not the case. At best our conjecture for a unique propagator can hold in the presence of interactions.

\section{Propagator in momentum space}
\label{Propagator in momentum space}

In momentum space one can decompose the propagator equation into separate ordinary linear differential equations for each momentum-mode. We employ this in order to construct the most general solution for the correlation function of a free massless scalar field in de Sitter space. It preserves the memory of initial conditions. We also construct the universal scaling correlation according to our conjecture. It equals the one found from the Bunch-Davies vacuum. In appendix A we discuss the Fourier transform to position space. We also identify the solution which corresponds to a well defined propagator in position space with de Sitter symmetry \cite{FSS}. The latter differs from the scaling correlation.
 
\subsection{Green's function in Fourier space}

Due to the translation symmetry in the $\vec x$-coordinate it is often convenient to make a Fourier transform in the three-dimensional subspace. On the other hand we keep the time $\eta$, for which a Fourier transform is for most cases not useful because of the lack of time-translation symmetry of the background geometry. With 
\be\label{42A}
\tilde\phi(\eta,\vec x)=\int_k e^{i\vec k\vec x}\tilde\phi(\eta,\vec k),
\ee
and the condition
\be\label{42B}
\tilde\phi^*(\eta,\vec k)=\tilde \phi(\eta,-\vec k)
\ee
for real fields $\tilde\phi(x)$, the Green's function for a real scalar field in Fourier space reads
\ba\label{42C}
G(\eta,\vec k;\eta',\vec k')=\langle\tilde\varphi(\eta,\vec k)\tilde \varphi^*(\eta',\vec k')\rangle
-\langle\tilde\varphi(\eta,\vec k)\rangle\langle \tilde \varphi^*(\eta',\vec k')\rangle.\nn\\
\ea
It is the Fourier transform of $G(\eta,\vec x;\eta',\vec y)$,
\be\label{F3}
G(\eta,\vec x;\eta',\vec y)=\int_k\int_{k'}
e^{i(\vec k\vec x-\vec k '\vec y)}
G(\eta,\vec k;\eta',\vec k ').
\ee
We use translation and rotation symmetry,
\be\label{F1A}
G(\eta,\vec k,\eta',\vec k ')=G(k,\eta,\eta')\delta(k-k'),
\ee
where $G(k,\eta,\eta')$ obeys eq. \eqref{A}. 

The matrix equation \eqref{4A} can be formulated in an arbitrary basis. In (three-dimensional) Fourier space one has, with $\bar e=e^{i\varphi}$,
\ba\label{F1}
&&\Gamma^{(2)}(\eta',\vec k ';\eta,k)=\\
&&\quad \delta(\eta'-\eta)\delta(k'-k)a^2
\big\{\bar ek^2-\frac{1}{\bar e} (\partial^2_\eta+2{\cal H}\partial_\eta)\}.\nn
\ea
The defining equation \eqref{17A} for the propagator reads
\be\label{F2}
a^2\big[\bar ek^2-\frac{1}{\bar e}(\partial^2_\eta+2\h\partial_\eta)\big]
G(k,\eta,\eta')=\delta(\eta-\eta').
\ee
In Fourier space the propagator equation can be solved separately for each $k$-mode. In particular, the inhomogeneous term on the r.h.s. of eq. \eqref{F2} restricts the normalization for each $k$-mode separately.The particular geometry of de Sitter space is given by
\be\label{47AA}
{\cal H}=-\frac1\eta~,~a=\frac{\h}{H_0}~,~\bar e=i.
\ee

For flat space the Fourier transform of the Minkowski propagator $G_0$ reads
\ba\label{F4}
G_0(k,\eta,\eta')&=&
\frac{1}{2a(\eta)a(\eta')k}
\Big\{\exp \big\{-\bar e k(\eta-\eta')\big\}\theta (\eta-\eta')\nn\\
&&+\exp \big\{-\bar e k(\eta'-\eta)\big\}\theta(\eta'-\eta)\Big],
\ea
The relation
\be\label{F5}
\bar e(k^2-\frac{1}{\bar e^2}\partial^2_\eta)
\big[a(\eta)a(\eta') G_0(k,\eta,\eta')\big]=\delta(\eta-\eta')
\ee
fixes the normalization of $G_0$ uniquely. For more general geometries the insertion of eqs. \eqref{F4}, \eqref{F5} into eq. \eqref{F2} yields the Fourier transform of eq. \eqref{23}
\ba\label{F6}
&&~(\partial_\eta^2+2\h\partial_\eta-\bar e^2 k^2)
\big (G(k,\eta,\eta')-G_0(k,\eta,\eta')\big)\nn\\
&&\qquad\qquad=\big (\h^2(\eta)+\partial_\eta\h(\eta)\big)G_0(k,\eta,\eta').
\ea

\subsection{Solution of homogeneous equation}
For solutions of the propagator equation \eqref{F2} we consider first the ansatz
\be\label{F7}
\tilde G(k,\eta,\eta')=w^-_k(\eta)w^+_k(\eta')\theta(\eta-\eta')+
w^+_k(\eta)w^-_k(\eta')\theta(\eta'-\eta),
\ee
with $w_k$ obeying the homogeneous linear differential equation
\be\label{F8}
\tilde D_kw^\pm_k(\eta)=0~,~
\tilde D_k=\partial^2_\eta+2\h\partial_\eta-\bar e^2k^2.
\ee
This ensures for $\eta\neq \eta'$ the homogeneous equation
\be\label{F9}
\tilde D_k\tilde G(k,\eta,\eta')=0,
\ee
in accordance with eq. \eqref{F2}. The ansatz \eqref{F7} is not the most general solution, which we will discuss later. For appropriate $w_k^{\pm}$ it will describe the scaling correlation. 

We explore for $w_k$ the form
\be\label{F10}
w^\pm_k(\eta)=\left(\frac{1}{a(\eta)\sqrt{2k}}+y^\pm_k(\eta)\right)
e^{\pm \bar e k\eta}
\ee
and realize that $G_0(k,\eta,\eta')$ in eq. \eqref{F4} obtains from $\tilde G$ for $y^\pm _k(\eta)=0$. The functions $y^\pm_k(\eta)$ obey the inhomogeneous differential equation
\ba\label{F11}
&&\partial^2_\eta y^\pm_k +2(\h\pm \bar e k)\partial_\eta y^\pm_k\pm 2\bar e k\h y^\pm_k\nn\\
&&\qquad\quad=\frac{1}{a\sqrt{2k}}(\partial_\eta\h+\h^2).
\ea
It is straightforward to verify that our ansatz solves eq. \eqref{F6} for $\eta\neq \eta'$. We may take $\eta>\eta'$ and evaluate the difference between l.h.s. and r.h.s. of eq. \eqref{F6},
\ba\label{F12}
&&\tilde D_k 
\left[\left(
\frac{y^-_k(\eta)}{a(\eta')\sqrt{2k}}+
\frac{y^+_k(\eta')}{a(y)\sqrt{2k}}+
y^-_k(\eta)y^+_k(\eta')\right)
e^{-\bar ek(\eta-\eta')}\right] \nn\\
&&\quad -\big(\h^2(\eta)+\partial_\eta\h(\eta)\big)
\frac{1}{2a(\eta)a(\eta')k}
e^{-\bar ek(\eta-\eta')}=0.
\ea
Thus eq. \eqref{F6} is indeed obeyed, and similarly for $\eta<\eta'$.

Simple solutions with constant $y^\pm_k$,
\be\label{F13}
y^\pm_k=\pm\frac{f}{\bar e\sqrt{2k^3}},
\ee
are found if the scale factor obeys
\be\label{F14}
\partial^2_\eta a=f\partial_\eta(a^2).
\ee
This is the case for de Sitter space with $\bar e=i,a=-1/(H_0\eta),f=H_0$, where
\be\label{F15}
w^-_k=\left(\frac{1}{a\sqrt{2k}}+\frac{iH_0}{\sqrt{2k^3}}\right)
e^{-ik\eta}~,~w^+_k=(w^-_k)^*.
\ee
The Green's function defined by eqs. \eqref{F7}, \eqref{F15} is precisely the one of the Bunch-Davies vacuum \cite{BD}.

For geometries obeying eq. \eqref{F14} the explicit form of our ansatz for the propagator reads for $\eta>\eta'$
\ba\label{F16}
\tilde G_>(k,\eta,\eta')&=&G_0(k,\eta,\eta')+\tilde G_1(k,\eta,\eta')
+\tilde G_2(k,\eta,\eta'),\nn\\
\tilde G_1(k,\eta,\eta')&=&
\left(
\frac{1}{a(\eta)}-\frac{1}{a(\eta')}\right)
\frac{fe^{-\bar ek(\eta-\eta')}}{2\bar ek^2},\nn\\
\tilde G_2(k,\eta,\eta')&=&-
\frac{f^2 e^{-\bar ek(\eta-\eta')}}{2\bar e^2k^3},
\ea
where we observe the relation
\be\label{F17}
\tilde G_1(k,\eta,\eta')=
-\frac{a(\eta)a(\eta')}{f\big[a(\eta)-a(\eta')\big]}
\partial_\eta\tilde G_2(k,\eta,\eta').
\ee
Using this explicit form we can now investigate if the ansatz \eqref{F10}, \eqref{F13} indeed obeys the defining eq. \eqref{F2}. We will see that this is not the case for general $f$. 

\subsection{Solution of inhomogeneous equation}
For the general investigation of the inhomogeneous term in the propagator equation \eqref{F2} we denote the propagator for $\eta>\eta'$ by $G_>$ and similarly use $G_<$ for $\eta<\eta'$,
\be\label{65A}
G\kk =G_>\kk \theta (\eta-\eta')+G_<\kk \theta(\eta'-\eta).
\ee
The symmetry of the Green's function implies
\be\label{65B}
G_<\kk=G_>(k,\eta',\eta).
\ee
For Minkowski signature the reality condition takes the form
\be\label{65C}
G^*_>\kk=G_>(k,\eta',\eta),
\ee
while for euclidean signature $G_>$ and $G_<$ are real.

We decompose $G_>$ into its symmetric and antisymmetric part with respect to an exchange of the time arguments,
\ba\label{F18}
G_>(k,\eta,\eta')&=&G_s(k,\eta,\eta')+G_a(k,\eta,\eta')\nn\\
G_s(k,\eta,\eta')&=&\frac12\big (G_>(k,\eta,\eta')+G_>(k,\eta',\eta)\big)\nn\\
G_a(k,\eta,\eta')&=&\frac12\big(G_>(k,\eta,\eta')-G_>(k,\eta',\eta)\big).
\ea
The propagator for arbitrary $\eta$ and $\eta'$ becomes then
\be\label{F19}
G\kk=G_s\kk+G_a\kk
\big(2\theta(\eta-\eta')-1\big).
\ee
Applying the operator $\tilde D_k$, eq. \eqref{F8}, yields
\be\label{F20}
\tilde D_kG=\tilde D_kG_s+(\tilde D_kG_a)\big(2\theta(\eta-\eta')-1\big)
+2\partial_\eta G_a\delta(\eta-\eta').
\ee
The first two terms vanish by virtue of eq. \eqref{F8}. Only the antisymmetric part $G_a$ can contribute to the inhomogeneous term in eq. \eqref{F2}. In particular, the equal time propagator $G(k,\eta)=G(k,\eta,\eta)$ only contributes to $G_s$ and therefore not to the inhomogeneous part. Besides the conditions
\be\label{72A}
\tilde D_k G_s=0~,~\tilde D_k G_a=0
\ee
the Green's function has to obey
\be\label{60A}
\partial_\eta G_a(k,\eta,\eta')_{|\eta=\eta'}=-\frac{\bar e}{2a^2}.
\ee

For the particular solution \eqref{F16} the time derivative of the antisymmetric part of $G_0$ already produces the correct coefficient of the term $\sim \delta(\eta-\eta')$ in eq. \eqref{F2}. For this solution \eqref{F16} eq. \eqref{60A} is therefore only obeyed if
\be\label{F21}
\partial_\eta(\g_1+\g_2)_{a|\eta=\eta'}=0. 
\ee
One has
\ba\label{F22}
(\g_1)_a&=&\frac{f}{4\bar ek^2}
\left(\frac{1}{a(\eta)}-\frac{1}{a(\eta')}\right)
\left(\e+e^{\bar ek(\eta-\eta')}\right)\nn\\
(\g_2)_a&=&-\frac{f^2}{4\bar e^2k^3}
\left(\e-e^{\bar ek(\eta-\eta')}\right),\nn\\
\ea
and, for $\eta=\eta'$,
\be\label{F23}
\partial_\eta(\g_1)_a=-\frac{f\h}{2\bar e k^2a}~,~
\partial_\eta(\g_2)_a=\frac{f^2}{2\bar ek^2}.
\ee
The condition \eqref{F21} is therefore obeyed only for $f=0$ or $f=\h/a$. The second relation is precisely realized for de Sitter space with $f=H_0$. We conclude that the Green's function of the Bunch-Davies vacuum indeed obeys eq. \eqref{F2}. De Sitter space is the only solution of eq. \eqref{F14} with constant $\h/a$. We conclude that for all other solutions of eq. \eqref{F14} the ansatz \eqref{F10} obeys the homogeneous eq. \eqref{F9}, but fails to be a solution of the inhomogeneous propagator equation \eqref{F2}. 

More generally, there are many solutions of the homogeneous equation \eqref{F9}. For example, any given solution of eq. \eqref{F8} for $w^\pm_k(\eta)$ can be multiplied by an arbitrary complex constant $\alpha^\pm_k$. Replacing in eq. \eqref{F7} the factors $w^\pm_k(\eta)$ by $\alpha^\pm_k w^\pm_k(\eta)$ still  leads to a solution of eq. \eqref{F9}. It is only the inhomogeneous term in eq. \eqref{F2} that fixes the values of $\alpha^\pm_k$.

\subsection{General solution of propagator equation in de}

~{\bf  Sitter space}

For de Sitter space the Green's function \eqref{F7}, \eqref{F15} is not the only solution of the propagator equation \eqref{F2}. For $\eta>\eta'$ the most general solution of eq. \eqref{F9} reads
\be\label{64A}
G_>(k,\eta,\eta')=b(k,\eta')w^-_k(\eta)+c(k,\eta')w^+_k(\eta),
\ee
with $w^\pm_k(\eta)$ given by eq. \eqref{F15} and $b(k,\eta'),c(k,\eta')$ arbitrary complex functions. For $\eta<\eta'$ we employ $G_<(k,\eta,\eta')=G_>(k,\eta',\eta)$. Applying eq. \eqref{F9} for $\eta<\eta'$ implies
\ba\label{64B}
&&b(k,\eta')=\alpha_+(k)w^+_k(\eta')+\alpha_-(k)w^-_k(\eta'),\nn\\
&&c(k,\eta')=\beta_+(k)w^+_k(\eta')+\beta_-(k)w^-_k(\eta'),
\ea
with arbitrary complex constants $\alpha_\pm(k),\beta_\pm(k)$. We infer

\ba\label{64C}
&&G_a(k,\eta,\eta')=\frac12\big[\alpha_+(k)-\beta_-(k)\big]\nn\\
&&\qquad\times\big[w_k^-(\eta)w_k^+(\eta')-w_k^+(\eta)w_k^-(\eta')\big],
\ea
and
\be\label{64D}
\partial_\eta G_a(k,\eta,\eta')_{|\eta=\eta'}=-\frac{i}{2a^2}\big (\alpha_+(k)-\beta_-(k)\big).
\ee
This fixes
\be\label{64E}
\alpha_+(k)-\beta_-(k)=1.
\ee
The general solution of the propagator equation \eqref{F2} therefore involves three free functions $\alpha_-(k),\beta_+(k)$ and $\beta_-(k)$, with $\alpha_+(k)$ fixed by $\alpha_+(k)=1+\beta_-(k)$. 

We next employ the reality condition \eqref{65C}. This imposes the restrictions
\be\label{77A}
\alpha_-(k)=\beta^*_+(k)~,~\alpha^*_+(k)=\alpha_+(k)~,~\beta^*_-(k)=\beta_-(k).
\ee
We are left with three real free functions that we denote by
\ba\label{77B}
\alpha(k)&=&\alpha_+(k)+\beta_-(k),\nn\\
\beta(k)&=&\beta_+(k)+\beta^*_+(k)\nn\\
\gamma(k)&=&-i\big(\beta_+(k)-\beta^*_+(k)\big).
\ea
The most general propagator in de Sitter space reads
\ba\label{77C}
&&G_s\kk=\frac12\alpha(k)\big\{\www \et\ww(\eta')+\ww (\eta) \www (\eta')\big\}\nn\\
&&\quad  +\frac12\beta(k)\big\{\ww\et\ww(\eta')+\www\et\www(\eta')\big\}\nn\\
&&\quad +\frac i2\gamma(k)\big\{\ww\et\ww(\eta')-\www\et\www(\eta')\big\},
\ea
and 
\be\label{77D}
G_a\kk=\frac12\big\{ \www\et\ww(\eta')-\ww\et\www(\eta')\big\}.
\ee

The equal time correlation function can be extracted from $G_s$,
\ba\label{77E}
&&G(k,\eta)=\alpha(k)\left\{\frac{1}{2a^2k}+\frac{H^2_0}{2k^3}\right\}\\
&&+\beta(k)\left\{\left(\frac{1}{2a^2k}-\frac{H^2_0}{2k^3}\right)\cos (2k\eta)+\frac{H_0}{k^2a}\sin (2k\eta)\right\}\nn\\
&&\gamma(k)\left\{-\left(\frac{1}{2a^2k}-\frac{H^2_0}{2k^3}\right)\sin(2k\eta)+\frac{H_0}{k^2a}\cos (2k\eta)\right\}.\nn
\ea
A requirement of positivity of $G(k,\eta)$ imposes some restrictions on $\alpha,\beta$ and $\gamma$ - the amplitude of the oscillating last two terms should not exceed the term $\sim\alpha$. The functions $\alpha(k)$, $\beta(k)$ and $\gamma(k)$ can be related to initial values of $G(k,\eta)$ and its time derivatives at large negative $\eta$. They therefore contain the information about the initial conditions, as we will discuss in more detail in sect. \ref{Effective quantum fields and initial conditions}. We show $G(k,\eta)$ for different values of $\alpha$, $\beta$ and $\gamma$ in fig. 1. At the time of horizon crossing the memory of initial conditions is not lost. Since initial conditions can be set independently for each $k$-mode the general solution \eqref{77E} retains memory of both the amplitude and the shape of the initial fluctuation spectrum. 

One may investigate a particular subset of initial conditions for which the correlation function is invariant under the symmetries of de Sitter space. We will show in appendix E that the symmetries of de Sitter space require the functions $\alpha,\beta$ and $\gamma$ to be independent of $k$. Nevertheless, we have no unique Green's function even at this stage, but rather find a three-parameter family of de Sitter invariant correlation functions. In the limit of modes far outside the horizon, $|k\eta|\ll 1$, the equal time correlation approaches
\be\label{77F}
\lim_{k\eta\to 0}G(k,\eta)=(\alpha-\beta)\frac{H^2_0}{2k^3}.
\ee
For $\alpha\neq\beta$ this has the same $k$-dependence and therefore the same spectrum of the fluctuations as for the usually assumed Bunch-Davies vacuum. Nevertheless, the amplitude of the fluctuations involves an undetermined factor $\alpha-\beta$. 

The relevant quantity for the observable cosmic fluctuation is the value of $G(k,\eta)$ at horizon crossing, $k/a=H_0,~k\eta=-1$, cf. sect.~\ref{Scalar correlation function in inflationary cosmology}. There we find
\be\label{87A}
G(k\eta=-1)=c_{hc}\frac{H^2_0}{k^3},
\ee
with 
\be\label{87B}
c_{hc}=\alpha-\beta\sin (2)+\gamma\cos (2)\approx \alpha-0.42\gamma-0.91\beta.
\ee
For $\alpha,\beta,\gamma$ of the order one this is only a moderate difference from the value for the Bunch-Davies vacuum $c_{hc}=1$. On the other hand, large values of $\alpha,\beta,\gamma$ can strongly modify the fluctuation amplitude. An example with $\beta=\gamma=0$ is discussed in sect. \ref{Impact of initial conditions on primordial fluctuations}.

If de Sitter symmetry is realized the spectrum $\sim k^{-3}$ is independent of the detailed values of $\alpha,\beta$ and $\gamma$. The main features of the observable spectrum follow already from symmetry and dimensionality, despite the fact that memory of initial conditions remains for the general solution \eqref{77E}. For symmetric correlation functions this memory affects the relation between the fluctuation amplitude and the Hubble parameter, which in turn is given by the inflaton potential. For the connection between the observed fluctuation amplitude and the inflaton potential it will be an important task to establish if there are good reasons for a choice $\alpha=1,\beta=\gamma=0$, or if the fluctuation amplitude remains to some extent an undetermined dynamical quantity which retains information about the ``initial state'' of the universe. In appendix A we show that a Fourier transform to position space only exists 
for $\alpha=\beta$. In 
this case the fluctuation amplitude is suppressed as compared to the scaling correlation by a factor around ten.  

\subsection{Scaling correlation}

We next formulate our conjecture for the scaling correlation that may be realized as an effective attractor solution for a large range of initial conditions. While within our present approximation no effective attraction to a universal propagator is found, attractor properties are possible beyond our approximation, as we will discuss in sect. \ref{Time evolution of correlation function}. If such an attractor exists there are several reasons to identify it with the scaling correlation that will be discussed next.

For the selection of the scaling correlation we propose an additional condition on the propagator which will make it unique (except for $k=0$). It is based on the behavior of the analytically continued Green's function in the infinite past for $\eta'\to-\infty$. The condition requires that $G$ should not diverge exponentially in this limit. This condition is the analogue of the $i\epsilon$-prescription for the Green's function in Minkowski space. It is our conjecture that for a long enough duration of inflation before horizon crossing of the observable fluctuations the time evolution of the propagator approaches this unique ``de Sitter propagator'' for rather arbitrary initial conditions. This should hold at least for the short-distance modes that become observable later.

In the limit of flat space our conjecture selects precisely the Minkowski propagator as obtained from the $i\epsilon$-prescription in sect. \ref{Correlation function for free massless scalar}. Furthermore, our conjecture can be applied independently for every $k$-mode. For a long enough duration of inflation before horizon crossing of the observable fluctuations the observable modes correspond to very large $k$, with $k/a\gg H$. In this range the propagator equation equals the one of flat space up to tiny corrections. As we have seen in sect. \ref{Correlation function for free massless scalar} our conjecture on the behavior of the analytically continued Green's function precisely selects the Lorentz-invariant propagator for flat space. This motivates our condition for the scaling correlation. For practical purposes it is actually sufficient that our conjecture holds for the large $k$-modes at the beginning of inflation. As we will see in sect. \ref{Time evolution of correlation function} the property characterizing our conjecture is preserved by the time evolution.

Even though the direct association with the flat space propagator yields a very good motivation for the selection of the scaling correlation, it remains a conjecture that the flat space propagator is indeed approached as a result of the time evolution. We recall that the relevant modes may have physical momenta $k/a$ much larger than the Planck mass at the beginning of inflation and could involve unknown ``trans-Planckian'' physics \cite{BM1,BM2}. 

Let us now apply our conjecture to the Green's function in de Sitter space. We need the mode functions $\w$ for a value of $\bar e$ slightly different from $\bar e=i$ for Minkowski signature. They are given by eq. \eqref{F10} for arbitrary $\bar e=\exp (i\varphi)$. For $\varphi=\frac{\pi}{2}-\epsilon$ one has to replace for $w^-_k$ in eq. \eqref{F15} the factor $e^{-ik\eta}$ by $e^{-ik\eta} e^{-\epsilon k\eta}$ and similarly for $w^+_k\sim e^{ik\eta}e^{\epsilon k\eta}$. For a given $\eta$ the contributions $\sim\alpha_-(k)$ and $\beta_-(k)$ in eqs. \eqref{64A}, \eqref{64B} diverge exponentially for $\eta'\to-\infty$. We require that the correlation between the value of the scalar field in the infinite past, $\tilde\varphi(\eta'\to-\infty,\vec x ')$, and the one at given time $\eta, \quad \tilde\varphi(\eta,\vec x)$, is not exponentially divergent. This implies $\alpha_-=\beta_-=0$. The function $\beta_+(k)$ remains undetermined by these considerations, such that our condition still allows for $\eta>\eta'$
\be\label{70A}
G(k,\eta,\eta')=w^-_k(\eta)w^+_k(\eta')+\beta_+(k)w^+_k(\eta)w^+_k(\eta').
\ee
This differs from the situation in flat space where the requirement that $G$ does not diverge exponentially in the infinite future for $\eta\to\infty$ also excludes the combination $w^+_k(\eta)w^+_k(\eta')$. For de Sitter space one has $\eta<0$ and the correlation \eqref{70A} remains finite in this range. 

The property excluding nonzero $\beta_+$ is the reality condition \eqref{65C}. Indeed,
\be\label{73A}
G^*(k,\eta,\eta')=w^-_k(\eta')w^+_k(\eta)+\beta^*_+w^-_k(\eta')w^-_k(\eta)
\ee
is compatible with eqs. \eqref{F15}, \eqref{65C} only for $\beta_+=0$. For the most general Green's function consistent with reality, as given by eqs. \eqref{77C}, \eqref{77D}, one infers from $\alpha_-=\beta^*_+=0$ that $\beta=\gamma=0$, while $\beta_-=0$ implies $\alpha_+=\alpha=1$. It is the combination of reality and the boundary behavior of the analytically continued Green's function that selects the unique scaling correlation for de Sitter space. This will be called ``de Sitter propagator''.

By virtue of our conjecture we are left (for $k\neq 0)$ with a unique propagator in momentum space, given for $\eta>\eta'$ by 
\ba\label{71A}
G(k, \eta, \eta')&=&\left\{\frac{1}{2a(\eta)a(\eta')k}-\frac{i H_0}{2k^2}\left(\frac{1}{a(\eta)}-\frac{1}{a(\eta')}\right)\right.\nn\\
&&+\left.\frac{H_0^2}{2k^3}\right\} \exp\left\{-ik(\eta-\eta')\right\}.
\ea
For $k=0$ the ambiguity of adding a constant in position space translates to the possibility of adding to $G(k,\eta,\eta')$ a term $\sim \delta(k)$. This ambiguity can only be resolved if we regulate $\Gamma^{(2)}$, e.g. by adding a small mass term. The absence of regularization translates to the absence of a well defined Fourier transformation to position space, as we discuss in appendix A. Indeed, the term $k^{-3}$ results in an infrared divergence of the Fourier integral. 

This infrared divergence of the de Sitter propagator shows also up in loop calculations \cite{MC2,VF,LI2,SW,LY,GHT,PJS,JS1}. This is of no direct concern here since loop effects are supposed to be already incorporated in the quantum effective action. For practical purposes the infrared singular behavior is often cut off by a physical infrared cutoff connected to a typical characteristic length scale for a given observation \cite{LY,GHT}.

The analytic continuation of the propagator \eqref{71A} reads explicitly
\ba\label{90A}
&&G_>(k,\eta,\eta')=w_k^-(\eta)w_k^+(\eta')\\
&&\qquad=\frac{1}{2k}\left(\frac{1}{a(\eta)}-\frac{H_0}{\bar{e}k}\right)\left(\frac{1}{a(\eta')}+\frac{H_0}{\bar{e}k}\right)e^{-\bar{e}k(\eta-\eta')}.\nn
\ea
For euclidean signature the analytic continuation of de Sitter space is a maximally symmetric space with negative curvature $(R=-12H_0^2/\bar{e}^2)$, not the sphere. The Green's function for this non-compact space is given by eq. \eqref{90A} with $\bar{e}=1$. We observe that it is real and turns negative for $a(\eta)H_0/k>1$. 

While appealing by its simplicity, our condition on the behavior of the analytically continued Green's function requires justification by the dynamics. Ultimately, one has to understand that dynamically a unique Green's function is singled out among a much more general class of time evolving correlation functions. This can at best be realized in the form of an asymptotic attractor solution that is approached rapidly enough such that the memory about the initial state is effectively forgotten. We will come back to this question in the next section and in sect. VIII.

\section{Time evolution of correlation function}
\label{Time evolution of correlation function}

The key question of this paper, namely if the observable cosmic fluctuations keep some memory of the details of initial conditions or not, can only be answered by following the time evolution of the correlation function from the infinite (or very remote) past to the moment when the wavelength of the fluctuation goes out of the horizon. We want to know if the propagator at horizon crossing takes a universal form, only dependent on the inflaton potential and other properties of the effective action, but independent of the details of initial conditions. Or else if there remains memory of the initial conditions. For this purpose we have to investigate large families of initial conditions for correlation functions and find out if they are all attracted to a universal propagator, and how long it takes until memory of the initial conditions is essentially lost.

A central tool will be appropriate evolution equations for the time dependence of Green's functions. In this section we derive these equations for a scalar field in a homogeneous and isotropic cosmological background. We consider interacting theories with an arbitrary inflaton potential in sect. \ref{Scalar correlation function in inflationary cosmology}. An exact evolution equation is found if three approximations are made for the general form of the effective action: (i) The effective action contains no explicit dependence on position and time due to boundary effects. (ii) Backreaction effects are neglected. In principle, the evolution of geometry and the inflaton field could be influenced by the correlation function. We assume that this effect is negligible. (iii) The second functional derivative of the effective action contains no more than two time derivatives. We find that with these approximations the memory of initial conditions is still present at horizon crossing. We briefly discuss possible symmetrization and equilibration effects that may arise from modifications of the effective action beyond these 
three approximations. 

For a given background geometry the propagator equation \eqref{F2} is a local differential equation, in the sense that it only involves properties for a given time $\eta$. Together with the boundary conditions it defines the propagator uniquely for all times. If the Green's function becomes universal and the detailed initial conditions play no role, one expects that $G(\eta,\eta')$ depends only on the properties of $\Gamma^{(2)}$ in a restricted time interval that comprises both $\eta$ and $\eta'$. In particular, the equal time Green's function $G(k,\eta)$ only needs knowledge of the background geometry and its time derivatives at time around $\eta$. This crucial feature of a universal propagator makes the correlation functions at a given time $\eta$ independent of the global form of the quantum effective action. Only its properties around $\eta$ are needed. This fact permits the computation of fluctuations even for rather complex geometries for which an overall determination as for Minkowski - or de Sitter 
space would seem very hard. In contrast, if memory of initial 
conditions is kept, more global features of $\Gamma^{(2)}$ would be needed. The range of time around $\eta$ for which $\Gamma^{(2)}$ needs to be known corresponds roughly to the time scale on which information on boundary conditions is lost.  

In the quantum mechanical operator formalism one distinguishes between in-in and in-out correlations. On the level of the quantum effective action this issue is related to the boundary conditions for the Green's function. In principle, boundary conditions could be imposed both in the far past and far future. In practice, however, a given set of initial conditions often severely restricts the allowed form of the correlation function in the far future. On the other hand, if one remains within the allowed range the future boundary conditions are not expected to play any role for the physics in the inflationary period. Within this restriction the in-in and in-out situation is physically almost equivalent. We impose boundary conditions only in the remote past or the beginning of inflation. Formally, this corresponds to an in-in situation. In case that a universal form of a scaling correlation is reached the boundary conditions play no role altogether.

\subsection{Evolution of mode functions and product} 

~{\bf form of propagator}

Our aim is the derivation of a time evolution equation for the equal-time correlation function $G(k,\eta)$. A given solution to this equation will typically be sufficient in order to reconstruct the unequal time correlation function as well. We may take de Sitter space as an example. A particular solution of the equal time correlation function fixes the functions $\alpha(k)$, $\beta(k)$ and $\gamma(k)$ in eq. \eqref{77E}. In turn, the unequal time correlation function \eqref{77C}, \eqref{77D} is fixed. 

Let us first assume that at some time $\eta_0$ the correlation function can be written in the product form \eqref{F7}. More precisely, this should hold if both $\eta$ and $\eta'$ are close to $\eta_0=(\eta+\eta')/2$. We will show that this product form remains preserved, with mode functions $w^\pm_k(\eta)$ solutions of the differential equation \eqref{F8}. Initial conditions for $w^\pm_k$ and $\partial_\eta w^\pm_k$ at $\eta_0$ are given by $w^\pm_k(\eta_0)$ and $\partial_\eta \w(\eta_0)$. The same holds for $\w(\eta')$. By definition of $\w(\eta)$ the Green's function constructed in this way obeys the homogeneous equation \eqref{F9} for $\eta\neq \eta'$. We therefore only need to establish that the normalization \eqref{60A} is preserved, such that our ansatz solves eq. \eqref{F2}. For a correlation function of the form \eqref{F7} this condition can be written in the form
\be\label{f1}
\partial_\eta\www(\eta)\ww(\eta)-
\partial_\eta\ww(\eta)\www(\eta)=-
\frac{\bar e}{a^2}.
\ee

We introduce rescaled functions
\be\label{f2}
v^\pm_k(\eta)=a(\eta)\w(\eta),
\ee
for which the normalization reads
\be\label{f3}
\partial_\eta v^-_k(\eta)v^+_k(\eta)-\partial_\eta v^+_k(\eta)v^-_k(\eta)=-\bar e.
\ee
The differential equation \eqref{F8} translates to
\be\label{f4}
\partial^2_\eta v^\pm_k=(\h^2+\partial_\eta\h+\bar e^2k^2)v^\pm_k.
\ee
Let us now assume that $v^\pm_k(\eta)$ are solutions of eq.~\eqref{f4} and compute the evolution of the normalization
\be\label{f5}
{\cal N}_k(\eta)=\partial_\eta v^-_k(\eta)v^+_k(\eta)-\partial_\eta v^+_k(\eta)v^-_k\et,
\ee
namely
\be\label{f6}
\partial_\eta{\cal N}_k\et=\partial^2_\eta v^-_k(\eta) v^+_k(\eta) -\partial^2_\eta v^+_k(\eta) v^-_k\et.
\ee
Here the mixed first derivatives have dropped out due to the relative minus sign of the two terms in eq. \eqref{f5}. Using eq. \eqref{f4} and noting that the r.h.s contains no derivatives of $v^\pm_k$ one obtains
\be\label{f7}
\partial_\eta{\cal N}_k\et=0.
\ee
The normalization does not change with time. If it is given by eq. \eqref{f3} for $\eta=\eta_0$ it will remain so for all $\eta$ as well. This establishes a simple exact evolution equation for propagators which have the product form \eqref{F7}. It is sufficient to solve the evolution equation for the factors $w^\pm_k$. We will see in sect. \ref{Effective quantum fields and initial conditions} how propagators with product form can be associated with effective pure quantum states. 

We have seen that the scaling correlation has the product form \eqref{F7}. Consider now a propagator which has the product form at time $\eta_0$, and further has the property that its analytic continuation for $\epsilon>0$ does not diverge for $\eta'\rightarrow-\infty$. Both the product form and the analyticity condition remain preserved for all later times $\eta$. For the scaling correlation that is uniquely defined by these conditions it is therefore sufficient that it is realized at some arbitrary time $\eta_0$. The correlation function will then be given by the scaling correlation for all later times $\eta>\eta_0$. This stability property is an important ingredient for any asymptotic solution to which other solutions may converge effectively. 

\subsection{Construction of scaling correlation by solution}

~{\bf of evolution equation}

The conservation of the properties of the scaling correlation by the time evolution provides us with a powerful tool for the construction of the scaling correlation at some time $\eta$. it is sufficient to find the scaling correlation at some early time, e.g. $\eta\to-\infty$. The scaling correlation for later times obtains by solving the evolution of $w^\pm_k(\eta)$. 

Assume one has found the general solution of eq. \eqref{f4} in some time neighborhood of $\eta$. This is to be understood in the sense that the overall geometry may be complex, but that in the vicinity of $\eta$ the behavior of geometry for much earlier or later times can be neglected for the general solution to the propagator equation. We need to select from the general solution the particular one $v_k^{\pm}$ that obeys the ``analyticity constraint'' for the scaling correlation. In general, it is a linear combination of two basis functions.  The correct choice of $v^\pm_k\et$ as linear combination of the basis functions of the general solution, including the correct normalization, can be established by connecting the local geometry smoothly to a geometry with known propagator in the past at $\eta_0<\eta$. The 
precise way how this interpolation is done is unimportant - the result is the same for all interpolations. This robustness with respect to choice of interpolation is a result of the uniqueness of the scaling correlation. 

As an example, we could have obtained the correct normalization of the de Sitter propagator for the different momentum modes along similar lines. For a given $k$-mode the inverse propagator becomes arbitrarily close to the free propagator in Minkowski space for times sufficiently far in the past, more precisely for $a\ll k/H_0$. For $a\to 0$ (or $\eta,\eta'\to-\infty)$ we can therefore use the propagator $G_0 \kk$ which takes indeed the form \eqref{F7} with 
\be\label{f8}
v^\pm_k(\eta\to-\infty)=\frac{1}{\sqrt{2k}}e^{\pm \bar e k\eta}.
\ee
This fixes the selection and normalization of $v^\pm_k(\eta)$ for all later times. The result agrees with the explicit construction in sect.~\ref{Propagator in momentum space}. 

The exact evolution equation for the scaling correlation provides for a solid basis for an approach that is often used in practice. One solves in Fourier space the time evolution of a small deviation $\delta\varphi (k,\eta)$ from the background field. The equal time propagator is taken as $G(k,\eta)\sim |\delta\varphi(k,\eta)|^2$. The proportionality factor is then determined by choosing ``initial conditions'' by comparison with the (approximative) correlation function for a quantized scalar field. This procedure is usually considered as an approximation. It is actually {\em exact} if the field equations are the ones derived from the quantum effective action. The functions $w_k\et$ are indeed proportional to $\delta\varphi(k,\eta)$, with a $k$-dependent proportionality factor. Both $w_k\et$ and $\delta\varphi(k,\eta)$ obey the same linear differential evolution equation.

\subsection{Evolution equation for equal time scaling}

~{\bf correlation }

We next turn to the general evolution equation for the equal time Green`s function for arbitrary initial conditions. It is formulated as a system of differential equations for three equal-time Green's functions for the field $\tilde{\phi}$ and its time derivative $\tilde{\pi} = \partial_\eta\tilde{\phi}$. We first derive here the evolution equation for the product form \eqref{F7} and show later that it actually holds for the most general propagator, imposing no longer the product form. 

For fixed $\eta'$ the evolution in $\eta$ obeys 
\be\label{88A}
(\partial^2_\eta+2\h \partial_\eta+k^2)G(k,\eta,\eta')=0.
\ee
The time evolution equation for the equal time correlation function $G(k,\eta)=G(k,\eta,\eta)$ has to take into account that derivatives act on both factors of
\be\label{88B}
G(k,\eta)=w^-_k(\eta)w^+_k(\eta)=G_{\varphi\varphi}(k,\eta).
\ee
We employ the definitions 
\ba\label{88D}
G_{\pi\varphi}(k,\eta)&=&\frac12 (\partial_\eta \www\et\ww\et+\www\et\partial_\eta \ww\et\big),\nn\\
G_{\pi\pi}\kt&=&\partial_\eta\www\et\partial_\eta\ww\et,
\ea
with
\be\label{88C}
\tilde \pi(\eta,\vec k)=\partial_\eta\tilde\varphi(\eta,\vec k).
\ee
They describe connected correlation functions of the type
\ba\label{88E}
&&\frac12\big(\langle\tilde\pi\ev\tilde\varphi^*(\eta,\vec k')\rangle_c+
\langle\tilde\varphi\ev\tilde\pi^*(\eta,\vec k')\rangle_c\big)\nn\\
&&\qquad=G_{\pi\varphi}\kt\delta(k-k'),\\
&&\langle\tilde\pi \ev\tilde \pi^* \eev\rangle_c=G_{\pi\pi}\kt\delta(k-k').\nn
\ea

The evolution equation for the equal time correlation function takes the form
\ba\label{88F}
\partial_\eta G_{\varphi\varphi}\kt&=&2G_{\pi\varphi}\kt,\nn\\
\partial_\eta G_{\pi\varphi}\kt&=&G_{\pi\pi}\kt-2\h G_{\pi\varphi}\kt-k^2 G_{\varphi\varphi}\kt,\nn\\
\partial_\eta G_{\pi\pi}\kt&=&-4\h G_{\pi\pi}\kt -2k^2 G_{\pi\varphi}\kt.
\ea
These equations bare certain similarities with the ones investigated in ref. \cite{ABW2} (see also refs. \cite{BW,ABW1}), which are based on exact evolution equations for the equal time effective action \cite{CWET1,CWET2}. There are two important differences, however. The present equations are based on the quantum effective action, which already includes the effects of quantum fluctuations. This explains why these equations can be exact for a simple given form of $\Gamma$, as in our case. In contrast, the equations in ref. \cite{ABW2} are based on the classical action and therefore comprise additional terms involving higher order vertices. A second characteristic feature is the Hubble damping related to $\h>0$, not present for investigations in Minkowski space. We will see below that eq. \eqref{88F} does not need the particular product form \eqref{88B} but rather follows directly from the definition of the equal time Green's function, together with eq. \eqref{88E}. 

\subsection{Time dependence of equal time Green's function}

~{\bf  in de Sitter space}

The system of equations \eqref{88F} permits us to address our question to what extent equal time correlation functions keep memory of initial conditions. We can specify initial conditions for $G_{\varphi\varphi}, G_{\pi\varphi}$ and $G_{\pi\pi}$ at some large negative $\eta_0$ and follow the evolution towards a later time $\eta$, for example at horizon crossing. In this section we will do so for de Sitter space with $\h=-1/\eta$, while we extend the discussion to more general geometries and the inclusion of an inflaton potential in sect.VIII.  

The combination
\be\label{88G}
\alpha^{-2}_G=G_{\varphi\varphi}G_{\pi\pi}-G^2_{\pi\varphi}
\ee
can be used to construct a conserved quantity. (We use the symbol $\alpha_G$ for reasons of consistency with earlier work on the time evolution of Green's functions. It should not be confounded with the parameter $\alpha(k)$ in the general propagator in de Sitter space \eqref{77C}). The combination $\alpha^{-2}_G$ evolves according to
\be\label{88H}
\partial_\eta\alpha^{-2}_G=-4\h\alpha^{-2}_G=\frac{4}{\eta}\alpha^{-2}_G,
\ee
with solution 
\be\label{88I}
\alpha^{-2}_G=\frac{c_\alpha}{a^4}.
\ee
The quantity $\alpha^{_2}_G a^4$ is conserved. One may verify this explicitly for $G_{\varphi\varphi}=\www\et\ww\et$ with $\w$ given by eq. \eqref{F15}. With 
\be\label{88J}
G_{\varphi\varphi}=\frac{1}{2a^2k}+\frac{H^2_0}{2k^3}~,~
G_{\pi\varphi}=-\frac{H_0}{2ak}~,~
G_{\pi\pi}=\frac{k}{2a^2},
\ee
one has
\be\label{88K}
\alpha^{-2}_G=\frac{1}{4a^4}.
\ee
We note that $\alpha_G(k)a^4$ is conserved for every $k$-mode. The existence of an infinite number of conserved quantities is a clear sign that memory is not lost. It prevents the approach to a scaling correlation.

In appendix B we analyze the behavior of the general solution in terms of a partial fixed point (scaling solution) and deviations from it. In this form the analysis can be generalized to a much wider context. It also highlights the role of the scaling correlation as a fixed point. If this fixed point is approached faster than by Hubble damping the memory of initial conditions would be lost asymptotically.

For our present setting we actually know already the most general form of the solution to the evolution equation \eqref{88F}: it is given by eq. \eqref{77E}. Indeed, it is straightforward to verify that eq. \eqref{77E} solves the evolution equation. The three free integration constants $\alpha,\beta$ and $\gamma$ correspond to the three initial values for $G\va,G_{\pi\varphi}$ and $G_{\pi\pi}$. The asymptotic value \eqref{XX1} reflects eq. \eqref{77F}. We conclude that for a free massless scalar field in a given geometrical background the information contained in the initial state is not lost. This information influences strongly the amplitude and spectrum of the fluctuations. The situation remains the same for a massive scalar field. This is similar to the behavior of the correlation function for a free scalar field in flat space. Infinitely many conserved quantities $\alpha_G(k)$ prevent a loss of memory \cite{BW}. In order to see equilibration (e.g. the approach to a thermal equilibrium state) one needs approximations for which $\alpha_G(k)$ are no longer conserved. For de Sitter space these conserved quantities are replaced by $\alpha^{-2}_G a^4$.

\subsection{Equilibration in de Sitter space}

In flat space, the situation changes profoundly in the presence of nonlinearities induced by scattering. For an interacting scalar field the correlation functions reach for asymptotic times the values of thermal equilibrium, provided certain conditions for the energy distribution in the initial state are met \cite{ABW1}. For intermediate times an approximate partial fixed point may be reached \cite{BW}, corresponding to ``prethermalization'' \cite{BBWPT, PT}. It seems plausible, even though not proven at present, that in the presence of interactions a similar phenomenon of equilibration could take place in de Sitter space. This would substantiate the picture of de Sitter space as a type of a ``thermal state'' with a ``de Sitter temperature''. We note in this context that interactions are not restricted to scalar self-interactions. In the presence of metric fluctuations every model contains gravitational interactions. 

A possible equilibration process contains different facets. As a first aspect one would expect that equilibration leads to a state consistent with the symmetries of de Sitter space. In this case the functions $\alpha(k),\beta(k)$ and $\gamma(k)$ would all become independent of $k$.  For example, the high momentum modes may equilibrate at early times to a scale invariant spectrum, which implies de Sitter symmetry (cf. appendix E). As a consequence, a (partially) equilibrated state with de Sitter symmetry would lead for later times, $aH\gg k$, to $G\va\sim k^{-3}$, except for very special initial conditions that lead to $\beta=\alpha$. The spectrum for $k/aH\to 0$ takes now a universal form, loosing the memory of the initial conditions. This also holds at horizon crossing where symmetry and dimensional analysis lead to a spectrum $G\sim H^2_0/k^3$, cf. eqs. \eqref{87A}, \eqref{87B}. The shape of the spectrum, in particular the spectral index of the observable primordial fluctuations, is determined in this case by the background geometry at the time of horizon crossing. In contrast, the fluctuation amplitude is not fixed by symmetry properties. It depends on the parameters $\alpha$,
$\beta$ and $\gamma$. An approach to de Sitter symmetry therefore looses only part of the initial information.

As a second aspect one may expect that an equilibrated state obeys some ``fluctuation-dissipation relation''. This links the amplitude of the symmetric part of the Green's function, as given by $\alpha,\beta$ and $\gamma$, to the one of the antisymmetric part which is fixed by the inhomogeneous equation. It is this type of relation that may finally fix the fluctuation amplitude. If equilibration leads to $\alpha-\beta=1$ the amplitude takes a universal form for $k/(aH)\ll 1$, while not necessarily being universal at horizon crossing. If, in addition, $\beta=0,\gamma=0$, the scaling solution corresponding to the Bunch-Davies vacuum will be reached. We briefly discuss the fluctuation-dissipation relation in appendix C. 

We conclude that equilibration in an interacting theory could lead to a loss of memory concerning the precise initial conditions. It is this phenomenon that is supposed to select the unique de Sitter propagator \eqref{71A}, as defined by requirements for the analytic continuation. What remains to be settled is an answer to the question if an equilibration happens, and on what time scales it is realized. Only after answering these crucial questions one can asses the questions raised in the introduction, namely if the cosmic fluctuation spectrum and amplitude is independent or not of the detailed initial conditions. 

A key ingredient for the memory of initial conditions is the linearity of the system of equations \eqref{88F}. We will see below that this linearity is maintained for an interacting scalar field, provided our assumptions on the absence of explicit boundary contributions to $\Gamma$ and the neglection of backreaction hold. A possible equilibration process has therefore to go beyond these approximations. 

\subsection{General structure of evolution equation}

For getting further insight into the initial value problem we should consider the quantum effective action in the presence of boundary conditions, imposed for $\eta\to-\infty$ or for some very large negative $\eta_{in}$. Not too far from the boundary the effective action $\Gamma$ is not supposed to have the simple universal form \eqref{12} - some information about the boundary is expected to be still present. For a different form of $\Gamma$ the flow equation for the equal time propagator will no longer have the simple form \eqref{88F}. We want to understand the structure of these modifications.

For this purpose we will consider an approximation where higher than second derivatives with respect to $\eta$ can be neglected in $\Gamma^{(2)}$, such that (omitting the unit matrix)
\ba\label{G1}
\Gamma^{(2)}&=&{\cal A} \tilde D_\eta,\nn\\
\tilde D_\eta&=&\partial_\eta^2+2{\cal C}\partial_\eta+{\cal B},
\ea
with ${\cal A},{\cal B}$ and $\C$ arbitrary real quantities that are computable for given $\eta$ and $k$. (The function ${\cal A}$ should have no zeros such that $1/{\cal A}$ exists. Generalizations beyond the homogeneous and isotropic setting are possible.) In particular, the functions ${\cal A},{\cal B}$ and $\C$ may depend themselves on other fields or on $G_{\pi\pi}$, $G_{\pi\varphi}$ and $G_{\pi\pi}$, such that the system of equations becomes non-linear. They may also involve the antisymmetric part $G_a$, enforcing a normalization of $G_s$ by the properties of $G_a$. We next derive evolution equations for the equal time Green's function for a general effective action that has a second functional derivative \eqref{G1}. 

We use the general form \eqref{65A} and concentrate on the symmetric part $G_s$ \eqref{F18}. It obeys
\be\label{G2}
\tilde D_\eta G_s(\eta,\eta')=0~,~\tilde D_{\eta'}G_s(\eta,\eta')=0,
\ee
where we omit the label $k$. For the second equation \eqref{G2} the differential operator $\tilde D$ acts on $\eta'$. It follows from the first equation by using the symmetry of the Green's function. The time evolution of $G_s$ involves only the functions ${\cal B}$ and ${\cal C}$. The function ${\cal A}$ appears in the normalization of the antisymmetric part $G_a$. We may assume a normalization of the scalar field for which ${\cal A}=ia^2\et$, cf. eq. \eqref{12a}, but this is not crucial. 

The general definition of $G_{\varphi\varphi},G_{\pi\varphi}$ and $G_{\pi\pi}$ is given by eqs. \eqref{42C}, \eqref{F1A} and \eqref{88E} , with $G_{\varphi\varphi}=G_s(\eta,\eta)$. The relation 
\be\label{G3}
\partial_\eta G_{\varphi\varphi}=2G_{\pi\varphi}
\ee
follows directly from the definition. Similarly, one obtains 
\ba\label{G4}
\partial_\eta G_{\pi\varphi}&=&G_{\pi\pi}+\frac12 \kl \partial^2_\eta\tilde\varphi\tilde\varphi^*+\tilde \varphi\partial^2_\eta\tilde\varphi^*\kr_c\\
&=&G_{\pi\pi}+\frac12\big(\partial^2_\eta G_s(\eta,\eta')+\partial^2_{\eta'}G_s(\eta,\eta')\big)_{|\eta'=\eta}.\nn
\ea
Employing eq. \eqref{G2} and the definition \eqref{G1} this yields
\ba\label{G5}
\partial_\eta G_{\pi\varphi}&=&G_{\pi\pi}-\big\{\C(\eta) \partial_\eta G_s (\eta,\eta')+\C(\eta')\partial_{\eta'} G_s(\eta,\eta')\nn\\
&&\hspace{-0.2cm}+\frac12{\cal B}(\eta) G_s(\eta,\eta')+\frac12{\cal B}(\eta') G_s(\eta,\eta')\big\}_{|\eta'=\eta}
\ea
or
\be\label{G6}
\partial_\eta G_{\pi\varphi}=G_{\pi\pi}-2\C G_{\pi\varphi}-{\cal B}G_{\varphi\varphi}.
\ee
A similar line of argument gives
\be\label{G7}
\partial_\eta G_{\pi\pi}=-4\C G_{\pi\pi}-2{\cal B} G_{\pi\varphi}.
\ee
Eqs. \eqref{G3}, \eqref{G6}, \eqref{G7} form for any given ${\cal B}$ and ${\cal C}$ in eq. \eqref{G1} a system of exact evolution equations for the equal time correlation functions. The ansatz \eqref{G7} covers a rather wide family of situations for interacting scalars. 

For $\C=\h$ depending only on $\eta$ and ${\cal B}=k^2$ we recover our previous setting \eqref{88F}. Eq. \eqref{88F} is therefore not linked to the particular product form of the propagator \eqref{88B}, but holds for the most general propagator.

The presence of interactions does not necessarily induce non-linearities in the evolution equation. We may include an arbitrary potential $V(\varphi)$ for the inflaton. The second derivative $\partial V/\partial\varphi^2$, evaluated for the background field $\bar\varphi(\eta)$, amounts to a mass term $m^2(\eta)$ that may depend on $\eta$ through the background field. This results in $B(\eta)=k^2+m^2(\eta)a^2(\eta)$, but leaves the linear structure of the evolution equation unchanged. We will discuss the correlation function of massive scalar fields in sections \ref{Scalar correlation function in inflationary cosmology} and \ref{Impact of initial conditions on primordial fluctuations}. 

The general form of the evolution equations for the equal time Green's function is given by eqs. \eqref{G3}, \eqref{G6} and \eqref{G7}. If higher derivative terms play no role the physics of a possible equilibration towards the unique de Sitter propagator is encoded in the two functions $\C(k,\eta)$ and ${\cal B}(k,\eta)$. There are several possible sources for non-linear effects in the evolution equation. First, $\Gamma^{(2)}$ typically depends on other fields. The evolution of these fields may depend on the value of $G$. An example is the energy momentum tensor associated to the scalar field fluctuations which involves $G$. It influences the evolution of the metric, inducing thereby an effective non-linearity in the system of equations. This effect is called ``backreaction''.

Second, the form of the effective action may be time-dependent itself as a result of explicit boundary contributions to $\Gamma$. The functions $\mathcal{B}$ and $\mathcal{C}$ could depend directly on $G(k,\eta)$, or more generally on the unequal time Green's function $G(k,\eta,\eta')$. Furthermore, higher order vertices may become effectively functions of time, and this results in a time dependent second functional derivative $\Gamma^{(2)}$ 
through a non-trivial background field $\bar\varphi(\eta,\vec x)$. Away from equilibrium $\Gamma^{(2)}$ could have an imaginary part. A third effect is a richer variety of background solutions. Even if $\Gamma^{(2)}$ remains unchanged, initial values may be such that the background solution is no longer homogeneous and isotropic. Since $\Gamma^{(2)}$ results from an expansion around the background the functions $\C$ and ${\cal B}$ reflect such a change of the background.

We note that the formal way how non-equilibrium effects modify the time-evolution of the propagator depends on the choice of fields. For example, the effect of radiation and matter in gravity may appear as an explicit source term in the background equations for the metric, in the form of the energy momentum tensor in the Einstein equation. Alternatively, one could introduce a composite field for the energy momentum tensor, which would then be part of the solution for the background field. The situation is similar for the correlation functions. The non-linearities are most easily visible if $G_{\varphi\varphi},G_{\pi\varphi}$ and $G_{\pi\pi}$ appear themselves as fields. This is realized in the two-particle irreducible formalism \cite{BC,AB,BR,AB2} or for the equal time effective action \cite{CWET1,CWET2}. In the two-particle irreducible formalism it is the whole propagator $G(x,y)$ that appears as a field. Non-linearities are directly visible in the effective action $\Gamma_{2PI}[\varphi,G]$. The quantum 
effective action discussed in the present paper obtains from $\Gamma_{2PI}$ by inserting a solution $G[\varphi]$ of the field equation $\partial\Gamma_{2PI}/\partial G=0$. 

For general $\C$ and ${\cal B}$ the evolution of $\alpha^{-2}_G$ obeys 
\be\label{144A}
\partial_\eta(\alpha^{-2}_Ga^4)=4(\h-\C)(\alpha^{-2}_Ga^4).
\ee
An approach to the unique de Sitter propagator requires that $\C$ differs from $\h$, except for a zero if $\alpha^{-2}_Ga^4=1/4$. The corresponding partial fixed point has to be attractive as $\eta$ increases. A non-vanishing r.h.s. of eq. \eqref{144A} is necessary in order to avoid conserved quantities which would be an obstacle to equilibration. The quantity ${\cal C}-\h$ stands for an effective irreversibility beyond the purely geometric one induced by the Hubble damping. A fixed point for this quantity requires non-linearity of the evolution equations and is part of the fluctuation-dissipation relation. 

\subsection{General structure of solution for propagator }

~{\bf equation}

Let us omit possible non-linearities and consider ${\cal B}$ and ${\cal C}$ as given functions of $\eta$ and $k$, with ${\cal A}=ia^2\et$. This should be a reasonable approximation on time scales shorter than the (unknown) equilibration time associated to the non-linearities. We can construct the general solution of the propagator equation for the unequal time correlation function for $\Gamma^{(2)}$ given by eq. \eqref{G1}. Since $G_a$ is fixed by the inhomogeneous term this amounts mainly to a solution of eq. \eqref{G2}. Since the evolution equation for the equal time correlation function follows from eq. \eqref{G2}, the general solution of the propagator equation automatically solves the evolution equations \eqref{G3}, \eqref{G6}, \eqref{G7} for the equal time correlation functions.

Following the line of arguments in eqs. \eqref{64A}, \eqref{64B} we can write the general solution in the form 
\ba\label{GS1}
&&G_>(k,\eta,\eta')=\alpha_+(k)\www\et\ww(\eta')+\beta_-(k)\ww\et\www(\eta')\nn\\
&&\qquad +\beta_+(k)\ww\et\ww (\eta')+\alpha_-(k)\www\et\www(\eta'),
\ea
with $w^\pm_k\et$ two basis functions from which the general solution to the homogeneous equation $\tilde D_\eta\psi_k\et=0$ can be constructed,
\be\label{GS2}
\psi_k\et=c_+\ww\et +c_-\www\et~,~\tilde D_\eta w^\pm_k\et=0.
\ee
Here $w^\pm_k\et$ need not yet to be complex conjugate to each other. The inhomogeneous equation constrains the combination $\alpha_+(k)-\beta_-(k)$, 
\ba\label{GS3}
\partial_\eta G_{a_{|\eta=\eta'}}&=&\frac12\big(\alpha_+(k)-\beta_-(k)\big)\big(\www\et\ww(\eta')\nn\\
&&-\ww\et\www(\eta')\big)_{|\eta=\eta'}\nn\\
&=&-\frac{i}{2a^2}.
\ea

We can choose for each $k$ complex multiplicative constants for $w^\pm_k$ such that these functions are normalized according to 
\be\label{GS4}
\partial_\eta\big(\www\et\ww(\eta')-\ww\et\www(\eta')_{|\eta=\eta'}=-\frac{i}{a^2(\eta)}.
\ee
With this normalization the inhomogeneous equation requires
\be\label{GS5}
\alpha_+(k)-\beta_-(k)=1.
\ee
If $\ww$ and $(\ww)^*$ are independent we can choose $\www=(\ww)^*$. If not, we can always choose linear combinations of the basis functions such that $\www=(\ww)^*$. This is compatible with a purely imaginary l.h.s. of eq. \eqref{GS4}. With this choice of mode functions $w^\pm_k$ the reality condition implies the constraint \eqref{77A}. The general solution to the propagator equation is therefore given by eqs. \eqref{77C}, \eqref{77D}. 

Similar general solutions arise from the discussion of free quantum fields in a geometrical background. The function $\psi_k\et$ plays there the role of a wave function. For the operator formalism to be efficient it is crucial that the quantum fields are non-interacting. Eigenstates of a quantum Hamiltonian for interacting theories are very complicated multi-particle objects and a perturbative expansion around the free field theory becomes rapidly rather involved. In contrast, our formalism covers directly interacting theories as well. Interaction effects are accounted for by the computation of the effective action $\Gamma$. The latter can have a rather complicated form, involving arbitrary potentials or other non-linearities. Nevertheless, we have found a structure analogous to free quantum fields, based only on the assumption that non-linearities from backreaction and explicit boundary terms are absent and that higher order derivatives with respect to $\eta$ play no role for the second functional 
derivative $\Gamma^{(2)}$. 

Our formalism covers a much wider range of physical situations than free quantum fields in a given background. In particular, it includes mixed quantum states. Moreover, our evolution equations for the correlation functions are exact once $\Gamma^{(2)}$ or, equivalently, the functions ${\cal A}, {\cal B}$ and ${\cal C}$, are known or assumed. The mode functions $w^\pm_k\et$ appear in our treatment just as a convenient way to solve the propagator equation, not as fundamental objects. Our setting can be realized in a rather wide context of statistical systems as discussed in sect. \ref{Effective action and analytic continuation}. It applies to euclidean signature as well if we replace the factor $-i$ in eqs. \eqref{GS3}, \eqref{GS4} by $-1$ and adapt the reality condition. 

\section{Effective quantum fields and initial conditions}
\label{Effective quantum fields and initial conditions}

The analogy to a free quantum field allows us to interpret $\psi_k\et$ as an effective quantum field even in the presence of interactions. (This resembles the use of a free Fermi gas for electrons in a solid despite the presence of strong interactions.) This interpretation requires certain restrictions on the functions $\alpha(k), \beta(k)$ and $\gamma(k)$, however. If we want to associate $G(k,\eta,\eta')$ with the propagator for a pure state of a free effective quantum field $\psi_k(\eta)$ it has to obey
\be\label{GS6}
G_>(k,\eta,\eta')=\psi_k\et\psi^*_k(\eta'). 
\ee
With eqs. \eqref{GS1}, \eqref{GS2} this implies (for all $k$) $\alpha_+=|c_-|^2$, $\beta_-=|c_+|^2$, $\beta_+=c_+c_-^*$, $\alpha_-=c_-c_+^*$. The reality condition is obeyed and eq. \eqref{GS5} implies
\be\label{GS7}
|c_-|^2=1+|c_+|^2.
\ee
For $\alpha$, $\beta$ and $\gamma$ one finds the ``pure state conditions''
\ba\label{GS8}
\alpha&=&|c_-|^2+|c_+|^2=1+2|c_+|^2,\nn\\
\beta&=&c_+c_-^*+c_+^*c_-, \quad \gamma=-i\left(c_+c_-^*-c_+^*c_-\right).
\ea
This implies, in particular, $\alpha\geqslant 1$. Other constraints apply for $\beta$ and $\gamma$, e.g.
\be\label{GS9}
\beta^2+\gamma^2=\alpha^2-1.
\ee
For example, the case $\alpha=\beta =1$, which leads to a well defined de Sitter invariant Green's function in position space, cf. eq.\eqref{80E}, is not compatible with the pure state condition. For pure state initial conditions all deviations from the scaling solutions are oscillatory, e.g. $\alpha>1$ implies $\beta^2+\gamma^2>0$. For de Sitter space the pure states correspond to the so called $\alpha$-vacua \cite{EM,BA} with $|c_-|^2=\cosh^2\hat \alpha,|c_+|^2=\sinh^2\hat\alpha$. 

There is, however, no need to impose pure state initial conditions. Remaining within the language of quantum mechanics we may explore ``mixed states'' where
\be\label{GS10}
G_>(k,\eta,\eta')=\sum_{i} p_i\psi^{(i)}_k\et (\psi^{(i)}_k\left(\eta')\right)^*,
\ee
with $p_i\geq 0,~\sum_{i} p_i=1$. This yields
\be\label{GS11}
\alpha_+=\sum_{i}
 p_i|c^{(i)}_-|^2~,~\beta_+=\sum_i p_i c^{(i)}_+ \left( c^{(i)}_-\right)^*,
\ee
and similar for $\alpha_-$ and $\beta_-$. One has again the restrictions $\alpha_+\geq 0$, $\beta_-\geq 0$ which imply that $\alpha$ is bounded from below by $1$,
\be\label{GS12}
\alpha=1+2\sum_{i} p_i|c^{(i)}_+|^2.
\ee
The relation for $\beta$ and $\gamma$ becomes now 
\be\label{GS13}
\beta^2+\gamma^2=4|\beta_+|^2=4\sum_{i,j}p_ip_j c^{(i)}_+c^{(j)^*}_+ c^{(j)}_-c^{(i)^*}_-.
\ee
It is easy to see that for arbitrarily large $\alpha$ one can realize $\beta=\gamma=0$. The correlation function needs no longer to be oscillatory for $\alpha>1$. This shows that deviations of the initial spectrum from the Bunch-Davies vacuum needs not to induce oscillatory features in the power spectrum (as assumed in ref. \cite{BM1,BM2}.) Modifications of the initial spectrum are not necessarily related to particle production or non-adiabatic processes. For mixed states one finds an upper bound for $\beta^2+\gamma^2$, such that the restrictions for a ``mixed state initial condition'' are
\be\label{GS14}
\alpha \kkl \geq 1~,~\beta^2\kkl +\gamma^2\kkl \leq \alpha^2\kkl -1. 
\ee
This still excludes states with $\alpha=\beta,\gamma=0$.

The relations \eqref{GS14} imply a minimal value for the equal time correlation function for each $k$-mode. We demonstrate this for de Sitter space and $\gamma\kkl =0$, where the local minima of oscillations in eq. \eqref{77E} occur for 
\be\label{GS15}
G_{min}\kt=\big(\alpha\kkl -\beta\kkl \big)
\left(\frac{1}{2a^2 k}+\frac{H^2_0}{2k^3}\right).
\ee
For a given $\alpha\kkl $ the minimal value of $G_{min}$ is realized for maximal $\beta\kkl=\sqrt{\alpha^2\kkl -1}$, corresponding to a pure state. The overall amplitude $\alpha-\beta=\alpha-\sqrt{\alpha^2-1}$ is larger than one for $\alpha>1$. The lower bound on $G\kt $ reflects the Heisenberg uncertainty relations associated to quantum systems - they imply a non-zero minimum of fluctuations. Among the pure states the one with minimal fluctuations is the Bunch-Davies vacuum with $\alpha=1,\beta=\gamma=0$. Thus our conjectured scaling correlation is also singled out as an effective quantum vacuum with minimal fluctuations. This holds even in the presence of interactions.  

It is not clear if for a general interacting theory the constraints \eqref{GS14} remain necessary. Formally, their violation only means that no description of the correlation function in terms of an effective free quantum field is possible. For a weakly interacting quantum field theory the ``fluctuation constraint'' \eqref{GS14} should hold at least approximately. We will in the following assume the validity of the fluctuation constraint. The de Sitter invariant Green's function in position space $(\beta=\alpha,\gamma=0)$ can then not be realized exactly. For large $\alpha\kkl$, that we may associate with high occupation numbers of modes, the role of the ``quantum uncertainty'' becomes negligible and the constraint becomes effectively $\beta^2+\gamma^2\leq \alpha^2$. This amounts simply to the constraint $G\kt \geq 0$.

\section{Scalar correlation function in\newline inflationary cosmology}
\label{Scalar correlation function in inflationary cosmology}

In this section we investigate the scaling correlation according to our conjecture for realistic inflationary models. The propagator in momentum space takes then the product from \eqref{F7}. For $\eta>\eta'$ it is expressed in terms of $v^\pm_k$ \eqref{f2} as 
\be\label{147A}
G_>(k,\eta,\eta')=\frac{1}{a(\eta)a(\eta')}
v^{-}_k(\eta)v^+_k(\eta').
\ee
In this form it is straightforward to compute the propagator for geometries different from de Sitter space. It is sufficient that for large negative $\eta$ and $\eta'$ the flat space propagator $G_0$ becomes a good approximation. We can then solve the evolution equation for $v^\pm_k(\eta)$ taking their values for flat space as initial condition. Furthermore, once the mode functions $\w(\eta)=v^\pm_k(\eta)/a(\eta)$ are computed it is also straightforward to construct the general solution \eqref{77C}, \eqref{77D}. For an arbitrary homogeneous and isotropic cosmology it differs from the one for de Sitter space only by the difference in the mode functions. 

\subsection{Green's function for geometries close to de Sitter }

\vspace{-0.5cm}
~~{\bf space}

Our starting point for the computation of the Green's function for a massless scalar field in a large class of inflation models is the evolution equation \eqref{f4} for $v^\pm_k(\eta)$ with $\bar e^2=-1$. This is a real equation such that both $v$ and $v^*$ are solutions. We make the ansatz
\be\label{I1}
v^-_k(\eta) =\frac{1}{\sqrt{2k}}b(y) e^{-ik\eta}~,~v^+_k\et=\big (v^-_k\et\big)^*,
\ee
with 
\be\label{I2}
y=\frac{k}{\h\et}.
\ee
Typically, $\h\et$ vanishes for $\eta\to-\infty$ or $a\to 0$ such that in this limit we have the propagator in Minkowski space. This fixes the normalization
\be\label{I3}
\lim_{y\to\infty}b(y)=1.
\ee

The evolution equation \eqref{f4} reads for $b(y)$
\be\label{I4}
\left[
\partial^2_y+\frac{1}{1+\nu}
\left(2i+\frac{\partial\nu}{\partial y}\right)
\partial_y-
\frac{2+\nu}{(1+\nu)^2y^2}\right]
b(y)=0,
\ee
with
\be\label{I5}
\frac{\partial y}{\partial \eta}=-(1+\nu)k.
\ee
Here $\nu$ is a slowly time dependent function defined by 
\be\label{I6}
\partial_\eta\h=(1+\nu)\h^2.
\ee
For the particular case of powerlaw inflation with 
\be\label{I7}
a=(-H_0\eta)^{-\frac{1}{1+\nu}}=a_0t^{-\frac1\nu}
\ee
one has a constant $\nu < 0$, while de Sitter space corresponds to $\nu=0$. For more general models of inflation $\nu$ is a small quantity and obeys in a Robertson Walker metric
\be\label{I7a}
\nu=\frac{\dot H}{H^2}.
\ee

In the Robertson-Walker metric one has 
\be\label{I8}
y=\frac{k}{aH},
\ee
such that modes far outside the horizon correspond to $y\to 0$ and horizon crossing occurs for $y=1$. For constant $\nu$ the asymptotic behavior for $y\to 0$ is easily found by neglecting in eq. \eqref{I4} the term $\sim\partial_y$, namely
\be\label{I9}
\lim_{y\to 0}b(y) =b_0 y^{-\frac{1}{1+\nu}}.
\ee
This solution is an attractor for $y\to0$, with general solution
\be\label{XA}
\lim_{y\to 0}b\y=b_0y^{-\frac{1}{1+\nu}}
\left(1-\frac{i}{1+\nu}y+\dots \right)+b_1y^{\frac{2+\nu}{1+\nu}},
\ee
and $b_1$ the second integration constant. Inserting the asymptotic solution \eqref{I9} into eqs. \eqref{f2}, \eqref{I1}, \eqref{F7} yields the correlation function for modes far outside the horizon
\be\label{I10}
k^3G(k,\eta)=\frac{|b_0|^2k^2}{2a^2}\left(\frac{k}{aH}\right)^{-\frac{2}{1+\nu}}=
\frac{|b_0|^2H^2}{2}
\left(\frac{k}{aH}\right)^{\frac{2\nu}{1+\nu}}.
\ee

For the amplitude we first note that $b_0$ is a dimensionless number that only depends on $\nu$, with $b_0=i$ for $\nu=0$. Indeed, eq. \eqref{I4} contains no scale if $\partial \nu/\partial\eta=0$. The constant $b_0$ obtains from the solution of this equation with the normalization \eqref{I3}. We may employ the ansatz
\be\label{I12}
b\y =\left(1+i\frac{f(y)}{y}\right)^{\frac{1}{1+\nu}},
\ee
which becomes exact for $\nu=0$ with $f\y =1$. This yields the equal time propagator \eqref{AI1} mentioned in the introduction. In the following we will discuss the properties of the function $f(y)$ in more detail. 

The correct asymptotic behavior of $b(y)$ both for $y\to 0$ and $y\to \infty$ obtains if $f\y$ reaches constant values in these limits, with 
\be\label{I13}
b_0=\left[if(0)\right]^{\frac{1}{1+\nu}}.
\ee
The differential equation for $f(y)$ reads
\ba\label{I14}
&&\left[
(f-iy)(1-iy)-iy\nu\right]
\partial_y f+\frac{\nu y}{2}(\partial_y f)^2\nn\\
&&-\frac{1+\nu}{2}y(f-iy)\partial^2_y f+\frac{y}{2}(f-iy)(f-y\partial_y f)\partial_y\nu\nn\\
&&=if(1-f)+y\left(1-f+\frac{\nu}{2}\right).
\ea
By multiplication with $i$ this becomes a real differential equation for the variable $iy$. For $y\to 0$ eq.\eqref{I14} simplifies to 
\be\label{I15}
\partial_yf=i(1-f),
\ee
with solution
\be\label{I16}
f=1+c e^{-iy}.
\ee
This tends for $y\to 0$ indeed to a constant,
\be\label{I17}
|b_0|^2=|1+c|^{\frac{2}{1+\nu}}.
\ee

\begin{figure}[h!tb]
\centering
\includegraphics[scale=0.8]{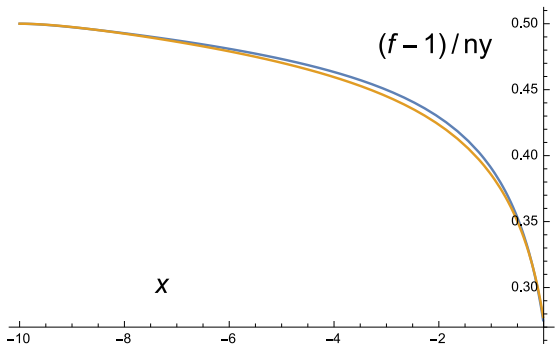}
\caption{Propagator near de Sitter space. We show $(f-1)/\nu$ as function of $x=iy$. The two curves for $\nu=0.01$ and $\nu=0.5$ are almost identical.}
\label{Figure_ff}
\end{figure}

The other limit for $y\to\infty$ reaches for constant $\nu$
\be\label{I18}
\partial_yf=\frac{1}{y}\left(f-1-\frac{\nu}{2}\right).
\ee
The only solution that is consistent with the normalization $b(y\to \infty)=1$ occurs for 
\be\label{I19}
f(y\to\infty)=1+\frac\nu2.
\ee
Thus both $b\y$ and $\partial_yb\y$ are fixed for $y\to\infty$ and we can integrate eq. \eqref{I4} unambiguously from some ``initial value'' for very large $y$ to $y\to 0$. We find that $(f-1)/\nu$ is almost independent of $\nu$, as shown in Fig.\ref{Figure_ff}. One obtains 
\be\label{167A}
f(y=0)=1+0.273\nu,
\ee
such that (for small $\nu$)
\be\label{167B}
|b_0|^2=\big(f(y=0)\big)^{\frac{2}{1+\nu}}\approx1+0.55\nu.
\ee

With eq. \eqref{I19} we can develop for $y\to\infty$ eq. \eqref{I14} iteratively, the next step being
\be\label{I20}
f=1+\frac\nu2-\frac{i\nu}{4y}
\left(1+\frac\nu2\right).
\ee
On the other end, for $y\to 0$, the solution \eqref{XA} implies
\be\label{X13}
f=-ib^{1+\nu}_0+
(i-b^{1+\nu}_0)y=f_0+iy(1-f_0),
\ee
where $f_0=- ib^{1+\nu}_0$ is real for constant $\nu$. For realistic inflationary models $|\nu|$ is small and $f(y)$ stays close to one. One therefore finds only a small correction, also in the case where $\nu$ depends on $\eta$,
\be\label{I21}
|b_0|^2=1+0(\nu).
\ee
If the $\eta$-dependence of $\nu$ remains moderate eq. \eqref{167B} is a good guide.

\subsection{Scalar Green's function far outside the horizon}

Eq. \eqref{f4} is valid for all times (as long as our assumptions hold and no other degrees of freedom play a role). We can therefore use its solution to compute the asymptotic behavior of the correlation function for a massless scalar for infinite time - or at least times long after horizon crossing. The case of massive scalars will be treated in appendix D. 

For geometries close to de Sitter space the time evolution for modes far outside the horizon is determined by the factor
\be\label{I22}
g\et=H^2(aH)^{-\frac{2\nu}{1+\nu}}=a^{-2}\h^{\frac{2}{1+\nu}}
\ee
in eq. \eqref{I10}. Inserting for constant $\nu$ eq.\eqref{I6} yields
\be\label{I23}
\partial_\eta\ln g\et=-2\h+\frac{2}{1+\nu}
\frac{\partial_\eta\h}{\h}=0.
\ee
We recover the known result that the fluctuation amplitude for modes far outside the horizon does not change anymore. To a good approximation we can evaluate the amplitude of the power spectrum at the time when a mode leaves the horizon, $y=1$, where eq. \eqref{I10} yields
\be\label{I24}
k^3G(k,a\to\infty)= \frac{|b_0|^2{\cal G}}{2}
H^2(y=1).
\ee
Here
\be\label{XC}
{\cal G}=\frac{g(a\to\infty)}{g(a=k/H)}
\ee
is a $k$-independent factor close to one. 

Corrections to this simple formula can be evaluated in our setting in an unambiguous way. For constant $\nu$ the relations \eqref{I10} and \eqref{I24} are exact since eq.\eqref{I4} has no explicit $k$-dependence and the asymptotic behavior \eqref{I9} fixes the $y$-dependence. 
For a varying $\nu(\eta)$ an additional $k$-dependence arises from $\partial_y\nu=-\partial_\eta\nu/[(1+\nu)k]$. 

The relation \eqref{I24} shows no explicit dependence on $k$ anymore, and it depends explicitly on $\nu$ only via the factor $|b_0|^2{\cal G}$. The $k$-dependence of the combination $k^3G(k,\eta)$ arises from the $k$-dependence of the value of the Hubble parameter at horizon crossing, $H(y=1)$. This is reflected by the shape of the pure scalar correlation, 
\ba\label{XX}
n_{\varphi}-1&=&\frac{\partial\ln (k^3 G(k,\eta))}{\partial\ln k}=2
\frac{\partial\ln H(y=1)}{\partial \ln k}\\
&=&2\frac{\partial \ln H}{\partial \ln (aH)}_{|y=1}
=2\frac{\partial\ln H/\partial t}{\partial \ln (aH)/\partial t}|aH=k.\nn
\ea
Using eq. \eqref{I7a} we find
\be\label{XY}
n_\varphi-1=\frac{2\nu}{1+\nu},
\ee
in accordance with the explicit $k$-dependence in eq. \eqref{I10}. This formula is exact for constant $\nu$. 

For realistic inflationary models eqs. \eqref{f4}, \eqref{I1}, \eqref{I4} cannot be used directly for the scalar fluctuations. We discuss below the modifications arising from the inflaton potential and the mixing with metric degrees of freedom. The tensor fluctuations of the massless graviton obey actually the same evolution equation as for a massless scalar, up to an overall normalization of the inhomogeneous term. Eqs.  \eqref{f4}, \eqref{I1}, \eqref{I4} can therefore be employed for the tensor power spectrum. Up to overall normalization eq. \eqref{AI1} describes then the graviton spectrum during inflation. This tensor power spectrum remains unchanged after the end of the inflationary epoch as long as the wavelength of the mode stays far outside the horizon. Once a tensor mode enters again the horizon in the radiation or matter dominated epoch the solution of the mode function will change qualitatively, leading to a fast decrease of the amplitude. 

\subsection{Inflaton potential}

Realistic models of inflation are often described in terms of an effective potential $V(\varphi)$ for the inflaton. This generalizes the effective action for the scalar field to
\be\label{m1}
\Gamma=\int_x e
\left\{\frac12\partial^\mu\varphi\partial_\mu\varphi+V(\varphi)\right\}.
\ee
The second functional derivative is extended to $(\bar{e}=i)$
\be\label{m2}
\Gamma^{(2)}=ia^2D+ia^4m^2(\bar\varphi),
\ee
where we have not written explicitly the unit matrix $\delta(x-y)$. The mass term,
\be\label{m3}
m^2(\bar\varphi)=\frac{\partial^2V}{\partial\varphi^2}\big(\bar\varphi(\eta)\big),
\ee
is evaluated for the homogeneous background scalar field $\bar\varphi(\eta)$ as determined by a solution to its field equation
\be\label{m4}
a^{-2}D\bar\varphi+\frac{\partial V}{\partial \varphi}(\bar\varphi)=0.
\ee

We emphasize that eq. \eqref{m2} holds for an arbitrarily non-linear inflaton potential $V(\varphi)$, thus including interactions. It is by no means restricted to a quadratic potential. The mass term $m^2$ may depend on time through the time dependence of the background field. For $m^2\neq 0$ the matrix $\Gamma^{(2)}$ is invertible if we restrict the function space according to our conjecture on the behavior of the analytic continuation or by imposing initial conditions. A constant field $\varphi_0$ no longer obeys $\Gamma^{(2)}\varphi_0=0$. A mass term is a simple way to regularize $\Gamma^{(2)}$.

The time evolution equation for the correlation function in momentum space remains valid, now with 
\be\label{m5}
(\partial^2_\eta+2\h\partial_\eta+k^2+a^2m^2)w_k(\eta) =0.
\ee
Let us consider models where for $\eta\to-\infty$, $a\to0$ the mass term remains bounded or increases less fast than $a^{-2}$. It can then be neglected for $\eta\to-\infty$, such that the normalization of $G$ can be computed from the ``initial value condition'' $G\kk\to G_0\kk$ for $\eta,\eta'\to-\infty$. As time increases, however, $a^2m^2$ increases and will finally overwhelm $k^2$. For a suitable potential inflation ends and the geometry extends to the range $-\infty<\eta<\infty$. 

Depending on the size of the dimensionless ratios
\be\label{m6}
\tilde m^2=\frac{a^2m^2}{\h^2}=\frac{m^2}{H^2}~,~\frac{\tilde m^2}{y^2}=\frac{a^2m^2}{k^2},
\ee
there are different regimes for the time evolution of $w_k$. We concentrate here on the case of constant (or slowly varying) $\tilde m^2$. The evolution equation for $v_k$ takes the form
\be\label{m7}
\big(\partial^2_\eta-(2+\nu-\tilde m^2)\h^2+k^2\big )v=0.
\ee
We therefore, replace in the last term in eq.\eqref{I4} the factor $2+\nu$ by $2+\nu-\tilde{m}^2$, and find for the leading behavior for $y\rightarrow0$
\be\label{M78}
\lim_{y\rightarrow0}b(y)=b_0y^{-1+\frac{\sigma}{2}},
\ee
with
\be\label{M7B}
\sigma=\frac{2\nu}{1+\nu}+\frac{3+\nu}{1+\nu}\left(1-\sqrt{1-\frac{4\tilde{m}^2}{(3+\nu)^2}}\right).
\ee
In consequence, the equal time correlation function obeys
\be\label{m8}
k^3G(k,\eta)=\frac{|b_0|^2H^2}{2}\left(\frac{k}{aH}\right)^{\sigma}.
\ee
For constant $\nu$ and $\tilde m^2$ this result is exact. It holds for arbitrary values of $\nu$ and $\tilde m^2$. For $\sigma>0$ the Fourier transform of $G(k,\eta)$ exists.  We obtain again 
\be\label{m9}
k^3G(k,a\rightarrow\infty)=\frac{|b_0|^2 \cal{G}}{2}H^2(y=1),
\ee
with $|b_0|^2$ and $\cal{G}$ adapted to eq.\eqref{m7}. The large time behavior of the correlation function for a massive scalar field is further discussed in appendix D. 

\subsection{Primordial cosmic fluctuations}

At this point we can make contact with the usual description of primordial scalar fluctuations in inflation. In the presence of gravitational fluctuations one  is interested in the fluctuations of the quantity
\be\label{m10}
{\cal R}_k=-\frac{H}{\dot{\bar\varphi}}w_k,
\ee
evaluated in zero spatial curvature gauge. This quantity is simply related to gauge invariant scalar fluctuations and has the property that it remains constant once a given $k$-mode has left the horizon. We can therefore employ the correlation function for $\cal{R}$, 
\ba\label{m11}
k^3 G_{{\cal R}}(k,y \rightarrow 0) 
=\frac{\tilde{c}H^4}{2\dot{\bar\varphi}^2}{|y=1},
\ea
with $\tilde{c}$ a factor close to one which is the analogue of $|b_0|^2\cal{G}$ in eqs.\eqref{I24}, \eqref{m9}
The scalar spectral index $n_s$ is defined by
\ba\label{m12}
n_s-1&=&\frac{\partial}{\partial\ln k}\ln\big[k^3G_{{\cal R}}(k,a\to \infty)\big]\nn\\
&=&4\frac{\partial\ln H}{\partial \ln(aH)}-\frac{\partial \ln\dot{\bar\varphi}^2}{\partial\ln (aH)}\nn\\
&=&\frac{4\nu}{1+\nu}-\frac{2\ddot{\bar\varphi}}{(1+\nu)H\dot{\bar\varphi}}.
\ea
All quantities in the last two lines of eq.\eqref{m12} are evaluated for $y=1$. 

The fluctuation amplitude is directly observable in the CMB. For the power spectrum
\be\label{190A}
P(k)=G_{\cal{R}}(k,\eta\rightarrow 0)
\ee
one uses conveniently the normalization
\be\label{190B}
\triangle^2(k)=\frac{k^3}{2\pi^2}P(k)=A_s(\frac{k}{k_s})^{n_s-1}.
\ee
The values observed by the Planck satellite are \cite{Planck}
\be\label{190C}
A_s=2.2\cdot10^{-9}, \quad k_s=0.05 Mpc^{-1}.
\ee
Comparing with eq.\eqref{m11} yields, approximating $\tilde{c}=1$,
\be\label{190D}
A_s=\frac{H^4}{4\pi^2\dot{\bar{\varphi}}^2} \big|y_s=1,
\ee
with $y_s=k_s/aH$.

For small $\nu$ and $\tilde m^2$ we recover the standard results of slow roll inflation. For a sufficiently flat potential eq. \eqref{m4} can be approximated by
\be\label{m13}
3H\frac{\partial\bar \varphi}{\partial t}+\frac{\partial V}{\partial \varphi}(\bar\varphi)=0,
\ee
and the gravitational field equation for Einstein gravity becomes (with $M$ the reduced Planck mass)
\be\label{m14}
3M^2H^2=V(\bar\varphi).
\ee
This allows us to express $\nu$ and $\tilde m^2$ in terms of the standard slow roll parameters $\epsilon$ and $\eta$
\be\label{m15}
\nu =\frac{\dot H}{H^2}=\frac{1}{6M^2H^3}\frac{\partial V}{\partial \varphi}\frac{\partial\bar\varphi}{\partial t}=-\frac{M^2}{2}
\left(\frac{\partial \ln V}{\partial \varphi}\right)^2=-\epsilon,
\ee
and 
\be\label{M16}
\tilde m^2=\frac{3M^2}{V}\frac{\partial^2V}{\partial\varphi^2}=3\eta.
\ee
With 
\ba\label{m17}
\dot{\bar\varphi}&=&-\frac{1}{3H}\frac{\partial V}{\partial \varphi},\nn\\
\ddot{\bar\varphi}&=&\frac{\nu}{3}\frac{\partial V}{\partial\varphi}-\frac H3\tilde  m^2\dot{\bar \varphi}
\ea
one finds the standard result
\be\label{m18}
n_s=1+6\nu+\frac23\tilde m^2=1-6\epsilon+2\eta.
\ee
Also for the amplitude the slow roll approximation yields the usual result
\be\label{196A}
\triangle^2(k)=\frac{V}{24\pi^2\epsilon M^4}\big|y_k=1.
\ee

We conclude that the solution of the propagator equation according to our conjecture on the behavior of the analytic continuation yields a scaling correlation that reproduces the usual discussion of fluctuations in inflationary cosmology. Results are often even exact and not only approximative. Nevertheless, we recall that the realization of the scaling correlation is not guaranteed. It has to be demonstrated that is is reached before horizon crossing, either by virtue of a non-linear equilibration process, or by restrictions on initial conditions. If not, the power spectrum will be modified as we discuss next.

\section{Impact of initial conditions on primordial fluctuations}
\label{Impact of initial conditions on primordial fluctuations}

In this section we turn back to the initial value problem and ask more quantitatively: how important is the influence of initial conditions for the primordial fluctuation spectrum in inflationary cosmology? We work here within an approximation of the effective action which keeps for the inflaton a covariant kinetic term and a potential. Possible non-linear symmetrization and dissipation effects, as discussed at the end of sect. \ref{Time evolution of correlation function}, are omitted. We investigate rather general initial conditions, that can be set many $e$-foldings before the observable fluctuations cross the horizon. Within this setting we find a substantial possible dependence of the fluctuation spectrum on the initial conditions. If  de Sitter symmetry is not imposed on the initial conditions rather arbitrary fluctuation spectra can be obtained for suitably chosen initial values. The inflaton potential no longer dictates the shape and amplitude of the primordial fluctuation spectrum. It rather distorts 
slightly a preexisting initial spectrum. This underlines that predictive power for inflationary cosmology either needs strong enough non-linear symmetrization and dissipation effects, or a detailed understanding of the selection of initial conditions. 

Our conclusions on the strong impact of initial conditions seems at first sight to differ from the suggestions in ref. \cite{AEH} that de Sitter geometry leads to an attraction towards universal physics even in the absence of non-linear terms. While the underlying equations are similar, the difference in conclusion originates from three observations. For a massive scalar field, we find indeed for every $k$-mode an asymptotic attraction to a universal form, in accordance with ref. \cite{AEH}. This asymptotic solution results from the exponential decay of one of the mode functions relative to the other. It does not fix the freedom of an independent amplitude for each $k$-mode. Still, the correlation function becomes unique asymptotically if one imposes scale or de Sitter symmetry. For realistic scalar masses in inflationary cosmology, however, this asymptotic attraction sets in only a long time after the fluctuations have crossed the horizon. At the time of horizon crossing, which is relevant for the observable fluctuation spectrum, the information from initial values is still present - memory is not yet lost. 

The second and third points concern the selection of initial values for a massless scalar field. Only a small subset of possible initial conditions is investigated in ref. \cite{AEH}. They are based on a particular assumption motivated by infrared finiteness. We do not impose such restrictions since the infrared behavior is not a severe problem for physical observables. Formally, it can always be cured by changing the initial spectrum for extremely small $k$ that are far outside the observable range. Since in our approximation the different $k$-modes follow separate evolution equations, a change of the spectrum in the extreme infrared does not affect the investigation for the observable fluctuations. As a third difference ref.\cite{AEH} considers pure quantum states, while our investigation admits mixed states as well. The setting in ref. \cite{AHR} is similar to ref. \cite{AEH}. Also ref. \cite{AHR} suggests a partial insensitivity of observables from the choice of initial conditions. This is again due to 
the selected observables and restricted initial conditions.

\subsection{Time evolution of correlation functions in} 

~~{\bf inflationary cosmology}

For a scalar field in a given geometry and with arbitrary inflaton potential one finds for the functions ${\cal A}$, ${\cal B}$ and ${\cal C}$ in the general form of the second functional derivative of the effective action \eqref{G1} 
\be\label{N1}
{\cal A}=1~,~{\cal B}=k^2+a^2m^2~,~{\cal C}=\h.
\ee
Here $m^2$ and $\h$ are functions of $\eta$ according to the particular geometry and cosmological solution for the inflaton mean value $\bar \varphi\et$. It is straightforward to solve the evolution equations for the correlation functions numerically. We employ dimensionless quantities $u=k\eta$ and 
\be\label{N3}
\tilde G_{\varphi\varphi}=2a^2 k G_{\varphi\varphi}~,~\tilde G_{\pi\varphi}=2a^2 G_{\pi\varphi}~,~\tilde G_{\pi\pi}=\frac{2a^2}{k}G_{\pi\pi}.
\ee
In these units the evolution equations become 
\ba\label{N4}
\partial_u\g_{\varphi\varphi}&=&-\frac{2\tilde h}{u}\g_{\varphi\varphi}+2\g_{\pi\varphi},\nn\\
\partial_u\g_{\pi\varphi}&=&\g_{\pi\pi}-\left(1+\frac{\hat m^2}{u^2}\right)\g_{\varphi\varphi},\nn\\
\partial_u\g_{\pi\pi}&=&\frac{2\tilde h}{u}\g_{\pi\pi}-2\left(1+\frac{\hat m^2}{u^2}\right)\g_{\pi\varphi},
\ea
with 
\be\label{N5}
\tilde h=-\h\eta=-\eta\frac{\partial \ln a}{\partial\eta}~,~\hat m^2=a^2\eta^2m^2.
\ee
For de Sitter space one has
\be\label{N6}
\tilde h=1~,~\hat m^2=\frac{m^2}{H^2_0}=\tilde m^2.
\ee

Eq. \eqref{N4} is a system of linear differential equations. In the limit of constant $\tilde h$ and $\hat m^2$ it is independent of $k$. In this case the dependence of the fluctuation spectrum on $k$ is completely described by the relations between $u$ and $\eta$ and by the relation \eqref{N3} between $\g$ and $G$. If $\tilde h$ and $\hat m^2$ depend on $\eta$ this induces an additional $k$-dependence through $\tilde h(u/k),\hat m^2(u/k)$. Typically, this type of dependence is only a small effect. For realistic inflationary models one has $\hat m^2/u^2\ll1$ before horizon crossing. With $\hat m^2/u^2$ of order one only a long time after horizon crossing the role of the mass is rather secondary for the issue if initial information is lost. 

For most of the time before horizon crossing $|\tilde h/u|$ is also small. In this range one has the general solution
\ba\label{N7}
\g_{\varphi\varphi}&=&\tilde \alpha(k)+\tilde \beta(k)\sin\big (2u+\delta(k)\big),\nn\\
\g_{\pi\varphi}&=&\tilde \beta(k)\cos \big(2u+\delta(k)\big),\nn\\
\g_{\pi\pi}&=&\tilde \alpha(k)-\tilde\beta(k)\sin\big(2u+\delta(k)\big).
\ea
The free functions $\tilde \alpha(k),\tilde\beta(k)$ and $\delta(k)$ specify the initial conditions. If initial conditions are set early enough, $\tilde\alpha(k)$ coincides with $\alpha(k)$ in the general solution \eqref{77E}. Similarly, $\beta(k)$ and $\delta(k)$ are related to $\beta(k)$ and $\gamma(k)$ by 
\be\label{N7A}
\beta(k)\cos (2u)-\gamma(k)\sin(2u)=\tilde\beta(k)\sin\big(2u+\delta(k)\big).
\ee
It is obvious that in the range $|u|\gg 1$ memory is not lost. The scaling correlation corresponds to $\tilde\beta=0$ and $\tilde \alpha=1$. Other initial conditions have rather arbitrary real functions $\tilde\alpha(k),\tilde\beta(k)$ and $\delta(k)$, obeying only mild restrictions. 

\subsection{Initial conditions}

If we set initial conditions sufficiently far in the past, i.e. for $|u|\gg 1$, we can use eq. \eqref{N7} for their characterization. Compatibility with the form of the leading expression $G_0(r,\eta,\eta')$ for the propagator in position space for $r\to 0,\eta\to\eta'$ (cf. eq. \eqref{20}) is achieved for 
\be\label{N8}
\lim_{k\to\infty}\tilde \alpha(k)=1,
\ee
provided the Fourier transform of the contributions from $\tilde \alpha(k)-1$ and from the term $\sim\tilde\beta(k)$ remain sufficiently subleading. The asymptotic form $G_0$ can be associated with the Hadamard condition for a well posed boundary problem in position space. (In momentum space the initial value problem es well posed anyhow.) For $\tilde \alpha(k)-1\sim k^{-f_1}$ the difference between $G(r,\eta)$ and $G_0(r,\eta)$ is proportional $r^{f_1-2}$ for $r\to 0$. It is subleading for $f_1>0$. Similarly, for $\tilde\beta(k)\sim k^{-f_2}$ and constant $\delta$ the contribution to $G(r)-G_0(r)$ remains subleading for $f_2>0$. We remark that these restrictions concern only the limiting behavior for $k\to\infty$, leaving much freedom in the range of finite $k$. For fixed $\eta$ the limit $k\to\infty$ corresponds to $u\to -\infty$ such that the approximation \eqref{N7} is indeed appropriate. 

A similar discussion can be done for the infrared behavior. For fixed $\eta$ the limit $k\to 0$ corresponds to $u\to 0$. Now the terms $\sim\tilde h$ and $\hat m^2$ matter for the issue of infrared finiteness. Specifying initial conditions for some large negative $\eta_{in}$ the modes relevant for the shape of the spectrum in the far infrared are already out of the horizon at this time. For these modes we can require suitable boundary conditions in order to have a well defined limit both in momentum and position space. The dynamics of such modes far outside the horizon is, however, of no concern to the fate of the modes with wavelength that become observable. The observable modes are all well within the horizon at $\eta_{in}$, with large $|u_{in}|$. Their horizon crossing occurs much later, when $u$ reaches $-1$. In a sense, the extreme infrared modes only form a background for the observable modes. Their influence can be incorporated effectively into the effective action $\Gamma$ that is relevant for the 
observable modes. 

Besides the conditions on the ultraviolet and infrared behavior which are not very restrictive for the observable modes, we may also require that the power spectrum should be positive at all times. The condition $\tilde G_{\varphi\varphi} \geq 0$ is obeyed for $|\tilde \beta(k)|<\tilde \alpha(k),\tilde \alpha(k)>0$. We will actually use the stronger ``fluctuation constraint'' \eqref{GS14} which amounts to 
\be\label{XXYY}
\tilde \alpha\kkl \geq 1~,~|\tilde\beta\kkl |^2\leq \tilde \alpha^2\kkl -1. 
\ee

\subsection{Processing of initial spectrum by geometry and}

~{\bf inflaton potential}

At a time close to horizon crossing, but still sufficiently before, say $u=-10$, the initial information is still fully preserved. The spectrum of fluctuations at this time can be associated with the initial spectrum. As a given mode goes through the horizon, say during the interval $-10<u<-0.1$, the details of the geometry and inflaton potential lead to a distortion of the initial spectrum. This typically violates scale symmetry since different $k$ modes reach the horizon for different values of $\eta$. Nevertheless, this ``processing'' of the initial spectrum near horizon crossing is a small effect since scale symmetry is approximately realized during inflation. Already a modest deviation of the initial spectrum from the ``universal form'' $\tilde \alpha=1,\tilde \beta=0$ can be comparable to the effect of this processing. In short, the physics relevant at horizon crossing does not ``produce'' the primordial fluctuation spectrum. It rather processes an ``initial spectrum'' that keeps memory of the 
``beginning of inflation''. The properties of the detailed geometry at the time of horizon crossing and the shape of the inflaton potential do only determine this processing. 

The situation is particularly simple for $\tilde\beta\kkl =0$. In this case every $k$-mode follows precisely the same evolution as for the scaling correlation. Only the amplitude is multiplied by $\tilde\alpha\kkl$. As a result, the correlation function reads
\be\label{N10}
G (\eta,k)=\tilde\alpha\kkl \bar G(\eta,k),
\ee
where $\bar G(\eta,k)$ denotes the scaling correlation \eqref{71A}. In consequence, the amplitude of the cosmic fluctuations
 is multiplied by $\tilde\alpha\kkl$. This can be a factor much larger than one. For the spectral index one finds
\be\label{N11}
n_s=\bar n_s+n_p,
\ee
with 
\be\label{N12}
n_p=\frac{\partial\ln \tilde \alpha\kkl}{\partial\ln k} 
\ee
and $\bar{n}_s$ the spectral index for $\tilde{\alpha}=1$.

\subsection{Example for initial conditions}

The initial conditions at the beginning of inflation may depend strongly on the particular model for the pre-inflationary era for our universe. It could be formation of a bubble in chaotic inflation, inflation could last forever since the infinite past, there may be a bounce and many more possibilities. Nevertheless, it seems useful to consider a definite example for an initial correlation function, such that issues of scales become apparent. 

For the beginning of inflation we consider a correlation function that approaches the propagator in Minkowski space for wavelengths much smaller than the Planck length, and has large classical fluctuations on length scales larger than the Planck length. Inbetween, there is a smooth crossover, with width $B$ in logarithmic $k$-space. Such a spectrum may be modeled by
\be\label{SA}
\alpha(k)=1+\frac{A}{2}(1-\frac{2}{\pi}arctg \hspace{0.1cm} x ), \quad x=B^{-1}\ln(\frac{k}{k_0}),
\ee
and $\beta(k)=\gamma(k)=0$. For the classical fluctuations on large length scales we assume an amplitude much larger than the one for a pure quantum state, $A\gg1$. Thus $\alpha(k)$ varies between one at short distances and $A+1$ at large distances.

We need the value of $x$ that corresponds to the observable fluctuations that leave the horizon around 60 e-foldings before the end of inflation. This value will depend on the number $N_{in}$ of e-foldings between the beginning of inflation and horizon crossing of the observable fluctuations. For the beginning of inflation we assume $k/k_0=\tilde{p}$, where $\tilde{p}$ is the physical momentum of a given fluctuation in units of the Planck mass $M$. We denote by $a_{in}$ the scale factor at the beginning of inflation, and $a_{hc}$ the one at horizon crossing. At the beginning of inflation one has
\be\label{SB}
\frac{k}{a_{in}}=\tilde{p}M,
\ee
while at horizon crossing one finds the physical momentum
\be\label{SC}
\frac{k}{a_{hc}}=H_0.
\ee
This yields
\be\label{SD}
\tilde{p}=\frac{H_0a_{hc}}{Ma_{in}}=e^{N_{in}}\frac{H_0}{M},
\ee
and therefore
\be\label{SE}
x=B^{-1}\left(N_{in}-\ln(\frac{M}{H_0})\right).
\ee
In terms of $a_{in}$ the crossover scale $k_0$ is given by
\be\label{SF}
k_0=a_{in}M.
\ee

With eq. \eqref{SE} for $x$ we can compute the amplitude and spectral index for the observable fluctuations. Assuming here for simplicity that the slow roll approximation is valid at horizon crossing one finds for the power spectrum
\be\label{SG}
\triangle^2(k)=\frac{\alpha(k)V}{24\pi^2\epsilon M^4}.
\ee
There is a substantial enhancement of the amplitude for $ x\lesssim A$, while for $x\gg A$ the Bunch-Davies spectrum becomes a good approximation. For the spectral index one finds
\be\label{SH}
n_s=1+n_p-6\epsilon+2\eta,
\ee
with
\be\label{SI}
n_p=\frac{\partial\ln\alpha(k)}{\partial\ln k}=-\left[B(1+x^2)(\frac{\pi}{2}-arctg \hspace{0.1cm} x)\right]^{-1}.
\ee
For $x\gg1$ this yields
\be\label{SJ}
n_p=-\frac{1}{B x}=-\left(N_{in}-\ln(\frac{M}{H_0})\right)^{-1},
\ee
with larger $|n_p|$ for smaller values of $x$. Small values $|n_p|\ll6\epsilon-2\eta$ require a large number of e-foldings $N_{in}$. Typically, a correction of less than 10\% of the standard result for $n_s-1$ needs for our example a rather large number of e-foldings between the beginning of inflation and horizon crossing $N_{in} \gtrsim 250$. This demonstrates the important role of an understanding of initial conditions for the quantitative interpretation of CMB-data in terms of the inflaton potential or similar quantities.

On the other hand, cosmologies with a very long, perhaps eternal, duration of the inflationary epoch of {\it our} universe can provide for an effective approach to the scaling correlation in a rather simple way. For $N_{in}\rightarrow\infty$ the effect of the initial conditions becomes negligible in our example. In this limit the relevant part of the initial spectrum is always the extreme ultraviolet tail. If this is given by the universal propagator for Minkowski space one finds at horizon crossing of the observable fluctuations the scaling correlation. This is in line with the general findings of sect. \ref{Time evolution of correlation function}. Initial values given by the scaling correlation lead to a scaling correlation for all later times. For observations it is sufficient that this holds for the relevant range in $k$. For $N_{in}\rightarrow\infty$ this relevant range is given by the ultraviolet limit $k\rightarrow\infty$, for which the universal propagator in Minkowski space can be 
motivated by various arguments. We may be able to look back to the beginning of inflation. What we may see, under certain circumstances, are simply the Lorentz-invariant fluctuations in flat space.

\section{Conclusions}
\label{Conclusions}

Do the observable primordial cosmic fluctuations keep memory of the ``beginning of the universe'' or the ``beginning of inflation''? The answer to this question is not yet entirely settled at present. This paper develops the conceptual setting for investigations that can clarify the situation. It addresses questions about the ``correct choice'' of the Green's function for de Sitter space, the ``correct choice'' of quantum vacua for time evolving situations or more generally the determination of a universal scaling correlation. 

The main unknown for a complete answer is the ``equilibration time'' it takes for an effective loss of memory of initial conditions for fluctuations in the observable range. At the present stage it is even not settled if equilibration takes place at all. Since interactions are small during inflation one expects, at least, that a possible equilibration time is very long, covering many e-foldings. Looking back towards the beginning of the universe, observations can detect signals from early stages of inflation, dating back for a time at least as long as the equilibration time. 

The key ingredient are time evolution equations for the correlation functions that permit to follow their fate for rather general ``initial conditions''. If memory of the detailed initial conditions is lost rapidly enough, a unique propagator is selected as an asymptotic attractor solution. Only in this case the choice of initial conditions is unimportant for an understanding of the cosmic fluctuations. Their properties will then reflect the physical situation in the epoch when the observable fluctuations leave the horizon. Observation of structure in the universe can therefore determine properties of the inflaton potential or similar quantities. If, in contrast, information about initial conditions still plays a role at horizon crossing, observations can only constrain some combination of physics at horizon crossing and ``initial conditions''.

The investigation of the present paper has been limited to a scalar field in a given time-evolving background geometry. Interactions are included in form of the inflaton potential. We have made an approximation where non-linear effects due to backreaction and an explicit dependence of the effective action on the correlation function are neglected. In this approximation we find that the information about initial conditions is not lost - equilibration time is infinite. Typically, infinitely many conserved quantities are an obstacle to equilibration. If our approximations are valid since the beginning of inflation, the observed anisotropies in the CMB reflect the initial fluctuation spectrum at the beginning of inflation, with only a modest processing at the time of horizon crossing. 

For a very long duration of inflation, perhaps even since the infinite past, we find it likely that our approximations do not remain valid and some equilibration occurs, at least for the range of wavelengths that are observable and which have been in the extreme ultraviolet in the remote past. For this case we formulate a conjecture that a unique correlation function is selected as an attractor solution. This is similar to the approach to thermal equilibrium. We give a criterion, based on the behavior of the analytically continued Green's function in the infinite past, which selects this universal scaling correlation uniquely for a given form of the effective action. For free scalar fields in de Sitter space this singles out the Bunch-Davies vacuum, while it can be extended to a large variety of situations, including interacting theories. 

A possible equilibration needs non-linear effects beyond our approximation, as backreaction or the explicit dependence of the effective action on $G$. 
Even without a detailed investigation of these effects some general features of the approach towards a unique correlation function become visible by our investigation. If the correlation function has reached a state with approximate de Sitter symmetry or scale symmetry already before horizon crossing, the shape of the power spectrum is close to the one assumed in the standard scenarios of inflation. This concerns, in particular, the spectral index. Symmetry is, however, not strong enough to fix the relation between the fluctuation amplitude and the Hubble parameter at the time of horizon crossing. For the latter some type of fluctuation-dissipation relation is needed. In the presence of de Sitter symmetry the possible deviations from the standard setting affect only the extracted value of the Hubble parameter at horizon crossing. We conclude that a rather rough approach of the correlation functions towards an equilibrated state is sufficient to realize the overall picture of fluctuations in inflationary cosmology. This is analogous to 
prethermalization \cite{PT} where only rough properties of correlation functions can reproduce some characteristic features of 
thermal 
equilibrium states. 

When it comes to precision and the determination of the inflaton potential by observation of the cosmic microwave background, the situation is less clear. The deviation of the spectral index $n_s$ from one is already a small quantity, such that rather modest memory of the initial state could distort it. For $n_s-1$ one has to compare the scaling violation arising from the time evolution of the Hubble parameter and the effective inflaton mass term (the parameters $\epsilon$ and $\eta$ in the slow roll approximation) with the size of effects induced by memory of the initial state. Only if the latter are negligible the usual determination of the inflaton potential by measurement of $n_s$ and the tensor to scalar ratio $r$ is justified.

Even if an approach towards a unique correlation function takes place (what we believe), the issue of the time scale for such an equilibration process remains important. For a scalar theory in a given background this ``equilibration time'' would diverge in the limit where the interactions go to zero. In the presence of metric fluctuations one has to asses the quantitative role of the gravitational interactions that are always present. For the interpretation of observations the time scale for the approach to a symmetric state (``symmetrization'' or ``generalized isotropization'' \cite{PT}) is particularly relevant.

Symmetrization and more generally equilibration is not expected to affect all wavelengths equally. The extreme infrared scales $k\rightarrow0$ are already outside the horizon very early in inflation and therefore not subject to equilibration processes. On the other hand, for the ultraviolet limit of the spectrum, $k\rightarrow\infty$, one may expect equilibration on comparatively short time scales. If couplings are taken dimensionless and $k/a$ exceeds all mass scales in the theory, one expects for the equilibration time a relation of the type $t_{eq}=\sigma/k$, with $\sigma$ a (possibly very large) dimensionless factor depending on couplings. This follows from dimensional analysis. (For vanishing interactions or in the presence of conserved quantities that obstruct equilibration $\sigma$ diverges.) In the extreme ultraviolet limit the evolution equation for the correlation function approaches the one for a free massless field in Minkowski space. It is reasonable to assume that for $k\rightarrow\infty$ the asymptotic solution is the free propagator in flat space. The factor $\sigma$ can be determined by a computation in flat space. 

If the number of e-foldings $N_{in}$ between the beginning of inflation and horizon crossing of the observable fluctuations tends to infinity, the above considerations select for the observable range of modes the scaling correlation according to our conjecture. This holds for any finite $\sigma$. For $N_{in}\rightarrow\infty$ the observable fluctuations correspond to $k\rightarrow\infty$. Also the time towards the beginning of inflation diverges, such that equilibration becomes arbitrarily accurate. On the other hand, for small $N_{in}$ the time since the beginning of inflation will not suffice to produce a strong equilibration. In this case the state of the universe at the beginning of inflation becomes observable. Inbetween, there will be a value ${N}^{(eq)}_{in}$ such that for $N_{in}\gg{N}^{(eq)}_{in}$ the memory of initial conditions is effectively lost, while for $N_{in}\lesssim{N}^{(eq)}_{in}$ observations are sensitive to the state of the universe at the beginning of inflation. The main open point for an answer to our question concerns the quantitative determination of ${N}^{(eq)}_{in}$. This may depend on the particular inflationary model. 

We have addressed the issue of the influence of initial conditions within the framework of the quantum effective action $\Gamma$. This offers the advantage that all effects of fluctuations are already incorporated in the effective action. The second functional derivative of $\Gamma$ yields an exact relation for the propagator and determines its time evolution. The complicated discussion of quantum vacua in a time-evolving situation and the ``correct choice'' of such a vacuum are avoided in our setting. Once the effective action is computed or assumed, the determination of the Green's function becomes an issue of ``classical'' field theory. Necessary approximations concern only the precise form of the effective action. 

Powerful functional methods for the computation of $\Gamma$ are available, as functional renormalization or the non-equilibrium two-particle irreducible effective action. We hope that the relevance of initial conditions for the cosmic fluctuations can be settled quantitatively by such methods.

\section*{Appendix A: Correlation function in{\newline{position space}}} 
\renewcommand{\theequation}{A.\arabic{equation}}
\setcounter{equation}{0}

In this appendix we discuss the propagator in position space. We first discuss the general short distance behavior. It is the same as for flat space. This limiting behavior for vanishing distance is often associated with the Hadamard property. We will see that the short distance behavior of the propagator does not fix it uniquely. All correlation functions discussed in this note actually share the property of a limiting Minkowski behavior, but differ from each other at larger distances. Next, we concentrate on the propagator of a massless scalar field in de Sitter space, discussing first the general short distance behavior and subsequently the Fourier transform of the scaling correlation in momentum space. The latter does not exist due to an infrared divergence of the momentum integral. We also investigate a proposed propagator that is well defined in position space and consistent with scale or de Sitter symmetry. It does not correspond to the scaling correlation. Even more, the associated initial conditions do not correspond to a mixed quantum state. 

\setcounter{subsection}{0}

\subsection{Short distance behavior of propagator}

The short-distance behavior of the propagator in position space is closely related to the one in Minkowski space. Indeed, the leading behavior of the Green's function $G(r,\eta,\eta')$ for $r\to 0$ and $\eta\to \eta'$ is given by
\be\label{20}
G_0(r,\eta,\eta')=
\frac{1}{4\pi^2 a(\eta)a(\eta')\big(r^2-(1-2i\epsilon)
(\eta-\eta')^2\big)},
\ee
which replaces in the free Green's function \eqref{18A} the factor $a^2$ by $a(\eta)a(\eta')$. With 
\ba\label{21}
G&=&G_0+\delta_1 G,\nn\\
D_0&=&(1+2i\epsilon)\partial^2_\eta-\Delta,\nn\\
D&=&D_0+2(1+2i\epsilon){\cal H}\partial_\eta,
\ea
we observe
\be\label{22}
i(1-i\epsilon)D_0
\big[a(\eta)a(\eta')G_0\big]=\delta(\eta-\eta')\delta^3(\vec r).
\ee
This yields
\ba\label{23}
D\delta_1 G&=&\frac{a(\eta')}{a^2(\eta)}
D_0\big[a(\eta)G_0\big]-D G_0\nn\\
&=&(1+2i\epsilon)({\cal H}^2+\partial_\eta{\cal H})G_0.
\ea
Thus $\delta G_1$ is determined by an inhomogeneous differential equation in terms of $G_0$. The r.h.s. is less singular than the $\delta$-distribution in eq.\eqref{13A}. The leading singularity is therefore taken care of by $G_0$.

For the particular geometry of a radiation dominated universe $(\eta>0)$, 
\be\label{24}
{\cal H}=\frac{1}{\eta}~,~a=c_1\eta=c_2\sqrt{t},
\ee
the r.h.s. of eq. \eqref{23} vanishes and $G_0$ is the exact propagator. For the general case we may explore near $r=0,\eta=\eta'$ a propagator of the form $G_0+G_1$, with
\ba\label{25}
G_1&=&\frac{K(\eta,\eta')}{32\pi^2 a(\eta)a(\eta')r}\ln
\left(\frac{r+(\eta-\eta')}{r-(\eta-\eta')}\right)^2,\nn\\
K(\eta,\eta')&=&{\cal H}(\eta)-{\cal H}(\eta')+
\int\limits^\eta_{\eta'} d\tilde\eta{\cal H}^2(\tilde\eta).
\ea
We note that $G_1$ is less singular for $r\to 0,\eta\to\eta'$ than $G_0$. (We have set $\epsilon=0$ for $G_1$, and similarly on the r.h.s. of eq. \eqref{23}.) Inserting eq. \eqref{25} into eq. \eqref{23} yields $(r\neq 0$ or $\eta\neq \eta'$)
\ba\label{26}
&&\delta_1G=G_1+\delta_2G,\nn\\
&&D(\delta_2 G)=({\cal H}^2+\partial_\eta{\cal H})G_0-DG_1\\
&&\quad =\big\{({\cal H}^2+\partial_\eta\h)K-2\h\partial_\eta\h-\partial^2_\eta\h\big\}
\frac{\ln\left(\frac{r+\eta-\eta'}{r-\eta+\eta'}\right)^2}{32\pi^2 a(\eta)a(\eta')r},\nn
\ea
where $\h$ stands for $\h(\eta)$. As compared to $D(\delta_1 G)$ in eq. \eqref{23} we see how the degree of divergence of $D(\delta_2 G)$ for $r\to 0,\eta\to\eta'$ has decreased further. We observe that $G_1$ vanishes for $\eta=\eta'$. The form of $G_1$ is not unique, however - the contribution $\delta_2G$ may have a similar divergence for $r\to 0,\quad\eta\to\eta'$. Furthermore, $DG_1$ contributes for $r=0,\eta=\eta'$ to an inhomogeneous term. This contribution has to be canceled by $\delta_2G$.

A unique short distance behavior in position space, as given by $G_0$, does not imply that the limit $k\rightarrow\infty$ in momentum space is unique. We will encounter examples where $G(k\rightarrow\infty, \eta, \eta')$ differs from $[2a(\eta)a(\eta')k]^{-1}$, while in position space $G_0$ is approached for $r\rightarrow0$, $\eta\rightarrow \eta'$.

\subsection{Short distance propagator in de Sitter space}
Let us consider the particular geometry of de Sitter space,
\be\label{36A}
\h=-\frac1\eta~,~a=-\frac{1}{H_0\eta},
\ee
where
\be\label{27}
K(\eta,\eta')=2\big(\h(\eta)-\h(\eta')\big).
\ee
(The range of $\eta$ is here $-\infty<\eta<0$, with $t=-\ln(-H_0\eta)/H_0$.) Eq. \eqref{26} simplifies to 
\be\label{28}
D(\delta_2G)=-\frac{H^3_0a(\eta)}{8\pi^2r}\ln
\left(\frac{r+(\eta-\eta')}{r-(\eta-\eta')}\right)^2.
\ee
For this case we may investigate
\be\label{M1a}
\delta_2 G=G_2+\delta_3 G,
\ee
with 
\ba\label{M1}
G_2(r,\eta,\eta')=&-&\frac{\h(\eta)\h(\eta')}{16\pi^2a(\eta)a(\eta')}
\left\{\ln\left(\frac{[r^2-(\eta-\eta')^2]^2}{r^4_0}\right)\right.\nn\\
&+&\frac{\eta-\eta'}{r}\ln
\left.\left(\frac{(r+\eta-\eta')^2}{(r-\eta+\eta')^2}\right)\right\}.
\ea
For the particular case of de Sitter space one finds $D(\delta_3G)=0$. 

The combination $\bar G=G_0+G_1+G_2$ is an exact solution of the propagator equation \eqref{13}, including the inhomogeneous term. The propagator
\ba\label{M2}
&&\bar G(r,\eta,\eta')=\\
&&\frac{H^2_0}{4\pi^2}\left\{\frac{\eta\eta'}{r^2-(\eta-\eta')^2}+\frac14\ln 
\left(\frac{r^4_0}{[r^2-(\eta-\eta')^2]^2}\right)\right\}\nn
\ea
involves an undetermined scale $r_0$ which corresponds to the possibility of a constant shift in $G$. The equal time propagator for $\eta'=\eta$ takes the simple form
\be\label{MB}
\bar G(r,\eta)=\frac{H^2_0}{4\pi^2}
\left\{\frac{\eta^2}{r^2}+\ln\left(\frac{r_0}{r}\right)\right\}.
\ee
The dependence on $\eta-\eta'$ is independent of the scale $r_0$. In particular, one has
\ba\label{M4}
&&G_2(r,\eta,\eta')-G_2(r,\eta,\eta)=-\frac{H^2_0}{16\pi^2}\times\\
&&\left\{\ln\left(\frac{[r^2-(\eta-\eta')^2]^2}{r^4}\right)+\frac{\eta-\eta'}{r}\ln
\left(\frac{(r+\eta-\eta')^2}{(r-\eta+\eta')^2}\right)\right\}.\nn
\ea

The function $\bar G(r,\eta,\eta')$ in eq. \eqref{M2} solves equation \eqref{13} - the contributions of $G_1$ and $G_2$ to the inhomogeneous term cancel. Nevertheless, $\bar G(r,\eta,\eta')$ may not be considered as an acceptable propagator for de Sitter space. In particular, $\bar G(r,\eta)$ turns negative for large $r$ and the Fourier transform is not defined. Furthermore, $\bar G(r,\eta,\eta')$ is not invariant under the symmetries of de Sitter space. One should therefore consider $\bar G(r,\eta,\eta')$ only as a good approximation for small $r$ and $(\eta-\eta')$. We will turn to this issue in sect. appendix E.

\subsection{Fourier transform of de Sitter propagator to}

~{\bf position space}

We next turn to the conjectured scaling correlation. The de Sitter propagator \eqref{71A} does not admit a Fourier transform to position space. If we look at the different pieces in eq. \eqref{F16} (with $f=H_0$) we see that the symmetric part of $\g_2$ diverges for $k\to 0$ as $k^{-3}$ such that the Fourier integral \eqref{A} becomes infrared divergent. Let us consider the form of the different pieces $G_0,\g_1$ and $\g_2$ in position space separately. The Fourier transform of $G_0$ reads for arbitrary $\bar e=\cos\varphi+i\sin \varphi$ and $\eta\geq \eta'$
\be\label{60}
G_0(r,\eta,\eta')=
\frac{1}{4\pi^2a(\eta)a(\eta')\big(r^2+\bar e^2(\eta-\eta')^2\big)},
\ee
in accordance with eq. \eqref{18B}. For $\g_1$ one obtains 
\ba\label{61}
&&\g_1(r,\eta,\eta')=\frac{H_0}{2\bar e}
\left(\frac{1}{a(\eta)}-\frac{1}{a(\eta')}\right)
\int_k
\frac{e^{i\vec k\vec r}\e}{k^2}\nn\\
&&=-\frac{iH_0}{16\pi^2\bar e r}
\left(\frac{1}{a(\eta)}-\frac{1}{a(\eta')}\right)\\
&&\quad \times\ln
\left(\frac{r^2+(\eta-\eta')^2+2r(\eta-\eta')\sin\varphi}{r^2+(\eta-\eta')^2-2r(\eta-\eta')\sin\varphi}\right)\nn\\
&&+\frac{H_0}{8\pi^2\bar er}
\left(\frac{1}{a(\eta)}-\frac{1}{a(\eta')}\right)
arctg
\left(\frac{2r(\eta-\eta')\cos\varphi}{(\eta-\eta')^2-r^2}\right).\nn
\ea
For $\varphi=\pi/2,\bar e=i$ the first term in eq. \eqref{61} equals $G_1$ in eq. \eqref{25}, whereas the second term vanishes.

We next consider the difference between $\tilde G_2(r,\eta,\eta')$ and its equal time counterpart $\tilde G_2(r,\eta,\eta)$,
\be\label{62}
\tilde G_2(r,\eta,\eta')-\tilde G_2(r,\eta,\eta)=\frac{H^2_0}{2}\int_k
\frac{e^{i\vec k\vec r}}{k^3}
(e^{-\bar ek(\eta-\eta')}-1).
\ee
This is a well defined Fourier integral for $Re(\bar e)\geq 0$. For Minkowski signature $\bar e=i$ the real part reads
\ba\label{63}
&&Re[\tilde G_2 (r,\eta,\eta')-\tilde G_2(r,\eta,\eta)]=\nn\\
&&\qquad\quad \frac{H^2_0}{16\pi^2}\left\{2\ln
\left(\frac{r^2}{(r-\eta+\eta')^2}\right)\right.\\
&&\qquad\quad \left.-\frac{r+\eta-\eta'}{r}\ln
\left(\frac{(r+\eta-\eta')^2}{(r-\eta+\eta')^2}\right)\right\}.\nn
\ea
This coincides with $G_2(r,\eta,\eta')-G_2(r,\eta,\eta)$ in eq. \eqref{M4}. For the equal time part of $\tilde{G}_2$ one has $(x=kr)$
\ba\label{65}
\g_2(r,\eta)&=&\tilde G_2(r,\eta,\eta)=-\frac{H^2_0}{4\pi^2 \bar e^{2}r}\int\limits^k_0
\frac{dk}{k^2}\sin (kr)\nn\\
&=&-\frac{H^2_0}{4\pi^2\bar e^2}\int\limits^\infty_0\frac{dx}{x^2}\sin x.
\ea
This integral is divergent, being formally independent of $r$ and $\eta$. Due to the lack of a well defined Fourier transform of $\tilde G_2$ the  propagator in position space can only be determined if $\Gamma^{(2)}$ is regulated. We will turn to this issue in appendix D, E. 

\subsection{Well defined propagator in position space?}

We do not consider the absence of a well defined propagator in position space as a very serious problem. For example, it disappears if we regularize the propagator by a small scalar mass. Also de Sitter space is presumably not realized exactly in nature, and modifications of the background geometry may solve the issue. Furthermore, the universal scaling correlation may not be realized in the far infrared. Nevertheless, it is reasonable to ask if a well defined propagator in position space can exist if we relax our condition based on analytic continuation and turn back to the more general family of propagators \eqref{77C}. We will take here $\alpha,\beta$ and $\gamma$ independent of $k$ in order to be consistent with the full continuous symmetry of de Sitter space. 

For general $\alpha$ and $\beta$ the problem will persist since the infrared behavior for $k\to 0$ is the same as for the propagator \eqref{71A} . There is, however, the special choice $\alpha=\beta$ (cf. eq. \eqref{77F}) for which the infrared divergence is canceled. Indeed, we find that for the propagator 
\ba\label{80E}
\hat G(k,\eta,\eta')&=&w^-_k(\eta)w^+_k(\eta')\\
&&+\frac12\big(w^+_k(\eta)w^+_k(\eta')+w^-_k(\eta)w^-_k(\eta')\big)\nn
\ea
the Fourier transform exists. For $\bar e=i$ it reads
\ba\label{80C}
&&\hat G(r,\eta,\eta')=\frac{1}{4\pi^2 a(\eta)a(\eta')}\nn\\
&&\qquad\times\left(\frac{1}{r^2-(\eta-\eta')^2}+\frac{1}{r^2-(\eta+\eta')^2}\right)\nn\\
&&\qquad+\frac{H^2_0}{16\pi^2}\ln
\left(\frac{\big (r^2-(\eta+\eta')^2\big)^2}{\big(r^2-(\eta-\eta')^2\big)^2}\right).
\ea
For the short distance behavior $r\to 0,\eta\to\eta'$ the Fourier transform of $\www(\eta)\ww (\eta')$ is given by $\bar G$ in eq. \eqref{M2}. The short distance behavior of the transform of $G_\beta=\big(\www\et\www(\eta')+\ww\et\ww(\eta')\big)/2$ obtains therefore as $\hat G-\bar G$, i.e.
\be\label{112Aa}
G_\beta(r,\eta,\eta')=\frac{H^2_0}{16\pi^2}
\left\{\ln \frac{16\bar \eta^4}{r^4_0}-1-\frac{1}{4\bar \eta^2}
\big(3r^2-(\eta-\eta')^2\big)\right\},
\ee
where $\bar\eta=(\eta+\eta')/2$. The role of $G_\beta$ is subleading, its dominant effect being to remove the free scale $r_0$ from $\hat G$. 

In contrast, we observe that $\hat G(k,\eta,\eta')$ has for small $k$ a behavior completely different from $G(k,\eta,\eta')$ in eq. \eqref{71A}. The equal time correlation reads
\ba\label{80F}
\hat G(k,\eta)&=&\frac{1}{2ka^2}\big(1+\cos (2k\eta)\big)+\frac{H^2_0}{2k^3}\big(1-\cos (2k\eta)\big)\nn\\
&&+\frac{H_0}{k^2a}\sin (2k\eta),
\ea
with a limiting behavior for modes outside the horizon $k\ll H_0 a,|k\eta|\ll 1$,
\be\label{80G}
\lim_{k\to 0}\hat G(k,\eta)=\frac{k^3}{9a^6H^4_0}.
\ee
The Green's function for $k\to 0$ looks completely different from the behavior $G\sim H^2_0/(2k^3)$ for the de Sitter propagator \eqref{71A}. On the other hand, at horizon crossing, $k/a=H_0,k\eta=-1$, one has
\be\label{100A}
\hat G(k\eta=-1)=\frac{H^2_0}{k^3}\big(1-\sin(2)\big)=0.091\frac{H^2_0}{k^3}.
\ee
This differs from the de Sitter propagator \eqref{71A} by  a substantial factor around ten, but leads to a spectrum with the same shape. It is the amplitude of fluctuations at horizon crossing that matters for the observable power spectrum, see sect.~\ref{Scalar correlation function in inflationary cosmology}. By virtue of the symmetries of de Sitter space the difference between the two propagators $G$ and $\hat G$ does not concern the shape of the spectrum. It only concerns the precise relation between the fluctuation amplitude and the Hubble parameter. We will discuss the propagator \eqref{80E} in more detail in appendix E.

Also the part $\sim\gamma$ in eq. \eqref{77C} is not infrared divergent and its Fourier transform exists. It may be added to the propagator $\hat G$.

\section*{Appendix B: Solution of evolution equation for equal time correlation function}
\renewcommand{\theequation}{B.\arabic{equation}}
\setcounter{equation}{0}

In this appendix we discuss the general properties of the solution of the evolution equation for the equal time Green's function. We concentrate on a massless scalar in de Sitter space. Our starting point is eq. \eqref{88F}. 

We can use the conserved quantity $\alpha_G$ in eq. \eqref{88I} in order to express $G_{\pi\pi}$ in terms of $G_{\varphi\varphi}$ and $G_{\pi\varphi}$,
\be\label{112A}
G_{\pi\pi}=\left(G^2_{\pi\varphi}+\frac{c_\alpha}{a^4}\right)G^{-1}_{\varphi\varphi},
\ee
such that 
\be\label{112B}
\partial_\eta G_{\pi\varphi}=(G^2_{\pi\varphi}+c_\alpha H^4_0\eta^4)G^{-1}_{\varphi\varphi}+\frac2\eta G_{\pi\varphi}-k^2 G_{\varphi\varphi}.
\ee
The remaining two equations \eqref{88F} can be combined in a non-linear second order differential equation for $G=G_{\varphi\varphi}$,
\be\label{112C}
\partial^2_\eta G-\frac2\eta \partial_\eta G-\frac12 
\frac{(\partial_\eta G)^2}{G}+2k^2 G-\frac{2c_\alpha H^4_0\eta^4}{G}=0.
\ee

\begin{figure}[h!tb]
\centering
\includegraphics[scale=0.8]{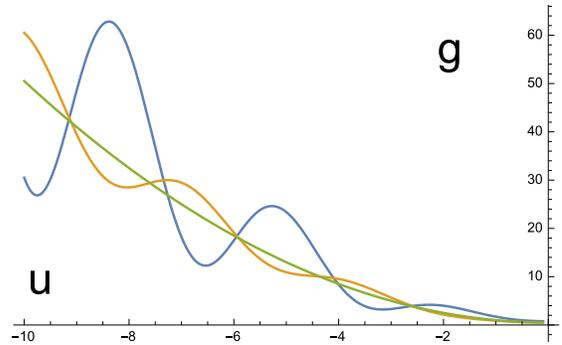}
\caption{Time evolution of $g$ as function of $u$, according to eq. \eqref{112E} with $Q=1$. For different initial conditions at $u=-10$ we observe oscillations around the scaling solution (central curve). Horizon crossing corresponds to $u=-1$.}
\label{propagatorflow_g2}
\end{figure}

\begin{figure}[h!tb]
\centering
\includegraphics[scale=0.8]{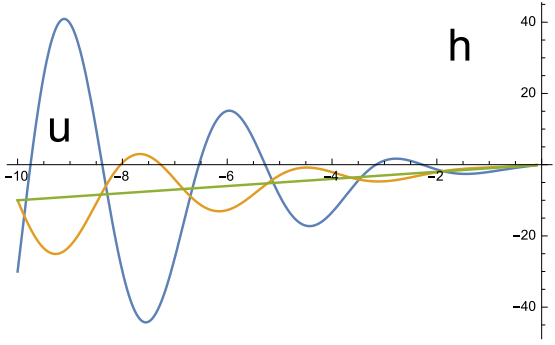}
\caption{Time evolution of $h$ as function of $u$. The straight line is the scaling solution.}
\label{Figure3}
\end{figure}

Using dimensionless variables and constants
\ba\label{112D}
g&=&G_{\varphi\varphi}k~,~h=2G_{\pi\varphi},\nn\\
u&=&k\eta=-\frac{k}{aH_0}~,~Q=\frac{4c_\alpha H^4_0}{k^4},
\ea
the evolution equation reads
\ba\label{112E}
\frac{\partial g}{\partial u}&=&h,\nn\\
\frac{\partial h}{\partial u}&=&\frac{2h}{u}-2g+\frac{h^2}{2g}+\frac{Q u^4}{2g}.
\ea
The dependence on the integration constant $c_\alpha$ or $Q$ can be absorbed by a rescaling
\be\label{112F}
\tilde g=\frac{g}{\sqrt{|Q|}}~,~\tilde h=\frac{h}{\sqrt{|Q|}}.
\ee
In terms of $\tilde g$ and $\tilde h$ eq. \eqref{112E} keeps the same form, with $Q$ replaced by $1$ (or $-1$ for negative $c_\alpha$). The numerical  solution shown in Figs. \ref{propagatorflow_g2}, \ref{Figure3} suggests an oscillating approach to the scaling solution 
\be\label{112G}
\tilde g=\frac{1+u^2}{2}~,~\tilde h=u,
\ee
which corresponds to eq. \eqref{88J}. For various different initial conditions for $\tilde g$ and $\tilde h$ at some initial $u_{in}$ (taken for Figs.\ref{propagatorflow_g2}, \ref{Figure3} as $u_{in}=-10$) the memory of the initial conditions seems at first sight to be partially lost for $|u|\lesssim 1$.

In order to investigate this issue more closely we introduce functions
\be\label{112H}
\hat g=\frac{2\tilde g}{1+u^2}~,~\hat h=\frac{\tilde h}{u},
\ee
such that (for $c_\alpha>0$)
\ba\label{112I}
\frac{\partial \hat g}{\partial u}&=&\frac{2u}{1+u^2}(\hat h-\hat g),\\
\frac{\partial \hat h}{\partial u}&=&u\left\{\frac{\hat h}{u^2}-\frac{(1+u^2)\hat g}{u^2}+\frac{\hat h^2}{(1+u^2)\hat g}+
\frac{u^2}{(1+u^2)\hat g}\right\}.\nn
\ea
One realizes the fixed point for $\hat h=\hat g=1$, which corresponds to the scaling solution. For any given mode with a given fixed $k$ the infinite past corresponds to large negative $u$. In the regime $u^2\gg 1$ one finds oscillations around the scaling solution with a damped amplitude. Indeed, for $u^2\gg 1$ we can approximate 
\be\label{112J}
\frac{\partial \hat g}{\partial u}=\frac{2}{u}(\hat h-\hat g)~,~\frac{\partial \hat h}{\partial u}=u
\left(\frac{1}{\hat g}-\hat g\right)
+\frac1u \left(\hat h+\frac{\hat h^2}{\hat g}\right).
\ee
During the oscillations $\tilde h$ and $\tilde g$ are of a similar order of magnitude and we employ
\be\label{112K}
\bar h=\frac{\hat h}{u}~,~\delta g=\hat g-1,
\ee
where
\ba\label{112L}
\partial_u\delta g&=&2\bar h,\nn\\
\partial_u\bar h&=&\frac{1}{1+\delta g}-(1+\delta g)+\frac{\bar h^2}{1+\delta g}.
\ea
Near the fixed point at $\delta g=0,\bar h=0$ we linearize the second equation \eqref{112L}, 
\be\label{112M}
\partial_u\bar h=-2\delta g,
\ee
and observe the oscillation with period $\Delta u=\pi$, 
\be\label{112N}
\delta g=c_g\sin (2u+\psi_g)~,~\bar h=c_g\cos (2u+\psi_g).
\ee
During the oscillation epoch the size of $\tilde g$ and $\tilde h$ decreases $\sim u^2$, reflecting the Hubble damping. Thus $G_{\varphi\varphi}\sim 1/(a^2k),G_{\pi\varphi}\sim 1/a^2$, with $G_{\pi\varphi}$ much larger in amplitude than the value \eqref{88J}. Except for the increase of $a$ this behavior resembles the one in flat Minkowski space. 

Once $u^2$ gets of the order one and smaller the oscillation stops and one encounters an overdamped behavior. Consider the combination
\be\label{112O}
\hat r=\frac{\hat h}{\hat g}-1
\ee
which obeys
\be\label{112P}
u\frac{\partial \hat r}{\partial u}=\hat r-\frac{u^2}{1+u^2}\hat r^2+\frac{u^4}{1+u^2}
\left(\frac{1}{\hat g^2}-1\right).
\ee
The solution for small $u^2$,
\be\label{112Q}
\hat r=c_r u~,~\hat h=(1+c_ru)\hat g
\ee
implies that the running of $\hat g$ stops for $u\to 0$,
\be\label{112R}
\frac{\partial\hat g}{\partial u}=2u(\hat h-\hat g)=2c_ru^2\hat g,
\ee
according to
\be\label{112S}
\hat g=\hat g_0\exp \left\{\frac{2 c_r u^3}{3}\right\}.
\ee
Numerically we observe indeed that $\hat g$ stops for $u\to 0$ at a value $\tilde g_0$ which depends on the initial conditions. The asymptotic behavior for $k\to 0$ is given by
\be\label{XX1}
G\va=\frac{\hat g_0\sqrt{c_\alpha}H^2_0}{k^3}.
\ee
The appearance of the factor $\hat g_0\sqrt{c_\alpha}$ demonstrates that the memory of initial conditions is not lost. 

The apparent approach of the numerical solution to the scaling solution visible in Figs.~\ref{propagatorflow_g2}, \ref{Figure3}  is mainly due to a combination of Hubble damping with the simple fact that $G\va$ approaches a constant for $u\to 0$. For the initial values shown this constant turns out to be only moderately different from the one in the scaling solution. The detailed memory of the initial conditions is not lost for the general solution. Since the integration constants $\hat g_0$ and $c_\alpha$ can depend on $k$, the spectrum of fluctuations needs not to be $\sim k^{-3}$. It retains information about the initial conditions. The same holds for the fluctuation amplitude.

The structure of partial fixed points and the qualitative behavior of deviations from them remains the same for more realistic inflationary geometries as the ones discussed in sect. \ref{Scalar correlation function in inflationary cosmology}. This demonstrates analytically that there is no loss of memory of initial conditions in our approximation of a derivative expansion.

\section*{Appendix C: Fluctuation-dissipation \newline{ relation}}
\renewcommand{\theequation}{C.\arabic{equation}}
\setcounter{equation}{0}

In this appendix we briefly discuss some aspects of the fluctuation-dissipation relation. The antisymmetric part of $G_>$ corresponds to the spectral function $\rho$
\be\label{FD1}
\rho(k,\eta,\eta')=2iG_a(k,\eta,\eta').
\ee
For a unique Green's function the symmetric part $G_s$ is related to $\rho$. Since $G_s$ describes the statistical information of the correlation function while $\rho$ contains information about decay rates of unstable particles, this connection is called a fluctuation-dissipation relation. For the most general solution of the propagator equation $G_s$ contains free parameters. Memory of these parameters has to be lost when the propagator approaches a universal form. The ``equilibration'' can therefore be seen as the dynamical establishment of a fluctuation-dissipation relation. 

Let us investigate this relation for the scaling correlation in de Sitter space \eqref{71A}, where
\ba\label{FD2}
\rho=H_0^2\left\{(\eta\eta'+\frac{1}{k^2})\frac{\sin(k(\eta-\eta'))}{k}\right.\nn\\
\left.-(\eta-\eta')\frac{\cos(k(\eta-\eta'))}{k^2}\right\},\nn\\
G_s=H_0^2 \left\{(\eta\eta'+\frac{1}{k^2})\frac{\cos(k(\eta-\eta'))}{2k}\right.\nn\\
\left.+(\eta-\eta')\frac{\sin(k(\eta-\eta'))}{2k^2}\right\}.
\ea
We introduce
\be\label{FD3}
\bar{\eta}=\frac{1}{2}(\eta+\eta'), \quad \eta_r=\eta-\eta', \quad \eta\eta'=\bar{\eta}^2-\frac{1}{4}\eta_r^2,
\ee
and make a ``finite interval Fourier transform'' with respect to the time difference $\eta_r$
\be\label{FD4}
\rho(k,\omega, \bar{\eta})=\int\limits_{-\epsilon}^{\epsilon}d\eta_re^{i\omega\eta_r}\rho(k,\bar{\eta}, \eta_r)
\ee
and similarly for $G_s$ The restrictions $\eta<0$, $\eta'<0$ for de Sitter space require $\epsilon<-2\eta$. One finds
\ba\label{FD5}
\rho&=&\frac{\pi H_0^2}{ik}\left(\bar{\eta}^2+\frac{1}{k^2}+\frac{1}{4}\frac{\partial^2}{\partial k^2}\right.\nn\\
&&\left. -\frac{1}{k}\frac{\partial}{\partial k}\right)(\hat{\delta}_{\epsilon}(\omega+k)-\hat{\delta}_{\epsilon}(\omega-k)),\nn\\
G_s&=&\frac{\pi H_0^2}{2k}\left(\bar{\eta}^2+\frac{1}{k^2}+\frac{1}{4}\frac{\partial^2}{\partial k^2}\right.\nn\\
&&\left.-\frac{1}{k}\frac{\partial}{\partial k}\right)(\hat{\delta}_{\epsilon}(\omega+k)+\hat{\delta}_{\epsilon}(\omega-k)),
\ea
with
\ba\label{FD6}
\hspace{-1.0cm}\hat{\delta}_{\epsilon}(\omega\pm k)&=&\frac{1}{2\pi}\int\limits_{-\epsilon}^{\epsilon}d\eta_re^{i(\omega \pm k)\eta_r}=\frac{\sin[\epsilon(\omega\pm k)]}{\pi(\omega\pm k)}.
\ea
One has the normalization
\be\label{FD6A}
\int\limits_{-\infty}^{\infty}d\omega\hat{\delta}_{\epsilon}(\omega\pm k)=1,
\ee
and for $\epsilon(\omega\pm k)\rightarrow\infty$ the function $\hat{\delta}_\epsilon$ becomes the $\delta$-distribution. With 
\ba\label{FD7}
G_>(k,\omega,\bar{\eta})&=&G_s(k,\omega,\bar{\eta})-\frac{i}{2}\rho(k,\omega,\bar{\eta}),\nn\\
G_<(k,\omega,\bar{\eta})&=&G_s(k,\omega,\bar{\eta})+\frac{i}{2}\rho(k,\omega,\bar{\eta}),
\ea
and the relations
\ba\label{FD8}
G_<(k,\omega,\bar{\eta})&=&G_>(k,-\omega,\bar{\eta}),\nn\\
G_s(k,-\omega,\bar{\eta})&=&G_s(k,\omega,\bar{\eta}),\nn\\
\rho(k,-\omega,\bar{\eta})&=&-\rho(k,\omega,\bar{\eta}),
\ea
one can easily reconstruct $G_s$ and $\rho$ from 
\be\label{FD9}
G_>(k,\omega,\bar{\eta})=\frac{\pi H_0^2}{k^3}\left(\bar{\eta}^2k^2+1+\frac{k^2}{4}\frac{\partial^2}{\partial k^2}-k\frac{\partial}{\partial k}\right)\hat{\delta}_{\epsilon}(\omega-k).
\ee

For $(\omega-k)^2\gg1/\epsilon^2$ the function $\hat{\delta}_{\epsilon}(\omega-k)$ oscillates fast, with period $2\pi/\epsilon$ and envelope decaying $\sim 1/(\pi(|\omega-k|)$. In this range the leading expressions for the derivatives are 
\ba\label{FD10}
\partial_k^2\hat{\delta}_{\epsilon}(\omega-k)&=&-\epsilon^2\hat{\delta}_{\epsilon}(\omega-k),\nn\\
\partial_k\hat{\delta}_{\epsilon}(\omega-k)&=&-\epsilon\hat{\delta}_{\epsilon}\left(\omega-k+\frac{\pi}{2\epsilon}\right).
\ea
For $k\epsilon\gg1$ the leading expression is
\be\label{FD11}
G_>=\frac{\pi H_0^2}{k}(\bar{\eta}^2-\epsilon^2)\hat{\delta}_{\epsilon}(\omega-k),
\ee
while for $k\epsilon\ll1$ one finds
\be\label{FD12}
G_>=\frac{\pi H_0^2}{k^3}(\bar{\eta}^2k^2+1)\hat{\delta}_{\epsilon}(\omega-k).
\ee
For fixed $\epsilon$ and $k$ the limit $\bar{\eta}\rightarrow-\infty$ yields the flat space result
\be\label{FD13}
G_>=\frac{\pi}{a^2(\bar{\eta})k}\hat{\delta}_{\epsilon}(\omega-k).
\ee
In the opposite limit $(\omega-k)^2\ll1/\epsilon^2$ one can expand
\be\label{FD14}
\hat{\delta}_{\epsilon}(\omega-k)=\frac{\epsilon}{\pi}\left(1-\frac{1}{6}\epsilon^2(\omega-k)^2+...\right),
\ee
such that
\ba\label{FD15}
G_>&=&\frac{\epsilon H_0^2}{k^3}(\bar{\eta}^2k^2+1)\nn\\
&-&\frac{\epsilon^3 H_0^2}{3k^3}\big\{2\bar{\eta}^2k^2(\omega^2-k^2)+2\omega^2-k^2\big\}.
\ea

For arbitrary initial conditions the fluctuation-dissipation relation \eqref{FD5} for the scaling correlation is not realized. Equilibration can thus be monitored by the degree to which the fluctuation-dissipation relation is dynamically realized. This corresponds to the dynamical realization of the particle number distribution in the approach to thermal equilibration.

\section*{Appendix D: Massive scalar propagator in \newline {de Sitter space}}
\label{Massive scalar propagator in de Sitter space}
\renewcommand{\theequation}{D.\arabic{equation}}
\setcounter{equation}{0}

The late time asymptotic behavior of the correlation function differs qualitatively between a massive or a massless scalar field. For a massive field the fast decrease of one of the mode functions leads to an effective loss of memory once the mass term becomes important. This loss of memory occurs already within our approximations, without the need of additional non-linear effects. We discuss this issue here for a de Sitter geometry. The qualitative behavior is similar for other homogeneous and isotropic cosmologies. 

\setcounter{subsection}{0}
\subsection{Scaling correlation}

The propagator for a massive scalar in de Sitter space can be obtained as a special case of the discussion in sect. \ref{Scalar correlation function in inflationary cosmology}. Indeed, taking in eq. \eqref{m7} the value $\nu=0$ and constant $\tilde{m}^2$, one obtains the mode equation for the correlation function of a free massive scalar in de Sitter space. In eq. \eqref{M7B} one has
\be\label{M18A}
\sigma=3-\sqrt{9-4\tilde{m}^2}.
\ee
For $\tilde{m}^2>0$ one has $\sigma>0$ such that the correlation function decreases as the scale factor $a$ increases. Eq.\eqref{m8} describes in this case the asymptotic behavior of the propagator of a massive particle in de Sitter space. 

This result is consistent with a simple direct computation. For $k/a\ll m$ we may use the field equation for a homogeneous scalar field perturbation in the Robertson-Walker metric,
\be\label{M18B}
(\partial_t^2+3H\partial_t+m^2)w=0.
\ee
For the variable $s=Ht=-\ln(-H\eta)=\ln a$ it takes the form
\be\label{M18C}
(\partial_s^2+3\partial_s+\tilde{m}^2)w=0.
\ee
For the general solution
\be\label{M18D}
w=c_1e^{\lambda_{1}s}+c_2e^{\lambda_2s},
\ee
with
\be\label{M18E}
\lambda_{1/2}=\frac{1}{2}\left(-3\pm\sqrt{9-4\tilde{m}^2}\right),
\ee
the leading behavior for $s\rightarrow\infty$ is given by the larger eigenvalue (with + sign in eq.\eqref{M18E}). The result
\ba\label{N18F}
|w|^2&=&|c_1|^2\exp\left\{(\sqrt{9-4\tilde{m}^2}-3)s\right\}\nn\\
&=&|c_1|^2a^{\sqrt{9-4\tilde{m}^2}-3}
\ea
agrees with eqs. \eqref{m8}, \eqref{M18A}.

For $\tilde m^2>0$ the momentum dependence of the equal time scalar Green's function in de Sitter space obeys for $k\to 0$
\be\label{202A}
G(k,\eta)\sim k^{\alpha-3}\sim k^{\frac{2\tilde m^2}{3}-3}
\ee
where we use $\tilde m^2\ll 1$ in the second expression. This cures the infrared divergence in eq. \eqref{65}. The Fourier transform of the de Sitter propagator exists for arbitrary small positive $\tilde m^2$. Such a mass term therefore regulates the propagator. 

\subsection{Time evolution}

In contrast to the massless case the scale invariant propagator of a massive scalar field in a de Sitter geometry is attracted towards a universal propagator for asymptotic time \cite{AEH}. Out of the three integration constants $\alpha,\beta,\gamma$ only one survives effectively. De Sitter symmetry or scale symmetry imply then that the spectrum is unique up to an overall constant. This behavior is in agreement with the findings of ref. \cite{AEH}. Without these symmetries one still has one free integration constant $\alpha(k)$ for every $k$-mode. 

The (partial) loss of memory is due to the fact that both eigenvalues $\lambda_1$ and $\lambda_2$ for the general solution \eqref{M18D} are real and negative. The component with the larger negative eigenvalue ($\lambda=-(3/2+\sqrt{9-4m^2}/2))$ dies out rapidly, such that a unique asymptotic solution for $w_k$ remains. The same holds for the general solution to the time evolution equation for the correlation function \eqref{88F}. We note, however, that this attractive behavior sets in once $k^2/a^2$ is of the order $m^2$. For small $\tilde{m}^2$ this happens only once modes are already outside the horizon. Before the term $k^2$ dominates in eq.\eqref{m7} and the solutions are oscillating similar to the solution for a free scalar field. For realistic inflationary cosmologies one has $\tilde m^2\ll 1$ and the asymptotic attractor property plays no role for the memory of initial conditions for the observable fluctuations \cite{CWneu}. 

\begin{figure}[h!tb]
\centering
\includegraphics[scale=0.6]{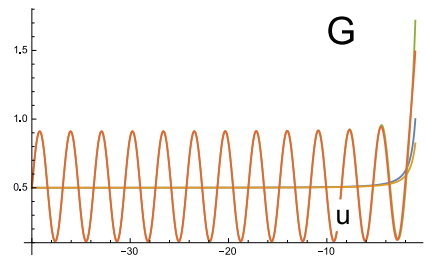}
\caption{Time evolution of correlation function G in units of $k^{-1}a^{-2}$, as a funtion of $u=k\eta$. The initial conditions correspond to the scaling correlation (smooth curves) and another (rather arbitrary) initial state (oscillating curves). The difference between the massive and massless case are only barely visible, except for the final increase for $u\rightarrow-1$.}
\label{masslarge1}
\end{figure}

\begin{figure}[h!tb]
\centering
\includegraphics[scale=0.6]{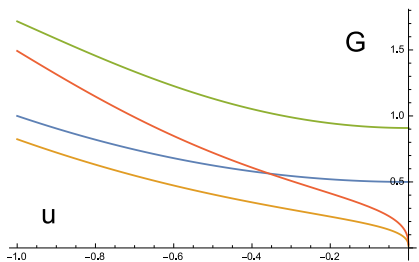}
\caption{Time evolution of G in units of $H_0^2/k^3$, for modes outside the horizon $(u \geqslant-1)$. Initial conditions are the same as for Fig. \ref{masslarge1}. The two curves for the massive scalar field approach the same asymptotic propagator, independent of initial conditions. In these units this propagator vanishes for $u\rightarrow0$. For a massless scalar field the propagator approaches a constant that depends on the initial condition.}
\label{masssmall1}
\end{figure}

We illustrate this behavior in Figs. \ref{masslarge1}, \ref{masssmall1} which show the evolution of $G$ as a function of $u=k\eta$. We plot two initial conditions both for a massless (blue and green) and a massive (red and orange) scalar field. One of the initial conditions corresponds to the scaling correlation (blue and orange), the other to a more arbitrary initial state (red and green). For $u\lesssim-5$ the effect of the mass term is completely insignificant, as visible in Fig.\ref{masslarge1} that shows $G$ in units of $k^{-1}a^{-2}$. In Fig.\ref{masssmall1} we focus on the behavior after horizon crossing $(u\geqslant-1)$ and plot $G$ in units of $H_0^2/k^3$. At horizon crossing, $u=-1$, the effect of the mass is visible for our choice of $\hat{m}^2=0.5$. The asymptotic behavior for $u\rightarrow0$ shows that for the massive scalar the correlation function becomes independent of initial conditions, in contrast to the massless case. For purposes of illustration we have taken a rather large mass term, $\hat{
m}^2=0.5$. 
For the smaller masses that are 
characteristic for realistic inflation both the separation between massive and massless propagators, and the approach of the massive propagator to the universal asymptotic form, set in for $u$ much closer to zero. The difference between the massive and massless scalar at horizon crossing would not be visible in the plot.

\section*{Appendix E: Symmetries}
\label{Symmetries}
\renewcommand{\theequation}{E.\arabic{equation}}
\setcounter{equation}{0}

In this appendix we discuss the consequences of symmetries. For a de Sitter background the scalar correlations that exhibit the symmetries of de Sitter space are partial fixed points of the time evolution. A possible ``symmetrization'' would describe the approach to these partial fixed points. We are therefore particularly interested in correlation functions with de Sitter symmetry. 

General covariance dictates the transformation properties of the correlation function under coordinate transformations. This restricts the functional dependence of $G(x,y)$ on the coordinates $x$ and $y$. In case of isometries of the background geometry one may require that the propagator is invariant under the symmetry transformations. Isometries can then impose rather severe restrictions on $G$. While Poincar\'e symmetry fixes $G$ uniquely for Minkowski space, this is not the case for the symmetries of de Sitter space. 

\setcounter{subsection}{0}

\subsection{Covariance under general coordinate}

~{\bf transformations}

The propagator in position space should not depend on the chosen coordinate system. A general coordinate transformation changes both the vierbein $e^m_\mu(x)$ (and the associated metric $\bar g_{\mu\nu}$) and the scalar field $\varphi(x)$. An infinitesimal transformation of the scalar translates directly to the correlation function which transforms as a scalar bilinear
\ba\label{S1}
\delta\tilde\varphi&=&-\xi^\mu(x)\partial_\mu\tilde\varphi(x),\\
\delta G(x,y)&=&-\left(\xi^\mu(x)\frac{\partial}{\partial x^\mu}+\xi^\mu(y)\frac{\partial}{\partial y^\mu}\right)G(x,y).\nn
\ea
This implies that $G(x,y)$ can only be a function of quantities that transform as scalar bilinears. An obvious example is the geodesic distance $s(x,y)$ between the two points $x$ and $y$, 
\ba\label{S2}
s(x,y)&=&\int\limits^{\tilde \tau(y)}_{\tilde \tau(x)}d\tilde \tau
\frac{\partial s(\tilde\tau)}{\partial\tilde\tau}~,~ds^2=\bar g_{\mu\nu}
\frac{\partial x^\mu(\tilde\tau)}{\partial\tilde\tau}\frac{\partial x^\nu(\tilde\tau)}{\partial\tilde\tau}d\tilde\tau^2,\nn\\
\delta s(x,y)&=&-\left(\xi^\mu(x)\frac{\partial}{\partial x^\mu}+\xi^\mu(y)\frac{\partial}{\partial y^\mu}\right)s(x,y),
\ea
with $x^\mu(\tilde \tau)$ on a geodesic parametrized by an invariant parameter $\tilde \tau$. The propagator in flat space \eqref{18B} has indeed the form 
\be\label{S3}
G(x,y)=\frac{1}{4\pi^2s^2}.
\ee

For a geometry with curvature there are other possible quantities transforming as scalar bilinears. For example, we may replace $ds$ by $d\tilde s$ where $\bar g_{\mu\nu}$ in the definition of $ds^2$ is replaced by $\bar R\bar g_{\mu\nu}$ or $\bar R_{\mu\nu}$ in the definition of $d\tilde s$. For the special case of a maximally symmetric space all geometric tensors are constructed from the metric, 
$\bar R_{\mu\nu}\sim\bar g_{\mu\nu},\bar R=$const, etc.. Then $s$ is the only independent bilinear function of $x$ and $y$, in the sense that all other covariant bilinears are functions of $s$. This applies to de Sitter space. A Green's function that only depends on geometrical quantities in de Sitter space (and not on boundary conditions or other fields) can therefore only depend on $s, G(x,y)=G(s)$.

\subsection{Global $SO(1,4)$-symmetry of de Sitter space}

If the background geometry admits isometries a unique scaling correlation must be invariant under the corresponding symmetry transformations. This is supposed to be realized as a consequence of the time evolution even if the initial correlation function does not exhibit this symmetry. Invariance under symmetry transformations constitutes an important restriction for the allowed function space for the Green's function. The geodesic distance is invariant under such symmetry transformations, e.g. the Poincar\'e transformations for flat space or $SO(1,4)$ transformations for de Sitter space. For $G(x,y)=G(s)$ the invariance of the propagator is guaranteed. 

For de Sitter space the four generators of the symmetry group $SO(1,4)$ beyond the three dimensional rotations and translations are \cite{FSS}
\ba\label{S4}
M_0&=&\eta\partial_\eta+x^j\partial_j,\\
M_i&=&H_0\big[2x_i\eta\partial_\eta+2x_ix^j\partial_j+(\eta^2-|\vec x|^2)\partial_i\big].\nn
\ea
A scalar bilinear $G(x,y)$ transforms under the corresponding infinitesimal transformation as 
\be\label{S5}
\delta G(x,y)=\big\{\alpha(M_0+\tilde M_0)+\beta_i (M_i+\tilde M_i)\big\}G(x,y),
\ee
with $\tilde M$ obtained from $M$ by $\eta\to\eta'$ $x_i\to y_i$. For functions $G(r,\eta,\eta')$ depending only on $r$ one has
\ba\label{S6}
M_0+\tilde M_0&=&\eta\partial_\eta+\eta'\partial_{\eta'}+r\delta_r,\nn\\
M_i+\tilde M_i&=&H_0(x_i+y_i)(M_0+\tilde M_0)+H_0(x_i-y_i)M_r,\nn\\
M_r&=&\eta\partial_\eta-\eta'\partial_{\eta'}+\frac{\eta^2-\eta'^2}{r}\partial_r.
\ea
An $SO(1,4)$ invariant Green's function must therefore obey $(M_0+\tilde M_0)G=0,~M_rG=0$. These relations hold for the combination
\be\label{S7}
z=\frac{r^2-(\eta-\eta')^2}{\eta\eta'},
\ee
e.g. $(M_0+\tilde M_0)z=0,M_rz=0$. The invariant $z$ is related to the geodesic distance in de Sitter space \cite{FSS}, \cite{SSV}
\ba\label{S8}
z=2(1-P)~,~P=
\left\{
\begin{array}{lll}
\cos(H_0s)&\text{for}&s^2\geq 0\\
\cosh(iH_0s)&\text{for}&s^2\leq 0
\end{array}.\right.
\ea
The propagator in de Sitter space is therefore only a function of $z,~G(r,\eta,\eta')=G(z)$. 

\subsection{De-Sitter symmetry in momentum space}

The action of the symmetry generators on scalar bilinears in Fourier space can be inferred by inserting eqs. \eqref{F3}, \eqref{F1A} into eq. \eqref{S5},
\ba\label{S9}
&&(M_0+\tilde M_0)G(k,\eta,\eta')=\hat M_0 G(k,\eta,\eta'),
\ea
and
\ba\label{S9A}
&&(M_i+\tilde M_i)\int\limits_k e^{i\vec k(\vec x-\vec y)}G(k,\eta,\eta')\nn\\
&&\quad=H_0\int\limits_k
e^{i\vec k(\vec x-\vec y)}
\big\{ (x_i+y_i)\hat M_0G\kk\nn\\
&&\quad \quad +i\frac{k_i}{k^2}\hat M_r G\kk,
\ea
with 
\ba\label{S10}
\hat M_0&=&\eta\frac{\partial}{\partial\eta}+\eta'\frac{\partial}{\partial\eta'}-k\frac{\partial}{\partial k}-3,\nn\\
\hat M_r&=&
\left(\eta\frac{\partial}{\partial\eta}-\eta'\frac{\partial}{\partial\eta'}\right)k
\frac{\partial}{\partial k}+
(\eta^2-\eta'^{2})k^2.
\ea
An invariant correlation function has to obey
\be\label{S11}
\hat M_0 G\kk=0~,~\hat M_r G \kk=0.
\ee
We observe that the symmetry transformations mix different $k$-modes due to the appearance of $\partial/\partial k$ in the generators. 

We may now discuss the symmetry transformations of the different pieces of the general propagator in eqs. \eqref{77C},\eqref{77D} with $w_k^{\pm}$ given by eq.\eqref{F15} . One finds that $G_a$ is invariant
\be\label{S12}
\hat M_0 \big[w^-_k(\eta)w^+_k(\eta')\big]=0~,~
\hat M_r\big[w^-_k(\eta)w^+_k(\eta')\big]=0.
\ee
(The invariance of $w^+_k(\eta)w^-_k(\eta')$ follows by complex conjugation.) With 
\be\label{215A}
\hat M_0 \big[\alpha(k)w^-_k(\eta)w^+_k(\eta')\big]=-k\frac{\partial\alpha(k)}{\partial k}w^-_k(\eta)w^+_k(\eta')
\ee
we find that the first term in $G_s$ in eq. \eqref{77C} is invariant provided that $\alpha$ does not depend on $k$. For the other pieces we compute $(\beta_+=(\beta+i\gamma)/2)$
\be\label{S13}
\hat M_0\big[\beta_+(k)w^+_k(\eta)w^+_k(\eta')\big]=-k
\frac{\partial\beta_+}{\partial k}w^+_k(\eta)w^+_k(\eta'),
\ee
such that symmetry requires a $k$-independent constant $\beta_+$. With
\ba\label{S14}
\hat M_r\big[w^+_k(\eta)w^+_k(\eta')\big]=0
\ea
we conclude that the combination $w^+_k(\eta)w^+_k(\eta')$ is also invariant under the symmetry of de Sitter space. By virtue of complex conjugation the same holds for $\www\et\www(\eta')$. We conclude that for $k$-independent $\alpha,\beta$ and $\gamma$ the general symmetric part $G_s$ in eq. \eqref{77C} is invariant under the de Sitter symmetry. In contrast to Poincar\'{e} symmetry for flat space, de Sitter symmetry does not fix the Green's function uniquely. One remains with a family of invariant propagators, parametrized by the three constants $\alpha$, $\beta$ and $\gamma$. 

\subsection{Scale invariance}

The generator $M_0$ corresponds to global scale transformations of coordinates. A scale invariant correlation function obeys
\be\label{242A}
(M_0+\tilde{M}_0)G=0.
\ee
In momentum space this simply means that $k^3G$ can only depend on the dimensionless combinations $k{\eta}$ and $k{\eta'}$. From eqs. \eqref{215A},\eqref{S13} it follows directly that scale symmetry alone implies that $\alpha$, $\beta$ and $\gamma$ are independent of $k$. Indeed, the combinations $k^3w_k^{\pm}(\eta)w_k^{\pm}(\eta')$ already depend only on $k{\eta}$ and $k{\eta'}$ and are therefore scale invariant. Any $k$-independence of $\alpha$, $\beta$, $\gamma$ would have to come along with a dependence on $\eta$ or $\eta'$ which is contradicting their role as integration constants. 

Since for $k$-independent $\alpha$, $\beta$, $\gamma$ the correlation functions are also independent under the full $SO(1,4)$ de Sitter symmetry, one concludes that scale symmetry, together with the symmetry properties and the propagator equation for a massless scalar, implies full $SO(1,4)$ invariance of the correlation function. (One should notice that this is not a pure symmetry property. There are many functions that depend only on $k{\eta}$ and $k{\eta'}$ and they all would lead to scale invariant $k^3G$. Only very particular ones are $SO(1,4)$ symmetric, however.) The impact of scale invariance extends to the propagator of a massive scalar field in de Sitter space. Scale symmetry implies that coefficients of the combinations $k^3w_k^{\pm}(\eta)w_k^{\pm}(\eta')$ must be $k$-independent, if $w_k^{\pm}=v_k^{\pm}/a$ are the two independent solutions of the homogeneous equation $Dw_k^{\pm}=0$, normalized according to eq.\eqref{147A} with $v_k^{\pm}=\tilde v^{\pm}(y)/\sqrt{2k}$. Indeed, for de Sitter space 
one 
has $y=-k\eta$ such that $k^3\w\et\w(\eta')$ is only a function of $k\eta$ and $k\eta'$. No further $k$-dependence of coefficients is allowed by coordinate scale symmetry.

We may recall here the different status of coordinate scale symmetry and dilatation symmetry. Coordinate scale symmetry is part of the general coordinate transformations with $\xi^{\mu}=\epsilon x^{\mu}$. The effective action is therefore invariant. This holds in the presence of arbitrary couplings with dimension of mass, as, for example, a scalar mass term. Coordinate scale symmetry can be broken, however, by the geometry and field values of a given cosmological solution. On the other hand, dilatation transformations are rescalings of fields according to their (canonical or anomalous) dimension. It is typically violated by couplings with dimension of mass since they single out a particular scale. Dilatation symmetry of the effective action restricts its form far beyond general coordinate invariance. The connection between the two symmetry transformations is that the field equations of a dilatation invariant effective action may have solutions that preserve dilatation symmetry (no spontaneous dilatation 
symmetry breaking). For such solutions coordinate scale symmetry is realized.

\subsection{Propagator in de Sitter space}

According to our conjecture the scaling correlation for a massless scalar field in de Sitter space has in momentum space (for $k\neq0$) the unique form eq.\eqref{71A}. The lack of regularization leads for $k\rightarrow0$ to an infrared behavior that does not admit a Fourier transform. In position space no propagator consistent with our conjecture and the symmetries of de Sitter space exists. On the other hand, one can find explicit solutions of the propagator equation in position space which are consistent with the symmetries. Since the issue of the correct choice of the propagator is debated \cite{EM,BA,BGO,HW,SSV,AP,EA,FSS} we sketch here a direct computation. This will reveal the connection to the general form of the propagator in momentum space. The problem of the absence of a propagator in position space for our conjectured unique scaling correlation disappears if one regularizes the inverse propagator by adding a small positive mass term. 

In position space the propagator can only depend on the invariant $z$, given by eq.\eqref{S7}, or the associated invariant
\be\label{PS1}
P=1-\frac{z}{2}=\frac{\eta^2+\eta'^2-r^2}{2\eta\eta'}.
\ee
For functions $G(P)$ depending only on $P$ one has 
\be\label{PS2}
DG=\frac{1}{\eta^2}\tilde{D}_P G,
\ee
with
\be\label{PS3}
\tilde{D}_P = (P^2-1)\partial^2_P+4P\partial_P.
\ee
If we add a mass term for the scalar field, $\tilde{m}^2=m^2/H^2_0$, the defining equation for the Green`s function reads
\be\label{PS4}
(\tilde{D}_P+\tilde{m}^2)G(P)=-iH^4_0\delta(\vec{r})\delta(\eta-\eta').
\ee

Let us first consider for $\vec{r}\neq 0$ or $\eta\neq\eta'$ the homogeneous equation
\be\label{PS5}
[(P^2-1)\partial^2_P+4P\partial_P+\tilde{m}^2]G(P)=0.
\ee
This equation is symmetric under the replacement $P\rightarrow-P$. It is convenient to employ
\be\label{PS6}
G=(1-P^2)^{-\frac{1}{2}}F, 
\ee
such that the homogeneous equation becomes 
\be\label{PS7}
\left\{(1-P^2)\partial^2_P-2P\partial_P+2-\tilde{m}^2-\frac{1}{1-P^2}\right\}F(P)=0.
\ee
This is the defining equation for the associated Legendre functions $P^1_{\nu}$ and $Q^1_{\nu}$, with
\be\label{PS8}
\nu(\nu+1)=2-\tilde{m}^2.
\ee
(The index $\nu$ of the associated Legendre functions should not be confounded with the symbol $\nu$ used in the preceding section for $\dot H/H^2$.)

The general solution of eq.\eqref{PS5} is therefore given by 
\be\label{PS9}
G=(1-P^2)^{-\frac{1}{2}}\left\{B_1P_{\nu}^1(P)+B_2Q^1_{\nu}(P)\right\}.
\ee
For a massless scalar field $(\tilde{m}^2=0)$ one has $\nu=1$ and the general solution of the homogeneous equation reads
\be\label{PS10}
\hat G=B_1+B_2\left\{\frac{P}{1-P^2}+\frac{1}{4}\ln\left(\frac{(1+P)^2}{(1-P)^2}\right)\right\},
\ee
corresponding to the proposal in ref. \cite{FSS}, see also \cite{AP}.
This solution has two poles at $P=1$ and $P=-1$, or $z=0$ and $z=4$. Comparing the behavior near $P=1$ with the free propagator yields $B_2=H^2_0/(4\pi^2)$, or
\be\label{PS11}
\hat G=\frac{H^2_0}{4\pi^2}\left\{\frac{1}{z}+\frac{1}{4}\ln\left(\frac{(4-z)^2}{z^2}\right)-\frac{1}{4-z}\right\}+B_1.
\ee
For $B_1=0$ this is identical with $\hat G$ in eq. \eqref{80C}. We observe $\hat G(z=2)=B_1$.

We next turn to the inhomogeneous equation 
\be\label{PS12}
\left\{(z-\frac{z^2}{4})\partial^2_z+(2-z)\partial_z\right\}G=\frac{iH^2_0\eta^4}{4}\delta(\vec{r})\delta(\eta-\eta').
\ee
For $\hat G$ the pole at $z=0$ generates the r.h.s. of eq.\eqref{PS12}, fixing the normalization \eqref{PS11}. The second pole at $z=4$ or $r^2=(\eta+\eta')^2$ produces a similar inhomogeneous term at $\eta+\eta'=0$. For the allowed range of $\eta<0,\eta'<0$ for de Sitter space this inhomogeneous term is not encountered. 
We conclude that $\hat G$ obeys the propagator equation. In the following we set $B_1=0$ such that eqs. \eqref{PS11} and \eqref{80C} coincide. The form of $\hat G$ in momentum space is given by eq. \eqref{80E}. The propagator $\hat G$ is not compatible with our conjecture for the scaling correlation. 

We have already discussed the power spectrum of primordial fluctuations for the propagator $\hat G$ in sect.~\ref{Propagator in momentum space}. It differs from the one of the Bunch-Davies vacuum by a suppression of the overall amplitude. There seems to be no obvious observational possibility to distinguish between $\hat G$ and the de Sitter propagator \eqref{71A}. The question which one of the propagators is ``correct'' is finally an issue of the time evolution of correlation functions as discussed in sect.~\ref{Time evolution of correlation function}. We observe that the form of $\hat G$ has its own diseases. The equal time correlation function turns negative in the region $r^2\lesssim4\eta^2\quad (z\lesssim4)$ and has a pole for $r^2=4\eta^2$. On the other hand the absence of a Fourier transform for the propagator \eqref{71A} disappears in the presence of a small scalar mass term. 

Indeed, the non-existence of the correlation function in position space is directly linked to the lack of invertibility of the inverse propagator $\Gamma^{(2)}$. For a massive scalar field, $\tilde{m}^2>0$, the inverse propagator has no longer a vanishing eigenvalue and the propagator exists both in position and momentum space. Adding an arbitrarily small mass term $\tilde{m}^2$ regulates the propagator. We have already discussed this propagator in momentum space in appendix D. For $0<\tilde{m}^2\ll1$ the behavior of the equal time Green's function for small $k$ is given by
\be\label{PS13}
\lim_{k\rightarrow0}G(k,\eta)=\frac{|b_0|^2H^2}{2k^3}\left(\frac{k}{aH}\right)^{\frac{2\tilde{m}^2}{3}}.
\ee
The Fourier transform $G(r,\eta)$ is well defined. Since the mass term respects the symmetries of de Sitter space $G(r,\eta,\eta')$ is only a function of $z$. It is a hypergeometric function, corresponding to the linear combination \eqref{PS9} with $\nu=1-\tilde{m}^2/3$. After imposing the normalization via the inhomogeneous term and the reality condition we remain with a family of de Sitter invariant propagators, labeled by a complex constant. They are associated with the ``$\alpha$-vacua'' \cite{EM,BA}.

For the scaling correlation the coefficients $B_1$ and $B_2$ should be chosen such that no pole at $P=-1$ occurs for $G(P)$, and the normalization corresponds to a leading behavior $G(P\rightarrow1)=H^2/(8\pi^2(1-P))$. A cancellation of the pole at $P=-1$ is not possible for $\tilde{m}^2=0$ since one of the basis functions for the most general homogeneous solutions is simply a constant.

\subsection{Special case $m^2=2H^2$}

An interesting special case for a massive propagator is $\tilde{m}^2=2$. In this case the general solution of the homogeneous equation \eqref{PS5} reads
\be\label{PS14}
G(P)=\frac{a_+}{P-1}+\frac{a_-}{P+1}, 
\ee
and the inhomogeneous equation \eqref{PS4} is obeyed for $a_-=0$, 
\ba\label{PS15}
G&=&\frac{H^2}{8\pi^2(P-1)}=\frac{H^2}{4\pi^2z}\nn\\
&=&\frac{1}{4\pi^2a(\eta)a(\eta')[r^2-(\eta-\eta')^2]}=G_0.
\ea
Indeed, the time evolution equation\eqref{m7} simplifies to 
\be\label{PS16}
({\partial_{\eta}}^2+k^2)v=0,
\ee
which is the same as for flat space. 



\newpage

\noindent

\bibliography{cosmic_fluctuations_from_quantum_effective_action}
\end{document}